\documentclass[tightenlines,superscriptaddress,showpacs,showkeys,preprint,
nofootinbib,floatfix]{revtex4}
\usepackage{graphicx} 
\usepackage{overpic} 
 
\begin{document}

\title{In-medium nuclear interactions of low-energy hadrons
\footnote{Physics Reports, in press}} 

\date{\today} 

\author{E.~Friedman\footnote{email: elifried@vms.huji.ac.il}}
\affiliation{Racah Institute of Physics, The Hebrew University, 
Jerusalem 91904, Israel\vspace*{1ex}}

\author{A.~Gal\footnote{email: avragal@vms.huji.ac.il}}
\affiliation{Racah Institute of Physics, The Hebrew University, 
Jerusalem 91904, Israel\vspace*{1ex}}

\begin{abstract} 

Exotic atoms provide a unique laboratory for studying strong 
interactions and nuclear medium effects at zero kinetic energy. 
Experimental and theoretical developments of the last decade in 
the study of exotic atoms and some related low-energy reactions 
are reviewed. The exotic atoms considered are of 
$\pi^-,~K^-,~{\bar p},~\Sigma^-$, and also the so far unobserved 
$\Xi^-$ atoms. The analysis of these atomic systems consists of 
fitting density dependent optical potentials 
$V_{\rm opt}=t(\rho)\rho$ to comprehensive sets of data of 
strong-interaction level shifts, widths and yields across the 
periodic table. These provide information on the in-medium 
hadron-nucleon $t$ matrix $t(\rho)$ over a wide range of 
densities up to central nuclear densities. For pions the 
review focuses on the extraction of the $\pi N$ in-medium $s$-wave 
interaction from pionic atoms, which include also the recently 
observed at GSI deeply bound $\pi^-$ atomic states in isotopes of 
Sn and Pb. Also included are recent measurements at PSI of elastic 
scattering of $\pi^{\pm}$ on Si, Ca, Ni and Zr at 21.5 MeV. 
The experimental results are analyzed in the context of chirally 
motivated $\pi$-nuclear potentials, and the evidence for partial 
restoration of chiral symmetry in dense nuclear matter is critically 
discussed. For antikaons we review the evidence from $K^-$ atoms, 
and also from low-energy $K^-p$ scattering and reaction data  
for and against a {\it deep} $\bar K$-nucleus potential 
of 150-200 MeV attraction at nuclear matter density. The case for 
relatively narrow deeply bound $K^-$ {\it atomic} states is made, 
essentially independent of the potential-depth issue. 
Recent experimental suggestions from KEK and DA$\Phi$NE (Frascati) 
for signals of $\bar K$-nuclear deeply bound states are reviewed, 
and dynamical models for calculating binding energies and widths 
of $\bar K$-{\it nuclear} states are discussed. 
For kaons we review the evidence, from $K^+$ total and reaction 
cross section measurements at the AGS (BNL) on Li, C, Si and Ca 
at $p_{\rm lab}=500-700$~MeV/c, for significant absorptivity of 
$t_{KN}(\rho)$ beyond that expected from $t_{KN}^{\rm free}$ 
within the impulse approximation. Attempts to explain the extra 
absorptivity for the relatively weak interaction of $K$ mesons 
in terms of a hypothetical exotic $S=+1$ pentaquark $\Theta^+$ 
strength are reviewed. For antiprotons the exceptionally broad data 
base due to the recent results of the PS209 collaboration at CERN 
are analyzed, together with results of radiochemical experiments. 
We discuss the dependence of the phenomenological $\bar p$-nucleus 
interaction on the model adopted for the neutron density, 
showing how the neutron densities favored by our comprehensive analysis 
are compatible with densities from other sources, including our own analysis 
of pionic atoms. It is also shown how the strong absorptivity of the 
$\bar p$-nucleus interaction, which leads to the prediction of saturation 
of widths in deeply-bound $\bar p$- atom states, also explains the observed 
saturation effects in low-energy $\bar p$ annihilation on nuclei. 
For $\Sigma$ hyperons we review the evidence, from continuum $\Sigma^-$ 
hypernuclear $(\pi^-,K^+)$ spectra obtained recently at KEK on C, Si, Ni, 
In and Bi, for substantial repulsion in the $\Sigma$-nucleus interaction, 
and the relationship to the inner repulsion established earlier 
from the density-dependence analysis of $\Sigma^-$ atoms and by analyses 
of past $(K^-,\pi^{\pm})$ AGS experiments. 
\newpage 
Lastly, for $\Xi$ hyperons we 
review prospects of measuring X-ray spectra in $\Xi^-$ atoms and thereby 
extracting meaningful information on the $\Xi$-nucleus interaction. The 
significance of the latter to the physics of $\Lambda \Lambda$ hypernuclei 
and to extrapolation into multistrange hypernuclei are briefly reviewed. 

\end{abstract} 

\maketitle 
 
\tableofcontents


\section{Introduction}
\label{sec:int} 

\subsection{Preview} 
\label{sec:preview} 

In 1997 we published together with Chris Batty a Physics Reports 
review {\it Strong Interaction Physics from Hadronic Atoms} 
\cite{BFG97} that has had a substantial impact on the progress made 
in the last decade on the study and understanding of in-medium nuclear 
interactions for various hadrons at low energy. The most spectacular 
advance on the experimental side has been perhaps the discovery and 
study of deeply bound pionic-atom states using the `recoil-free' 
$({\rm d},{^3{\rm He}})$ reaction near the pion production threshold on 
isotopes of Sn and Pb at GSI. Another potentially promising advance, 
very recently, concerns the as yet weak evidence for the existence of 
deeply bound antikaon-nuclear states gathered from stopped $K^-$ reactions 
studied at KEK and at DA$\Phi$NE, Frascati. 
These advances were triggered by, and have stimulated related theoretical 
work in which low-energy in-medium hadronic properties were considered in 
terms of a systematic chiral-perturbation expansion. This holds not only 
for pions, owing to their relatively small mass, but also for antikaons 
where the dominant effect of the $S=-1$ subthreshold quasi-bound state 
$\Lambda$(1405) was treated by a unitarized coupled-channel approach based 
on chiral-perturbation expansion. Significant advances in the study of 
nuclear interactions of other hadrons at low energy, for kaons, antiprotons 
and $\Sigma$ hyperons, have also been made. A good evidence for the wide 
interest within the nuclear physics and hadronic physics communities 
in these subjects is provided by the large number of recent topical reviews 
devoted to the overall theme of in-medium nuclear interactions of hadrons, 
including energy regimes higher than considered here. 
A representative list of such Reviews during the last five years covers 
the following subjects: {\it chiral symmetry in nuclei and dense nuclear 
matter} \cite{BRh02}, {\it pions in nuclei, a probe of chiral symmetry 
restoration} \cite{KYa04}, {\it chiral symmetry and strangeness at SIS 
energies} \cite{Lut04}, {\it medium modifications of hadrons - recent 
experimental results} \cite{Met05}, {\it kaon production in heavy ion 
reactions at intermediate energies} \cite{Fuc06}, {\it nucleon and hadron 
structure changes in the nuclear medium and impact on observables} 
\cite{STT07}. 

In the present Review we discuss and summarize the developments in 
understanding the in-medium properties of several hadron-nucleon systems 
at low energy as unraveled by our recent phenomenological studies and related 
ones by other authors in this field. This brief Preview subsection is 
followed by brief introductory subsections on wave equations and optical 
potentials, on nuclear densities, and on in-medium interactions.

\subsection{Wave equations and optical potentials} 
\label{sec:pot} 

The interaction of hadrons in nuclear medium of density $\rho$ is 
traditionally described by a dispersion relation based on the Klein-Gordon 
(KG) equation 

\begin{equation}\label{eq:disper} 
E^2 - {\bf p}^2 - m^2 - \Pi(E,{\bf p},\rho) = 0 \;, \;\;\;\;\;\; 
\Pi = 2EV_{\rm opt} \;, 
\end{equation} 
where $\Pi(E,{\bf p},\rho)$ is the hadron self-energy, or polarization 
operator and $V_{\rm opt}$ is the optical potential of the hadron in 
the medium \cite{EWe88}. Here $m$, ${\bf p}$ and $E$ are the rest mass of 
the hadron, its three-momentum and energy, respectively. For finite nuclei, 
and at or near threshold as applicable to most exotic-atom applications, 
Eq.~(\ref{eq:disper}) gives rise to the following KG equation: 

\begin{equation}\label{eq:KG} 
\left[ \nabla^2  - 2{\mu}(B+V_{{\rm opt}} + V_c) + (V_c+B)^2\right] \psi = 0~~
~~(\hbar = c = 1) 
\end{equation} 
where $\mu$ is the hadron-nucleus reduced mass, $B$ is the complex binding 
energy and $V_c$ is the finite-size Coulomb interaction of the hadron with 
the nucleus, including vacuum-polarization terms, added according to the 
minimal substitution principle $E \to E - V_c$. A term $2V_cV_{\rm opt}$ 
and a term $2BV_{\rm opt}$ were neglected in Eq.~(\ref{eq:KG}) with respect 
to $2{\mu}V_{\rm opt}$; the term $2BV_{\rm opt}$ has to be reinstated in 
studies of deeply-bound states. 

The simplest class of optical potentials 
$V_{\rm opt}$ is the generic $t\rho(r)$ potential, which for underlying 
$s$-wave hadron-nucleon interactions assumes the form: 

\begin{equation}\label{eq:Vopt} 
2\mu V_{\rm opt}(r) = - 4\pi(1+\frac{A-1}{A}\frac{\mu}{M})
\{b_0[\rho_n(r)+\rho_p(r)] + \tau_z b_1[\rho_n(r)-\rho_p(r)] \} \;. 
\end{equation} 
Here, $\rho_n$ and $\rho_p$ are the neutron and proton density 
distributions normalized to the number of neutrons $N$ and number of protons 
$Z$, respectively, $M$ is the mass of the nucleon and $\tau_z = +1$ for the 
negatively charged hadrons considered in the present Review{\footnote{$\tau_z 
= - 2t_z$ for isodoublets and $-t_z$ for isotriplets, where $t_z$ is the 
value of the $z$~th projection of isospin for the hadron considered.}}. 
In the impulse approximation, $b_0$ and $b_1$ are minus the hadron-nucleon 
isoscalar and isovector scattering lengths, respectively. 
Generally these `one-nucleon' parameters are functions of the density $\rho$, 
but often the density dependence may be approximated by fitting effective 
values for $b_0$ and $b_1$ to low-energy data. Extensions to situations which 
require `two-nucleon' terms representing absorption and dispersion on pairs 
of nucleons, or which are motivated by $p$-wave hadron-nucleon interactions, 
will be dealt with in the next section, for pionic atoms. 

For scattering problems, the applicable form of the KG equation is given by: 
\begin{equation} 
\label{eq:KGs}
\left[ \nabla^2  + k^2 - (2 \varepsilon^{(A)}_{\rm red}
(V_{\rm c} + V_{\rm opt}) - {V_{\rm c}}^2) \right] \psi = 0 
\end{equation} 
in units of $\hbar = c = 1$, where $k$ is the wave number in the 
center-of-mass (c.m.) system. 
For the simplest possible $t\rho$ $s$-wave term, the optical 
potential $V_{\rm opt}$ is of the form 
\begin{equation} 
\label{eq:Voptk} 
2 \varepsilon^{(A)}_{\rm red} V_{\rm opt}(r) = 
-4\pi F_A \{b_0[\rho_n(r)+\rho_p(r)] + \tau_z b_1[\rho_n(r)-\rho_p(r)] \}  ~~,
\end{equation} 
where $\varepsilon^{(A)}_{\rm red}$ is the c.m. reduced energy, 
\begin{equation} 
\label{equ:kin1} 
(\varepsilon^{(A)}_{\rm red})^{-1}=E_p^{-1}+E_A^{-1} 
\end{equation} 
in terms of the c.m. total energies $E_p$ for the projectile and $E_A$ for 
the target, and 
\begin{equation} 
\label{equ:kin2}
F_A = \frac {M_A \sqrt{s}}{M(E_A+E_p)}
\end{equation}
is a kinematical factor resulting from the transformation 
of amplitudes between the hadron-nucleon and the hadron-nucleus c.m. systems. 
Here $M_A$ is the mass of the target nucleus and $\sqrt s$ the total 
projectile-nucleon energy in their c.m. system. These forms of the potential 
and the equation take into account $1/A$ corrections, which is an important 
issue when handling light nuclear targets. The kinematical factor $F_A$ 
reduces at threshold to the kinematical term $(1+(1-1/A)\mu/M)$ 
appearing in Eq.~(\ref{eq:Vopt}) for the one-nucleon $s$-wave potential 
term in exotic-atom applications. 

For Fermions, such as antiprotons or $\Sigma$ hyperons, one might ask why 
the KG equation is here used in going into the relativistic domain 
instead of the Dirac equation. Indeed when interpreting experimental 
transition energies, in order to extract the strong interaction effects, 
it is essential to use the Dirac equation with finite size nuclear
charge distribution and vacuum polarization terms, e.g. Ref.~\cite{SHE98} 
for antiprotonic atoms. However, strong interaction effects are normally 
given as proper averages over the fine structure components.
The use of the KG equation rather than the Dirac equation is numerically 
justified when fine-structure effects are negligible or are treated in 
an average way, as for the X-ray transitions considered here.
The leading $j$ dependence ($j = l \pm \frac{1}{2}$) of the energy for 
solutions of the Dirac equation for a point-charge $1/r$ potential 
goes as $(j + \frac{1}{2})^{-1}$, and on averaging it over the projections 
of $j$ gives rise to $(l + \frac{1}{2})^{-1}$ which is precisely the 
leading $l$ dependence of the energy for solutions of the KG equation. 
The higher-order contributions to the spin-orbit splitting are suppressed 
by O$(Z \alpha /n)^2$ which, e.g., is of order 1$\%$ for the high-$n$ X-ray 
transitions encountered for antiprotons. It was checked numerically for few 
typical cases that the spin-orbit averaged shifts and widths thus obtained 
differ by less than 1\% from the $(2j+1)$-average of the corresponding 
quantities obtained by solving the Dirac equation. This difference is 
considerably smaller than the experimental errors placed on the measured 
X-ray transition energies and widths. 

\subsection{Nuclear densities} 
\label{sec:nucldens} 

The nuclear densities are an essential ingredient of the optical potential. 
The density distribution of the protons is usually considered known as
it is obtained from the nuclear charge distribution~\cite{FBH95} by
unfolding the finite size of the charge of the proton. The neutron 
distributions are, however, generally not known to sufficient accuracy. 
A host of different methods have been applied to the extraction of 
root-mean-square (rms) radii of neutron distributions in nuclei but the 
results are sometimes conflicting, e.g. 
Refs.~\cite{BFG89,GNO92,SHi94,KFA99,CKH03,JTL04}. 
For many nuclei there is no direct experimental information whatsoever 
on neutron densities and one must then rely on models. 
To complicate things further we note that there is a long history of
conflict between values of neutron rms radii derived 
from experiments using hadronic projectiles and neutron rms radii obtained 
from theoretical calculations. For that reason we have adopted a
semi-phenomenological approach that covers a broad range of possible
neutron density distributions.

\begin{figure}[t] 
\centerline{\includegraphics[scale=0.7,angle=0]{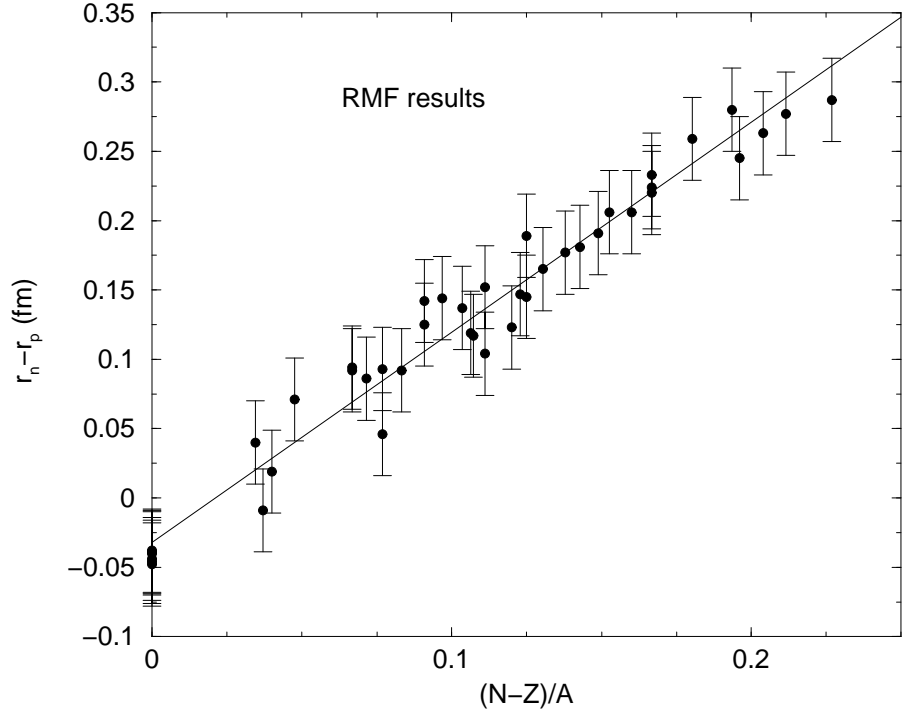}}
\caption{Fit of a linear expression in the asymmetry parameter
to RMF values of $r_n-r_p$.}
\label{fig:RMF}
\end{figure}

Experience with pionic atoms showed \cite{FGa03} that the feature of 
neutron density distributions which is most relevant in determining 
strong interaction effects in pionic atoms is the radial extent, as 
represented e.g. by $r_n$, the neutron density rms radius. Other features
such as the detailed shape of the distribution have only minor effect. 
For that reason we chose the rms radius as the prime parameter in the 
present study. Since $r_p$, the rms radius for the proton density 
distribution, is considered to be known, we focus attention on values 
of the difference $r_n-r_p$. A linear dependence of $r_n-r_p$ 
on $(N-Z)/A$ has been employed in $\bar p$ studies \cite{TJL01,JTL04,FGM05}, 
namely 
\begin{equation} \label{eq:RMF} 
r_n-r_p = \gamma \frac{N-Z}{A} + \delta \; ,
\end{equation}
with $\gamma$ close to 1.0~fm and $\delta$ close to zero. 
The same expression with $\gamma$ close to 1.5~fm was found \cite{FGa03} 
to represent well the results of relativistic-mean-field (RMF) 
calculations \cite{LRR99} for stable nuclei, as shown in Fig.~\ref{fig:RMF}, 
but these values of $r_n-r_p$ are larger by about 0.05-0.10 fm than the 
`experimental' values in medium-weight and heavy nuclei used in 
recent relativistic Hartree-Bogoliubov (RHB) versions of mean-field 
calculations \cite{NVF02,LNV05}. Expression (\ref{eq:RMF}) has been 
adopted in the present work and, for lack of better global information
about neutron densities, the value of $\gamma$ was varied over a reasonable
range in fitting to the data. This procedure is based on the expectation 
that for a large data set over the whole of the periodic table some local 
variations will cancel out and that an average behavior may be established. 
Phenomenological studies of in-medium nuclear interactions are based on 
such averages. 

In order to allow for possible differences in the shape of the neutron
distribution, the `skin' and `halo' forms of Ref.~\cite{TJL01} were
used, as well as an average between the two. Adopting a two-parameter
Fermi distribution both for the proton (unfolded from the charge distribution)
and for the neutron density  distributions 
\begin{equation} 
\label{eq:2pF} 
\rho_{n,p}(r)  = \frac{\rho_{0n,0p}}{1+{\rm exp}((r-R_{n,p})/a_{n,p})} \; ,
\end{equation}
then for each value of $r_n-r_p$ in the `skin' form the same diffuseness 
parameter for protons and neutrons, $a_n=a_p$, was used and the 
$R_n$ parameter was determined from the rms radius $r_n$. In the `halo' 
form the same radius parameter, $R_n=R_p$, was assumed and $a_n^{\rm h}$ 
was determined from $r_n$. In the `average' option the diffuseness parameter 
was set to be the average of the above two diffuseness parameters, 
$a_n^{{\rm ave}}=(a_p+a_n^{\rm h})/2$, and the radius parameter $R_n$ was 
then determined from the rms radius $r_n$. In this way we have used three 
shapes of the neutron distribution for each value of its rms radius all 
along the periodic table. These shapes provide sufficient difference 
in order to be tested in global fits. 

\begin{figure}
\includegraphics[scale=0.6,angle=0]{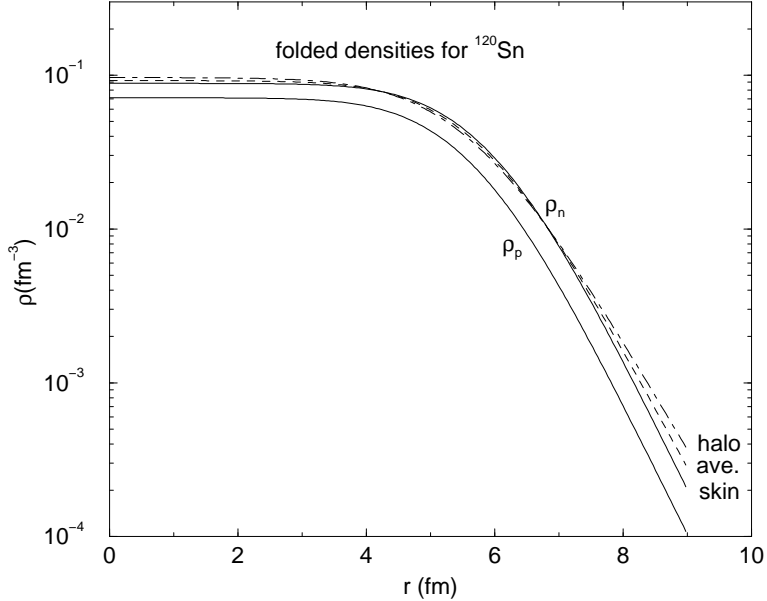} 
\caption{Proton and neutron finite-range folded densities for $^{120}$Sn 
with $\beta=0.85$~fm. Neutron densities are calculated for $\gamma=1.2$~fm, 
see Eqs.~(\ref{eq:RMF},~\ref{eq:fold}).} 
\label{fig:rhoSn}
\end{figure} 

Another sensitivity that may be checked in global fits is to the radial 
extension of the hadron-nucleon interaction when folded together with 
the nuclear density. The resultant `finite range' density is defined as 
\begin{equation}
\label{eq:fold} 
\rho ^{\rm F}(r)~~=~~\int d{\bf r}' \rho({\bf r}') \frac{1}{\pi ^{3/2} \beta^3}
e^{-({\bf r}-{\bf r'})^2/\beta^2}~~,
\end{equation}
assuming a Gaussian interaction. Other forms such as a Yukawa function
may also be used. 
Figure \ref{fig:rhoSn} shows for example finite-range folded proton and 
neutron densities in $^{120}$Sn, calculated using the three models 
listed above for generating neutron densities. The difference between these 
three models becomes pronounced from about 8~fm on, a radial extent to which 
$\bar p$ atoms are particularly sensitive. 

\subsection{In-medium interactions}
\label{sec:inmedium} 

\begin{figure}[t] 
\centerline{\includegraphics[scale=0.6,angle=0]{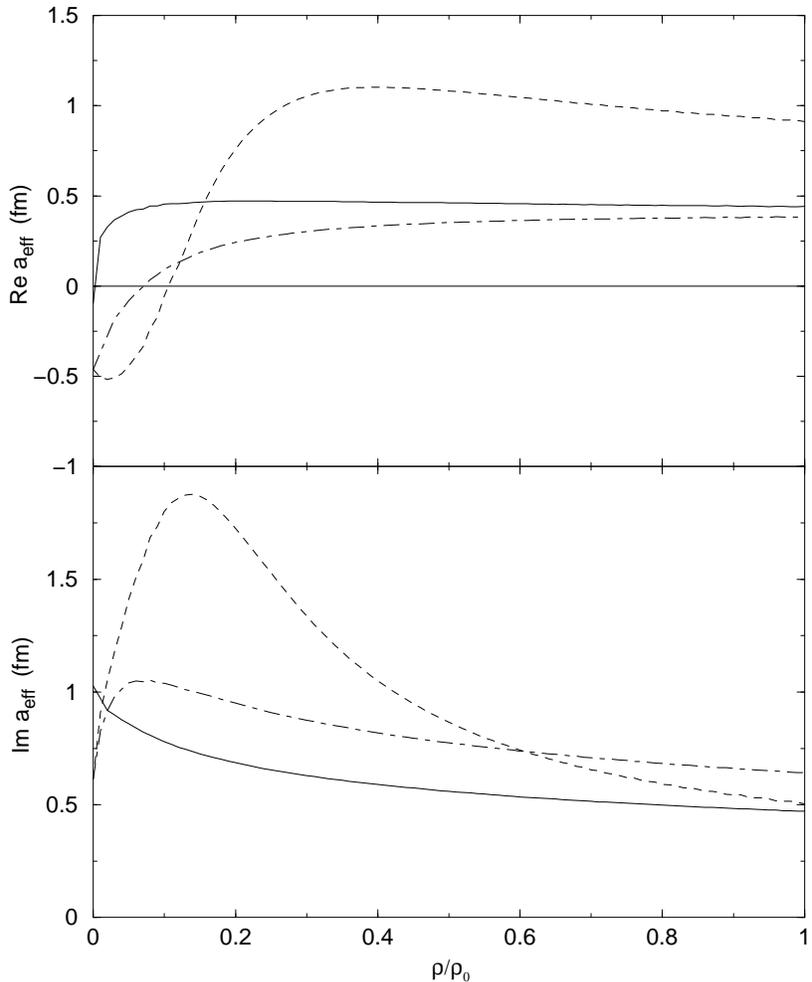}} 
\caption{Real (top) and imaginary (bottom) parts of the in-medium 
isospin-averaged $K^- N$ (effective) threshold scattering amplitude 
as function of density, calculated in Ref.~{\protect \cite{CFG01}} 
including the effects of the Pauli principle (dashed lines), 
plus the effect of self consistency for $K^-$ propagation 
(dot-dashed lines), and also the effect of self consistency for 
$N$ propagation (solid lines).} 
\label{fig:arho} 
\end{figure} 

The $t\rho$ form of the optical potential, where $t$ is the two-body 
hadron-nucleon $t$ matrix and $\rho$ is the nuclear density (more precisely, 
the nucleon-center distribution density), holds at high collision energy 
where most of the nuclear medium effects such as the Pauli principle are 
negligible. At low energy, and particularly near threshold, nuclear medium 
effects may and often do assume special importance. To demonstrate the 
scope of medium effects we use as an example the case of $K^-$ nuclear 
interaction near threshold (${\sqrt s} \sim m_{K^-} + M_p = 1432$~MeV), 
for which the underlying two-body $\bar K N$ system is strongly coupled 
to the $\pi \Sigma$ and $\pi \Lambda$ reaction channels, all in $s$ waves.  
In particular, the interaction in the $I=0$ $\bar K N$-$\pi \Sigma$ 
coupled-channel system is so strong as to generate a subthreshold quasibound 
$\bar K N$ state about 27 MeV below threshold with a width approximately 50 
MeV. This $\Lambda(1405)$ is observed as a resonance in $\pi \Sigma$ 
final-state interaction spectra~\cite{DDe91}. The $K^-$ nuclear optical 
potential in the large $A$ limit, from Eq.~(\ref{eq:Vopt}), assumes the form: 
\begin{equation} 
\label{eq:Ks} 
2\mu V_{{\rm opt}}(r) = -4\pi(1+\frac{\mu}{M})[a_{K^-p}(\rho)\rho_p(r)+ 
a_{K^-n}(\rho)\rho_n(r)] ~,
\end{equation} 
where 
\begin{equation} 
\label{eq:b2a} 
a_{K^-p}~ =~ b_0~-~b_1~, ~~~ a_{K^-n}~ =~ b_0~ +~b_1~  
\end{equation} 
in the particle basis. The density dependent effective 
scattering amplitudes $a_{K^-p}(\rho),~a_{K^-n}(\rho)$ are complex due to 
the coupling to the reaction channels. For $\rho \to 0$, the low-density 
limit asserts their limiting values $a_{K^-p},~a_{K^-n}$ respectively, 
where the latter are (strictly speaking minus) the corresponding scattering 
lengths. Figure \ref{fig:arho} shows the density dependence of the 
{\it effective} isoscalar threshold scattering amplitude 
$a_{\rm eff} = \frac{1}{2}[a_{K^-p}(\rho) + a_{K^-n}(\rho)]$ for three cases: 
$(i)$ no medium effects beyond Pauli blocking are included (dashed line); 
$(ii)$ a self-consistent calculation including the $\bar K$ self energy 
(dot-dashed line); and $(iii)$ including also the nucleon self energy 
(solid line). The change of the sign of Re$a_{\rm eff}$ from negative 
to positive corresponds to the transition from an apparently repulsive 
free-space interaction to an attractive one in the nuclear medium. 
The underlying physics is that the Pauli principle suppresses the contribution 
from Pauli forbidden intermediate states, thus weakening the in-medium 
$t$ matrix which no longer supports a subthreshold quasibound state; 
thus, the $\Lambda$(1405) gets pushed up above threshold \cite{Koc94,WKW96}. 
The inclusion of $\bar K$ and $N$ (self energy) medium modifications pushes 
this transition to a lower density, as first shown by Lutz \cite{Lut98}, 
but the free-space ($\rho = 0$) 
threshold scattering amplitude remains negative, reflecting the dominance 
of the $\Lambda$(1405) $I=0$ subthreshold resonance. The figure shows that 
apart from the very low density regime, the $K^-$ optical potential 
$V_{\rm opt}=t(\rho)\rho$ evaluated within such self consistent models 
is well approximated over a wide range of densities by a $t_{\rm eff}\rho$ 
form, where $t_{\rm eff} = -(2\pi/\mu)(1+\mu/M)a_{\rm eff} = {\rm const.}$ 
A genuine $\rho$ dependence of $t_{\rm eff}$ appears only at very low 
densities. The strength of Re$a_{\rm eff}$ is seen to be reduced to about 
$50\%$ of its initial value upon imposing self consistency. 
This is due to the suppressive effect of Im$t_{\rm eff}$ in the $K^-$ 
propagator of the Lippmann-Schwinger equation for $t(\rho)$: 
\begin{equation}\label{eq:sc} 
t=v+v\frac{1}{E-H^{(0)}_{\rm mB}-t\rho-V_N+{\rm i}0}t ~. 
\end{equation} 
Here $v$ and $t$ are coupled-channel meson-baryon (mB) potential and 
$t$ matrix, respectively, and $H^{(0)}$ is the corresponding kinetic energy 
operator which depends implicitly on the density $\rho$ through the imposition 
of the Pauli principle in $\bar K N$ intermediate states. The $K^-$ optical 
potential $t\rho$ and the nucleon potential $V_N$ act only in $\bar K N$ 
intermediate states. A sizable Im$t$ leads to an exponential decay 
of the propagator $(E-H^{(0)}_{\rm mB}-t\rho-V_N+{\rm i}0)^{-1}$, so that 
$t \approx v$ thus losing the cooperative coupling effect to the $\pi Y$ 
channels in higher-order terms of $v$.


\section{Experimental background}
\label{sec:exp}

In this section we outline experimental results, relevant to the topic of 
the present work, that have been obtained since the publication of 
Ref. \cite{BFG97} which was focused on the strong interaction physics
involved in exotic atoms of mostly medium weight and heavy nuclei.
As before, most of the information on the interaction of low energy
hadrons with nuclei which provides insight to in-medium properties,
comes from strong interaction effects in exotic atoms. Here we include
only a brief reminder of exotic atoms, referring to other reviews
for more details \cite{BFG97,Got04}. 

An exotic atom is formed  when a negatively charged particle
stops in a target and is captured by a target atom into an
outer atomic orbit. It will then emit Auger electrons and
characteristic X-rays whilst cascading down its own sequence of atomic
levels until, at some state of low principal quantum number $n$,
the particle is absorbed due to its interaction with the nucleus.
The lifetimes of all the particles considered here, namely
$\pi ^-$, $K ^-$, $\bar p$ and $\Sigma ^-$, are much longer than
typical slowing down times and
atomic time scales. Therefore, following the stopping of
the hadron in matter, well-defined states of an exotic atom
are established and the effects of the hadron-nucleus strong interaction 
can be studied. The overlap of the atomic orbitals
with the nucleus covers a  wide range of nuclear densities thus
creating a unique source of information on the density dependence
of the hadronic interaction.

In the study of strong interaction effects in exotic
atoms, the observables of interest are the shifts ($\epsilon$)
and widths ($\Gamma$) of the atomic levels caused by the
strong interaction with the nucleus. These levels are shifted and 
broadened relative to the electromagnetic case but the shifts and 
widths can usually only be measured directly for one, or possibly 
two levels in any particular exotic atom. The broadening due to the 
nuclear absorption usually terminates the atomic cascade at a low value 
of the radial quantum number $n$, thus limiting the experimentally 
observed X-ray spectrum. In some cases the width of the next higher 
$n+1$ `upper' level can be obtained indirectly from measurements of 
the relative yields of X-rays when they  depart from their
purely electromagnetic values. Shifts and widths caused
by the interaction with the nucleus may be calculated by adding
an optical potential to the Coulomb interaction. The study of the
strong interaction in exotic atoms thus becomes the study of
this additional potential.
On the experimental side, studies of strong interaction effects
in exotic atoms have been transformed over the years with the introduction
of increasingly more advanced X-ray detectors and with increasing
the efficiency of stopping the hadrons, such as with a  cyclotron 
trap \cite{Got04}. In recent years exotic atom physics has turned
into precision science.

With the present topic of in-medium interactions of low energy hadrons
we include not only data on exotic atoms but also
data on the interaction of hadrons with nuclei at low
kinetic energies where the interaction models have similarities to the
models used with exotic atoms. In such cases features of
the interaction may be studied across threshold, thus enhancing
our knowledge of the hadron-nucleus interaction. 
Moreover, the respective free hadron-nucleon interactions at very low energies
are obviously the reference to which the in-medium interactions
have to be compared. Therefore key experiments on those more elementary
reference systems will also be mentioned in the present section.

Starting with pions, recent years have seen the continuation of 
experiments at PSI on pionic hydrogen and pionic deuterium, with
ever increasing sophistication and efficiency, reaching accuracies
which are limited by theoretical corrections \cite{Sim05}.
Pionic atoms of deuterium are obviously the source of knowledge
on the pion-neutron interaction, which is a pre-requisite for studies of 
heavier targets and for obtaining separately the isoscalar and the isovector
interactions. The use of deuterium for this purpose inevitably introduces
some dependence on models in the extraction of the two basic interactions. 

Turning to heavier pionic atoms, the last decade has been dominated by the
experimental observation of `deeply-bound' pionic atom states
in the recoil-free (d,$^3$He) reaction \cite{YHI96} which populates such 
states from `inside'
the nucleus. With this technique one avoids 
the cut-off imposed by nuclear absorption
on the usual process of X-ray emission during 
the atomic cascades. 
The observation of
these states in isotopes of Pb and Sn 
\cite{SFG04}    made it
possible to  test predictions of interaction models which were based
on data for conventional pionic atom states. Alternatively, 
it became possible to derive interaction 
parameters from deeply bound atomic states, 
for comparisons with models based on X-ray
data. Moreover, studies of several isotopes of a single element
have the promise of providing information on the role played by neutron
density distributions in the pion-nucleus interaction.

Motivated by the renewed interest in the pion-nucleus interaction
at very low energy, caused by the observation of deeply-bound states 
and by the possibility of linking the long-standing $s$-wave `anomaly'
with aspects of chiral-symmetry restoration (see below), 
differential cross sections for the
elastic scattering of both $\pi ^+$ and $\pi ^-$
by several target nuclei were measured at 21~MeV. A dedicated experiment 
\cite{FBB04,FBB05} where both charge states of the pion were measured
with the same setup and where absolute normalizations were provided by 
muon scattering, yielded angular distributions
which could be analyzed with the same interaction models as used for
pionic atoms, thus providing tests across thresholds of various
characteristics of the pion-nucleus interaction.

Turning to kaonic atoms, significant progress has been made with kaonic 
hydrogen, due to experiments with precision greatly exceeding that of the 
earlier generation experiments that removed the so-called kaonic hydrogen 
`puzzle', where the strong interaction shift appeared initially to be 
attractive, contrary to expectations. The experiments at KEK~\cite{IHI97} 
showed that the shift of the 1$s$ level in kaonic hydrogen is repulsive, 
as expected from earlier phase-shift analyses. More recent results from 
DA$\Phi$NE~\cite{BBC05}, using the unique low energy $K^-$ and $K^+$ from 
$\phi$ decay, are barely consistent with the KEK results within error bars. 
In addition, experiments at KEK on kaonic atoms of $^4$He \cite{OBB07} 
seem to produce results quite different from the previous ones, which were 
at variance with predictions of most calculations. 
With the exception of the above two examples, 
the world's data on kaonic atoms have not been expanded in the last decade,
because low energy $K^-$ beams of sufficient quality are not available.

Outside the realm of kaonic atoms, there have been experimental 
indications \cite{SBF04,SBF05,KHA05,ABB05}
of possible existence of strongly bound antikaon states in {\it nuclei}.
These caused renewed interest in the question of the depth of the real
part of the $K^- $-nucleus potential at threshold, where `deep' 
real potentials are known from $\chi ^2$ fits to kaonic atom data
and `shallow' potentials are obtained from chiral approaches. Although 
the experimental situation is not settled at the present time, some 
understanding of the antikaon-nucleus interaction is being promoted 
thanks to studies inspired by the speculations on strongly bound states.

The interaction of $K^+ $ mesons with nuclei has not been discussed
in our previous Review~\cite{BFG97}. This topic is included in
the present Review since further analyses of previous transmission
experiments have clearly demonstrated \cite{FGW97,FGM97a,FGM97b} that the 
elementary $KN$ interaction in the range of 500-700~MeV/c is modified 
in the nuclear medium. In common with the case of $K^-$ mesons, the recent 
renewed interest in the topic of medium-modification of the elementary 
$K^+$ interaction was motivated by speculations based on experimental 
indications, in this case on the possible existence of the $\Theta^+$ 
pentaquark.

The large cross sections for annihilation of antiprotons on nucleons
set the scene for the interaction of ${\bar p}$ with nuclei at low
energies. In terms of optical potentials that means dominance of the
imaginary part which complicates the issue of the connection between the 
free ${\bar p} N$ interaction and the interaction in the nuclear medium.
It also means that ${\bar p}$ do not penetrate deeply into nuclei.
On the other hand, one of the experimental consequences is the ability
to measure annihilation  cross sections at extremely low
energies. Indeed such measurements have been made in recent years
at momenta as low as 40-50~MeV/c on light nuclei \cite{Zen99a,BBB00a}
and on hydrogen \cite{ZBB99}. Total cross sections for ${\bar n}p$
down to 50~MeV/c have also been reported \cite{Iaz00}.

The experimental situation with ${\bar p}$ atoms has changed significantly 
in the last decade with the publication by the PS209 collaboration \cite{TJC01}
of high-quality data for several sequences of isotopes along the periodic
table. For most target nuclei strong-interaction level shifts and widths 
have been measured for two atomic levels of the same
antiprotonic atom, with two examples where information is available for 
three levels. In addition close to 20 target nuclei have been studied
by the radiochemical method \cite{LJT98,SHK99}, observing the
production of nuclei differing from the target nucleus by the removal
of one neutron or one proton. Such data provide unique information
on the absorption of ${\bar p}$ by a neutron or by a proton, repectively, 
at about 2.5-3~fm outside of the nuclear surface, and in particular
the ratios between the probabilities for the two processes
are determined quite reliably. The two kinds of data, namely,
level shifts and widths on the one hand and 
ratios from radiochemical data on the other,
were shown to lead to consistent results and could also be analyzed
together. 

Finally we mention the interaction of low energy $\Sigma ^-$ with
nuclei. No additional data on $\Sigma ^-$ atoms have been produced
in recent years and the only relevant new 
experimental information was obtained from the
$(\pi ^-,K^+)$ reaction on nuclear targets \cite{SNA04} which showed 
some features in common with the $\Sigma ^-$ atom potential derived more than  
a decade earlier.


\section{Pions}
\label{sec:pions}

\subsection{The pion-nucleus potential}
\label{sec:pipotl}

At zero energy the interaction of pions with nucleons is
rather weak and consequently a $t\rho$ approach would be
expected to yield a reasonably good optical potential or at least
provide a theoretically motivated {\it form} for the potential.
The interaction of low energy pions with nucleons is affected 
by the (3,3) resonance at about 180 MeV, and this is manifested 
by a significant $p$ wave term in the $\pi N$ interaction which 
in turn is reflected in the form of the optical potential~\cite{EWe88}. 
The Kisslinger potential, where this $p$ wave interaction leads to 
gradient terms in the pion nucleus potential, was introduced more 
than half a century ago \cite{Kis55} and model-independent analyses 
of elastic scattering of pions by nuclei showed \cite{Fri83} that 
indeed the local-equivalent potential has all the features expected 
for the Kisslinger potential (see also \cite{JSa96}).
A $t\rho$ potential at zero energy is real because pions
cannot be absorbed at rest by a single nucleon, although they can
be absorbed by the nucleus. For that reason
Ericson and Ericson \cite{EEr66} introduced  $\rho ^2$ terms into
the potential which describe schematically the absorption of
$\pi^-$ on pairs of nucleons. The potential for $\pi^-$ mesons 
then becomes, in its simplest form,
\begin{equation} \label{eq:EE1}
2\mu V_{\rm opt}(r) = q(r) + \vec \nabla \cdot \alpha(r) \vec \nabla
\end{equation}

\noindent
with

\begin{eqnarray} \label{eq:EE1s}
q(r) & = & -4\pi(1+\frac{\mu}{M})\{b_0[\rho_n(r)+\rho_p(r)]
  +b_1[\rho_n(r)-\rho_p(r)] \} \nonumber \\
 & &  -4\pi(1+\frac{\mu}{2M})4B_0\rho_n(r) \rho_p(r),
\end{eqnarray}

\begin{eqnarray} \label{eq:EE1p}
\alpha (r) & = & 4\pi(1+\frac{\mu}{M})^{-1} \{ c_0[\rho_n(r)+\rho_p(r)]
  +c_1[\rho_n(r)-\rho_p(r)] \}   \nonumber  \\
 & & +4\pi(1+\frac{\mu}{2M})^{-1}4C_0\rho_n(r)\rho_p(r),
\end{eqnarray}

\noindent
where $\rho_n$ and $\rho_p$ are the neutron and proton density
distributions normalized to the number of neutrons $N$ and number
of protons $Z$, respectively, $\mu$ is the pion-nucleus reduced mass
and $M$ is the mass of the nucleon.
In this potential $q(r)$ is referred to
as the $s$ wave potential term and $\alpha(r)$ is referred to
as the $p$ wave potential term.
The real coefficients $b_0$ and $b_1$ are minus the pion-nucleon
isoscalar and isovector $s$ wave
scattering lengths, respectively, whilst the real coefficients $c_0$ and
$c_1$ are the pion-nucleon
isoscalar and isovector $p$ wave scattering volumes,
respectively. The parameters $B_0$ and $C_0$ represent $s$ wave and $p$ wave
absorptions, respectively, and as such have imaginary parts.
Dispersive real parts are found to play a role in pionic atom potentials.
The $\rho _n \rho _p$ in the absorption terms represent
two-nucleon absorption which takes place predominantly on neutron-proton
pairs. These were originally written \cite{EEr66} as 
$B_0 \rho _m^2$ and $C_0 \rho _m^2$,
with $\rho_m=\rho _n+\rho _p$ without distinguishing between neutrons
and protons. The factor 4 is introduced above to make the coefficients
$B_0$ and $C_0$ comparable in the two formulations.
In practice when parameters are obtained from fits to the data
the two forms yield practically the same results.

In the above expressions the terms linear in the nuclear densities
are associated, in the $t\rho$ approach, with the interaction between
pions and free nucleons. Ericson and Ericson showed 
that the $p$ wave dipole
interaction is modified in the nuclear medium in a way analogous
to the Lorentz-Lorenz effect in electrodynamics, replacing
the above expression for $\alpha (r)$ as follows:

\begin{equation} \label{eq:LL1}
\alpha(r) \longrightarrow \frac {\alpha(r)}{1+\frac{1}{3} \xi \alpha (r)}
\end{equation}
\noindent
where $\xi$ is a constant of the order 1. This effect is generally
referred to as the Lorentz-Lorenz-Ericson-Ericson (LLEE) effect and it
results from short range repulsive correlations between nucleons.
This modification should apply only to
the linear part of $\alpha$, and Eq. (\ref{eq:EE1p}) is re-written as
\begin{equation} \label{LL2}
\alpha (r) = \frac{\alpha _1(r)}{1+\frac{1}{3} \xi \alpha _1(r)}
 + \alpha _2(r)
\end{equation}
\noindent
with
\begin{equation} \label{eq:alp1}
\alpha _1(r) = 4\pi (1+\frac{\mu}{M})^{-1} \{c_0[\rho _n(r)
  +\rho _p(r)] +  c_1[\rho _n(r)-\rho _p(r)] \}
\end{equation}

\begin{equation} \label{eq:alp2}
\alpha _2(r) = 4\pi (1+\frac{\mu}{2M})^{-1} 4C_0\rho _n(r) \rho _p(r).
\end{equation}
Another complication arises due to the parameter $b_0$ 
being exceptionally small. Hence second order
effects in the construction of the isoscalar $s$ wave potential
term in $q(r)$ become important \cite{EEr66} and this
causes $b_0$ to be replaced by
\begin{equation} \label{eq:b0b}
\overline{b}_0 = b_0 - \frac{3}{2\pi}(b_0^2+2b_1^2)k_F,
\end{equation}
where $k_F$ is the Fermi momentum calculated for the local nuclear density.

Finally there is another relatively small term of a kinematical origin
whose presence is supported by fits to pionic atom data. This is the
so-called angle transformation term \cite{FGa80,SCM80} which is given by

\begin{eqnarray} \label{eq:angtr}
2\mu\Delta V_{\rm opt} & = & -4\pi \{\frac{\mu}{2M}(1+\frac{\mu}{M})^{-1} \nabla^2
  [c_0(\rho _n+\rho _p)+c_1(\rho _n-\rho _p)]   \nonumber \\
 & &+\frac{\mu}{M}(1+\frac{\mu}{2M})^{-1}\nabla^2[C_0\rho _n\rho _p] \}.
\end{eqnarray}

The above potential is inserted into the KG equation (\ref{eq:KG})
to obtain a complex eigenvalue. The Coulomb potential due to the finite size
charge distribution as well as 
the Uehling $\alpha(Z\alpha )$ vacuum polarization potential \cite{FRi76}
are also included. The strong interaction effects are the differences
between these eigenvalues with and without the above 
potential (Eq.~(\ref{eq:EE1})), respectively.

\subsection{Pionic atom data}
\label{sec:piatdata}

\begin{table}
\caption{Data for 1$s$ states in pionic atoms}
\label{tab:piat_s}
\begin{ruledtabular}
\begin{tabular}{lccc}
     &  shift (keV) & width (keV) & Ref. \\ \hline
   $^{20}$Ne& $-$32.17 $\pm$  0.77  &   15.43 $\pm$  0.41  & \cite{OBF82} \\
   $^{22}$Ne& $-$40.42 $\pm$  0.50  &   12.7$\pm$   3.5  & \cite{OBF82} \\
   Na       & $-$50.6 $\pm$  1.0  &   17.1$\pm$   1.6  & \cite{BBM87} \\
   $^{24}$Mg& $-$60.2 $\pm$  1.2  &   24.3 $\pm$  1.6  & \cite{TDH90} \\
   $^{28}$Si& $-$95.1 $\pm$  2.0  &   41.0 $\pm$  4.0  & \cite{TDH90} \\
   $^{115}$Sn& $-$2402 $\pm$  24 &     441 $\pm$  87&  \cite{SFG04} \\
   $^{119}$Sn& $-$2483 $\pm$ 18 &     326 $\pm$  80&  \cite{SFG04} \\
   $^{123}$Sn& $-$2523 $\pm$ 18 &     341 $\pm$  72&  \cite{SFG04} \\
   $^{205}$Pb& $-$5354 $\pm$ 61 &     764 $\pm$  165&  \cite{GGG02} \\
\end{tabular}
\end{ruledtabular}
\end{table}

Experimental results on pionic atoms covering the whole of the periodic table
have been published for a few decades, with improved accuracy over
the years and with increased use of separated isotopes for targets.
The most extensive analysis has been that of Konijn~et~al.~\cite{KLT90}
who analyzed 140 data points covering states
from 1$s$ in $^{10}$B to 4$f$ in $^{237}$Np. We note that 
a different definition of the strong interaction shift is used
in Ref. \cite{KLT90}, namely, 
the difference between the complex binding energy for the full potential,
including finite size Coulomb and vacuum polarization potential, and the 
binding energy for the point Coulomb potential. This is different from the
conventional definition used here and, in any case, it breaks down for
1$s$ states when the charge of the nucleus is $Z~>~137/2$. Such data are
available now for the 1$s$ state in $^{205}$Pb. 

\begin{table}
\caption{Data for 2$p$ states in pionic atoms}
\label{tab:piat_p}
\begin{ruledtabular}
\begin{tabular}{lccc}
     &  shift (keV) & width (keV) & Ref. \\ \hline
 $^{24}$Mg & 0.129 $\pm$  0.004  &  0.0725   $\pm$0.0018 & \cite{TDH90} \\
 $^{26}$Mg & 0.126 $\pm$  0.004  &  0.0811   $\pm$0.0019& \cite{CBB85} \\
 $^{28}$Si & 0.286 $\pm$  0.010  &  0.192   $\pm$0.009& \cite{TDH90} \\
 $^{30}$Si & 0.281 $\pm$  0.010  &  0.196   $\pm$0.008& \cite{CBB85} \\
 S     & 0.615  $\pm$  0.022  &  0.430   $\pm$0.021& \cite{BBH79} \\
 $^{40}$Ca & 1.941 $\pm$  0.080  &  1.590   $\pm$0.023& \cite{BBF79} \\
 $^{42}$Ca & 1.650 $\pm$ 0.080 &  1.65   $\pm$0.15& \cite{PWH80} \\
 $^{44}$Ca & 1.583  $\pm$ 0.080  &  1.60   $\pm$0.07& \cite{PWH80} \\
 $^{48}$Ca & 1.295  $\pm$ 0.115  &  1.64   $\pm$0.11& \cite{PWH80} \\
 $^{46}$Ti & 2.490  $\pm$ 0.140  &  2.39   $\pm$0.15& \cite{PWH80} \\
 $^{48}$Ti & 2.290  $\pm$ 0.130  &  2.62   $\pm$0.15& \cite{PWH80} \\
 $^{50}$Ti & 1.937  $\pm$ 0.110  &  2.15   $\pm$0.27& \cite{PWH80} \\
 $^{50}$Cr & 3.560  $\pm$ 0.210  &  4.3   $\pm$0.4& \cite{KPH83} \\
 $^{52}$Cr & 3.120  $\pm$ 0.190  &  3.85   $\pm$0.21& \cite{KPH83} \\
 $^{54}$Cr & 2.827  $\pm$ 0.190  &  3.84   $\pm$0.29& \cite{KPH83} \\
 Fe      &  4.468 $\pm$  0.340  &  6.87   $\pm$0.21& \cite{BBH79} \\
 Ge      &  5.5 $\pm$  0.9  &  18.5   $\pm$2.5& \cite{ABG77} \\
 As      &  4.6 $\pm$  0.9  &  14.5   $\pm$4.0& \cite{ABG77} \\
 Nb      &  2.1 $\pm$  3.0  &  64  $\pm$ 8& \cite{TDH90} \\
 Ru      &  $-$27.6  $\pm$ 7.0  &  77  $\pm$24& \cite{TDH90} \\
 $^{205}$Pb &  $-$835 $\pm$ 45 &    321   $\pm$61& \cite{GGG02} \\
\end{tabular}
\end{ruledtabular}
\end{table}

In the present work we are mostly concerned with global
properties of the pion-nucleus interaction and its dependence on the
nuclear density. This is effected by performing {\it global} 
fits to pionic atom
data, handling together all relevant data. In order to avoid 
some distortion of the emerging picture we excluded from the analysis several
deformed nuclei and also several very light nuclei, where the concept
of an optical potential could be questionable. Moreover, we study
extensively the dependence of strong-interaction effects
on the neutron densities $\rho _n$,
using two-parameter Fermi distributions. For that reason
we excluded from the data base also nuclei with $Z~\leq~8$ where other
densities such as a modified harmonic oscillator are more appropriate.

The data used in the present work are summarized in the following
four tables. The number of data points is 100, compared to 54 points
used in our previous Review \cite{BFG97}. Ten of the points are due to
the recently observed deeply bound states, as can be
seen from the last four entries in Table~\ref{tab:piat_s} and the
last line of Table~\ref{tab:piat_p}. These states are discussed below.

\begin{table}
\caption{Data for 3$d$ states in pionic atoms}
\label{tab:piat_d}
\begin{ruledtabular}
\begin{tabular}{lccc}
     &  shift (keV) & width (keV) & Ref. \\ \hline
 Nb &  0.761   $\pm$0.020  &  0.402   $\pm$0.016 &\cite{TDH90} \\
 Ru  &   1.428   $\pm$0.080  &  0.75    $\pm$0.08 &\cite{TDH90} \\
 Ag  &   1.99   $\pm$0.05  &  1.43   $\pm$0.04 &\cite{TDH90} \\
 Cd  &   2.23   $\pm$0.09  &  1.65   $\pm$0.07 &\cite{TDH90} \\
 Ba  &   5.44   $\pm$0.27  &   4.3   $\pm$0.9 &\cite{JPK70} \\       
 $^{140}$Ce &  7.05   $\pm$0.30 & 5.6 $\pm$1.0 &\cite{JPK70} \\
$^{142}$Ce &   7.23   $\pm$0.30 &   6.5   $\pm$0.9 &\cite{JPK70} \\
$^{148}$Nd &   7.85   $\pm$0.40 &   8.8   $\pm$1.2 &\cite{KPP77} \\
$^{150}$Nd &   7.77   $\pm$0.40 &   9.2   $\pm$1.1 &\cite{KPP77} \\
$^{166}$Er &  17.4   $\pm$1.0 &   19.7   $\pm$0.9 &\cite{TSS84} \\
$^{168}$Er &  16.3   $\pm$0.5 &   19.4   $\pm$1.0 &\cite{TSS84} \\
 Pt &   28.8  $\pm$1.9   & 37   $\pm$5 &\cite{LDH91} \\
$^{208}$Pb &   32.25   $\pm$2.40  &   47   $\pm$4 &\cite{LDH91} \\
 Bi  &  30.5  $\pm$3.0   &  52   $\pm$4 &\cite{LDH91} \\
\end{tabular}
\end{ruledtabular}
\end{table}

\begin{table}
\caption{Data for 4$f$ states in pionic atoms}
\label{tab:piat_f}
\begin{ruledtabular}
\begin{tabular}{lccc}
     &  shift (keV) & width (keV) & Ref. \\ \hline
   $^{168}$Er&  0.351  $\pm$0.020  &   0.22   $\pm$0.03 &\cite{BBR81} \\
    Re &  0.76  $\pm$0.04  &   0.41   $\pm$0.05 &\cite{LDH91} \\
    Pt &  1.13  $\pm$0.04  &   0.59   $\pm$0.05 &\cite{LDH91} \\
   $^{208}$Pb&  1.72  $\pm$0.04  &   0.98   $\pm$0.05 &\cite{LDH91} \\
   Bi &  1.83  $\pm$0.06  &   1.24   $\pm$0.09 &\cite{LDH91} \\
   U &  5.08  $\pm$0.20  &   3.65   $\pm$0.65 &\cite{BBH79} \\
\end{tabular}
\end{ruledtabular}
\end{table}

\newpage

\subsection{Deeply bound pionic atom states}.
\label{sec:deep}

As mentioned in Sec.~\ref{sec:exp}, the last decade has 
been dominated, for pionic atoms, 
by the observation of deeply bound states, which 
contributed to the revival of interest in pionic atoms in general
and in the $\pi N$ in-medium interaction in particular. We therefore 
include here a brief outline of this topic.

\begin{figure}
\includegraphics[scale=0.7,angle=0]{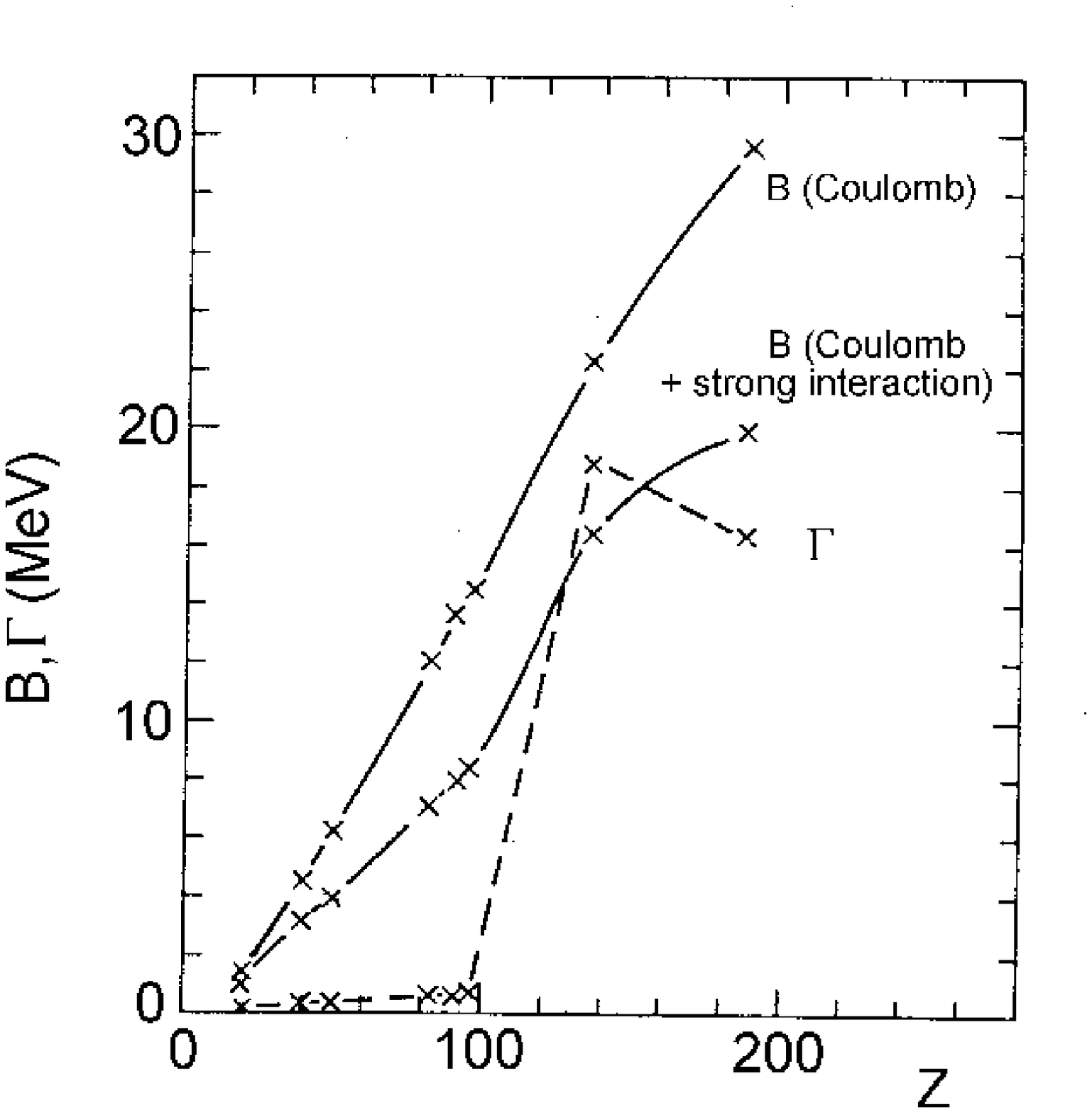}
\caption{Calculated binding energies and widths of 1$s$ states in
pionic atoms (from Ref.~\cite{FSo85}).}
\label{fig:deep1}
\end{figure}

The term deeply bound pionic atoms refers to 1$s$ and 2$p$ states in heavy 
pionic atoms which cannot be populated via the X-ray
cascade process because upper levels such as the 3$d$  are already
broadened by the strong interaction to the extent that the radiative yield
becomes exceedingly small.
The first to show that 1$s$ and 2$p$ states in heavy pionic atoms,
perhaps surprisingly, are so narrow  as to make them well defined, were
Friedman and Soff~\cite{FSo85}. They calculated numerically 
binding energies and widths
for pionic atom states well beyond the experimentally reachable
region using optical potential parameters \cite{BFG83} which reproduced
very well experimental results over the entire periodic table.
Figure~\ref{fig:deep1} is the original figure from 1985 (Ref.~\cite{FSo85})  
showing the calculated
binding energies ($B$) and widths ($\Gamma$) for 1$s$ states 
of pionic atoms as function of the atomic
number $Z$. It is seen that up to the top end of
the periodic table the widths of the states are relatively small.

The explanation of that unexpected results was also given by Friedman and 
Soff \cite{FSo85} in terms of the overlap of the atomic wavefunction
with the nucleus. The atomic wavefunctions of these deeply bound 
states are excluded from
the nucleus due to the repulsive $s$ wave part of the potential
such that their overlap with the imaginary
part of the potential becomes very small. In this context it is
instructive to note that for a Schr\"odinger equation
the width of a state is given {\it exactly} (i.e. not perturbatively)
by the following expression
\begin{equation} \label{eq:gamma}
\Gamma=-2 \frac{\int \left|\psi\right|^2 {\rm Im} V_{\rm opt}d{\bf r}}
{\int \left|\psi\right|^2 d{\bf r}}.
\end{equation}
A slightly different expression is obtained 
for the KG equation. It is, therefore, easy to see that 
the reduced overlap of the atomic wavefunction with the
nucleus, due to the repulsive real part of the $s$-wave term of the
potential Eq.~(\ref{eq:EE1s}), provides a natural explanation of the 
numerically observed saturation of widths. Only for larger charge
numbers, beyond the range of stable nuclei, the Coulomb attraction 
overcomes the $s$-wave repulsion, resulting in very large increase
in the calculated widths.

Three years later Toki and Yamazaki also concluded that deeply bound
pionic atom levels would be sufficiently narrow and therefore could
be observed, and in addition
discussed experiments that could populate such
states \cite{TYa88}. After several unsuccessful attempts in various
laboratories, it was realized that the key to success was the creation
of a pion at rest (it appears that the first to suggest this approach,
although in a different context, were Ericson and Kilian~\cite{EKi84}). 
That requires `recoil-free' kinematics, which for 
the (d,$^3$He) reaction on Pb 
creating a bound pion means a beam energy 
around 600~MeV. The experiment at GSI used the fragment separator
in order to achieve the required resolution and reduction of background
and the first observation of the 1$s$ state in pionic atoms of $^{207}$Pb
by Yamazaki et al.~\cite {YHI96} was a clear demonstration of
the ability to study such states. Further experiments by the same
group achieved improved accuracies for the 1$s$ and 2$p$ levels in
pionic atoms of $^{205}$Pb using a $^{206}$Pb target \cite{GGG02}
and  studies of the 1$s$ state in isotopes of Sn \cite{SFG04}
followed soon after.
Full details of this fascinating project
can be found in a Review by Kienle and Yamazaki~\cite{KYa04}. 

\begin{figure}
\includegraphics[scale=0.65,angle=0]{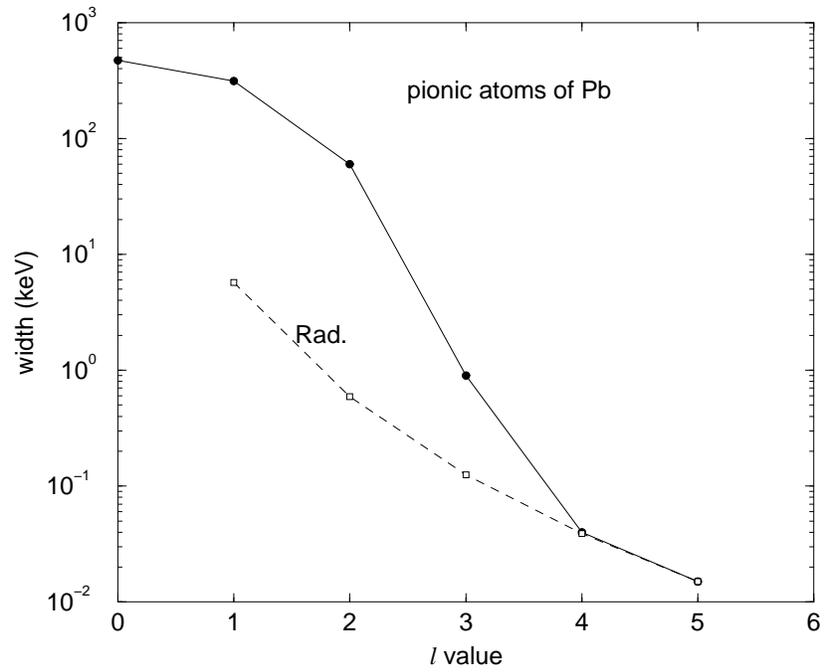}
\caption{Calculated widths of `circular' states in
pionic atoms of Pb, showing radiative (dashed) and total widths (solid line).}
\label{fig:Pbwidths}
\end{figure}

\begin{figure}
\includegraphics[scale=0.65,angle=0]{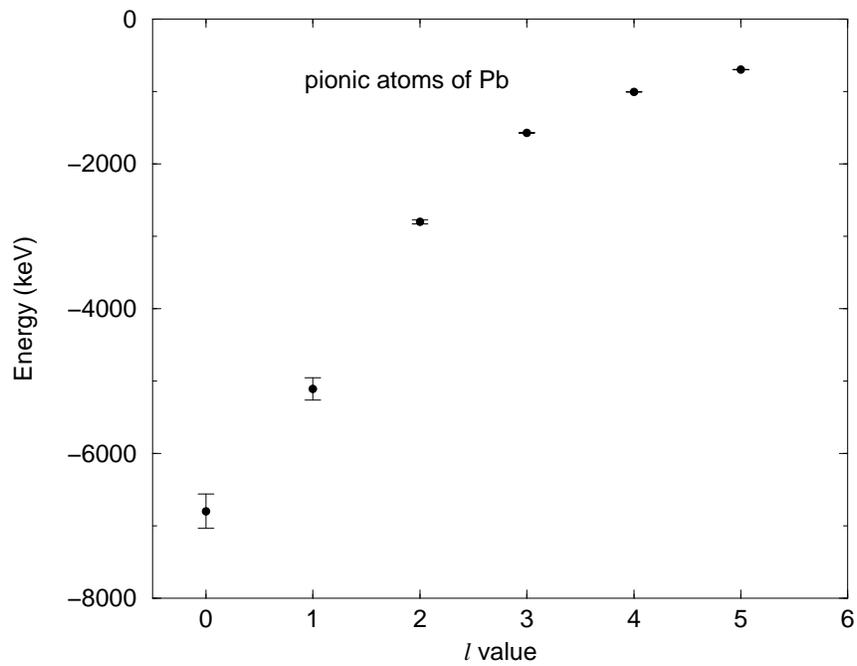}
\caption{Calculated energy levels of `circular' states in
pionic atoms of Pb. The bars show the widths of the states.}
\label{fig:Pbspect}
\end{figure}

To conclude this section, Fig.~\ref{fig:Pbwidths} illustrates the phenomenon 
of saturation of widths in pionic atoms by showing calculated radiation and 
total widths of `circular' states (i.e. states with radial number
$n=l+1$) in pionic atoms of Pb, using for the strong interaction the 
present-day best fit potential (see below). It is seen that for large
$l$ values the widths are essentially the radiative widths and X-ray
transitions will be observed. As the $l$-value decreases the strong
interaction width becomes significant and eventually, for low $l$ values,
the radiative transitions will be suppressed. However, the initial rise 
of the strong interaction width with  decreasing $l$ is saturated, 
resulting in the relatively narrow widths of deeply 
bound states. This is seen in Fig.~\ref{fig:Pbspect} showing the energy
spectrum for circular states of pionic atoms of Pb. Comparing with
Fig.~\ref{fig:Pbwidths}, it is clear that the energy levels are
well-defined only thanks to the saturation of the widths.


\subsection{Fits to pionic atom data}
\label{sec:pifit}

\subsubsection{General}
\label{sec:general}

With nine parameters in the  pion-nucleus optical potential 
it is not a straightforward task to get meaningful information
on the potential from $\chi ^2$ fits to the data. Moreover,
there is the question of the proton $\rho _p$
and neutron $\rho _n$ density distributions which are an essential
ingredient of the potential. 
The proton density distributions  are known quite well from
electron scattering and muonic X-ray experiments
\cite{FBH95}, and can be obtained
from the nuclear charge distributions by numerical unfolding of the 
finite size of the proton. In contrast, the neutron densities 
are not known to sufficient accuracy and their uncertainties must be
considered when extracting parameters from fits to the data.

Parameters of the pion-nucleus potential Eq.~(\ref{eq:EE1}) had been 
obtained by performing $\chi ^2$ fits to the data already in the
late 1970s and early 1980s and with the additional accumulation of
data they were found to possess good predictive power of yet unmeasured
quantities, with little dependence on the details of the potential
form used to fit the data.
Two approaches have been made to the problem of large number of
free parameters: (i) assumptions can be made in order to reduce the number
of free parameters, such as introducing `effective' density in the
quadratic terms \cite{SMa83} thus making them linear in the density
and avoiding correlations with the corresponding genuinely linear terms; 
(ii) performing {\it global} fits to very large sets of data where
the effects of correlations between parameters are reduced.
In most cases the dependence on neutron densities have not been
studied and `reasonable' densities have been assumed in the analysis. 
Only  Garc\'{\i}a-Recio, Nieves and Oset \cite{GNO92} varied also neutron 
densities and obtained parameters
for neutron distributions from global fits to pionic atoms data.
Using three different versions for the pion-nucleus potential,
they showed that the neutron rms radii were determined to good accuracy
and were in reasonable agreement with values deduced from Hartree-Fock
calculations.
Below we present  detailed studies of the dependence of
derived potential parameters on neutron densities \cite{Fri07}, with emphasis
on the `$s$-wave anomaly' problem (see below).

Addressing first the $p$-wave part of the potential (Eq.~(\ref{eq:EE1p})),
we note that due to its gradient nature it is expected to be effective
only in the surface region of nuclei and that 
large medium or density-dependent effects are  not  expected
beyond the `trivial'
LLEE correction Eq.~(\ref{eq:LL1}). As discussed in Ref.~\cite{BFG97}, 
the dependence of $\chi ^2$ on the parameters of the $p$-wave term
is rather weak, with the free pion-nucleon scattering volumes
of $c_0$=0.21$m_\pi ^{-3}$ and $c_1$=0.165$m_\pi ^{-3}$ being 
consistent with the best-fit values, when the `classical' 
value of $\xi$=1.0 is being used. We therefore adopt these three
values and focus most of the attention on the $s$ wave part of
the potential.   

The role of the deeply bound states in the global picture of
pion-nucleus interaction was an obvious question 
when the first such states were observed, and it was found~\cite{FGa98} 
that shifts and widths of deeply bound states agreed with
predictions made with potentials based on `normal' states. 
That is fully understood by considering the overlaps between
atomic and nuclear densities, which are very similar for normal
and for deeply bound states, as can be seen e.g. from Fig.~17 of
Ref.~\cite{KYa04}. However, the experimental study of deeply bound states 
provides information on 1$s$ states throughout the periodic table and
has enriched our undertanding of the pion-nucleus interaction at low
energies. Studies of isotopes have the potential of providing
additional valuable information.
In what follows we combine the data on deeply bound states with data
on normal states, as is seen in Tables \ref{tab:piat_s} and \ref{tab:piat_p}.

\subsubsection{The role of neutron densities}
\label{sec:neutdens}

As discussed in Sec.~\ref{sec:nucldens},
$r_p$, the rms radius for the proton density
distribution, is considered to be known and  we therefore 
focus attention on values
of the difference $r_n-r_p$, using the  linear dependence of $r_n-r_p$
on $(N-Z)/A$ given by Eq.~(\ref{eq:RMF}).
In order to allow for possible differences in the {\it shape} of the neutron
distribution, the `skin' and the `halo' forms of Ref.~\cite{TJL01} were
used, as well as their average, see Sec.~\ref{sec:nucldens}.

Parameters of the potential Eq.~(\ref{eq:EE1}) were determined by
minimising the $\chi ^2$ in fits to the combined  data given in 
tables \ref{tab:piat_s} to \ref{tab:piat_f}. The linear $p$-wave parameters
were held fixed at their respective free $\pi N$ values of
$c_0$=0.21$m_\pi ^{-3}$ and $c_1$=0.165$m_\pi ^{-3}$ and the LLEE parameter
was held fixed at $\xi$=1. The quadratic $p$-wave parameters were 
found to be close to
Re$C_0$=0.01$m_\pi ^{-6}$  and Im$C_0$=0.06$m_\pi ^{-6}$. The neutron
radius parameter $\delta $ was fixed at $-$0.035 fm and we scanned over the
other radius parameter $\gamma$.

\begin{figure}
\includegraphics[scale=0.7,angle=0]{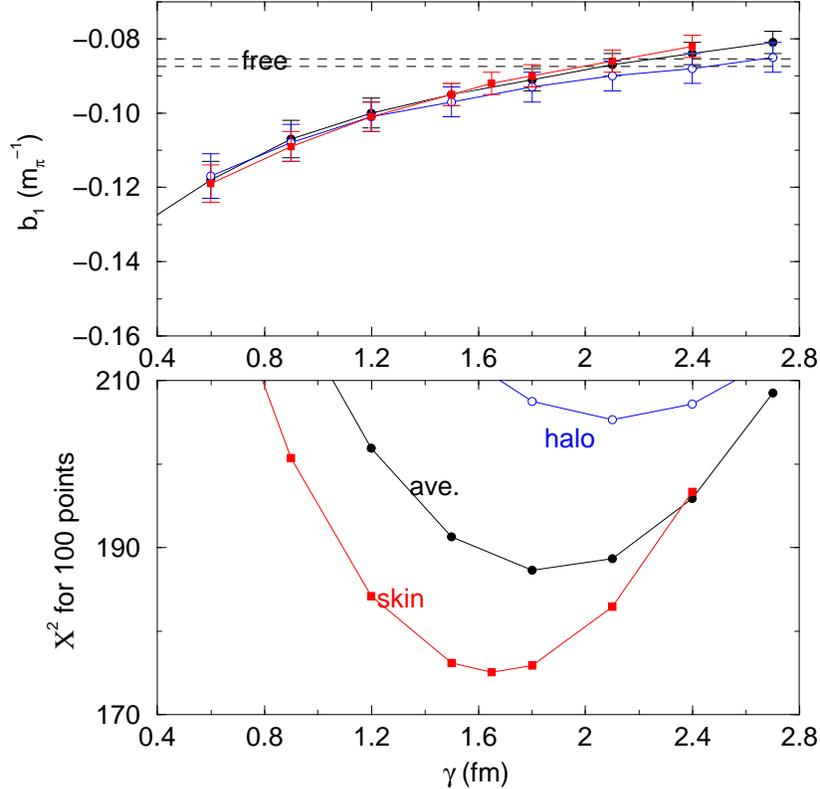}
\caption{Results of global fits to pionic atom 
data for different values of the neutron
radius parameter $\gamma $ of Eq.~(\ref{eq:RMF}). 
Lower part - values of $\chi ^2$
for 100 data points for three shapes of the neutron density $\rho _n$.
Upper part - the corresponding isovector parameter
$b_1$ in comparison to its free $\pi N$ value (marked `free').}
\label{fig:piZRC}
\end{figure}

Figure \ref{fig:piZRC} shows results of global fits to the 100 data points
for pionic atoms, as discussed above. The lower part shows values
of $\chi ^2$ as function of the  parameter $\gamma $, as defined 
by Eq.~(\ref{eq:RMF}). It is seen that the quality of fit depends
on the shape of the neutron distribution, where the `skin' shape is
definitely preferred. The existence of quite well-defined minima
is gratifying. The upper part shows the corresponding values of the
isovector parameter
$b_1$ in comparison with its free $\pi N$ value (marked `free').
It is clear that the resulting values of $b_1$ are almost independent
of the shape of the neutron density used in the fit, depending mostly
on the values of the rms radius $r_n$, as represented by the parameter
$\gamma$. From the figure one might conclude that for the best fit value
of $\gamma $=1.6 fm the value of $b_1$ turns out to be in reasonable agreement
with the free $\pi N$ value. 

\begin{figure}
\includegraphics[scale=0.7,angle=0]{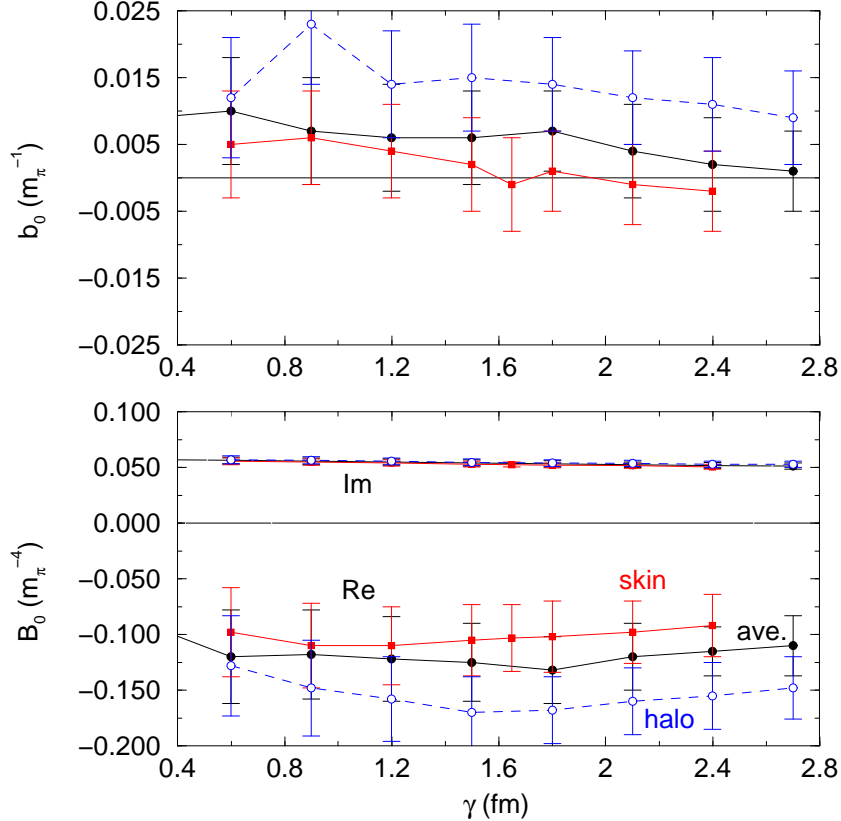}
\caption{Results of global fits for different values of the neutron
radius parameter $\gamma $ of Eq.~(\ref{eq:RMF}). 
Lower part - the complex parameter $B_0$. 
Upper part - the isoscalar parameter $b_0$.}
\label{fig:piZRC2}
\end{figure}

Figure~\ref{fig:piZRC2} shows 
in the upper part values of the isoscalar parameter $b_0$. 
As noted above, this parameter is exceptionally small and the errors 
on the best-fit results are relatively large. 
The lower part of the figure shows 
the resulting values of the quadratic parameter $B_0$. The values of
Im$B_0$ are determined to a very good accuracy and are independent of 
the neutron radius parameter $\gamma$ and of the shape of the distribution.
The values of Re$B_0$ are not as accurate but they are clearly not zero,
representing non-negligible repulsion in addition to the repulsion
provided by the linear term.

The rms radii of neutron distributions implied by the global fits
to pionic atom data need some attention, particularly in view of the
fairly strong dependence of the derived {\it in~medium} values
of the isovector $s$-wave parameter $b_1$
on the rms radius assumed for $\rho _n$. As mentioned above, values
of the  difference $r_n-r_p$ between rms radii obtained 
for medium-weight and heavy nuclei with $\gamma$=1.5 fm are too
large by 0.05 to 0.1 fm in comparison with recent RHB 
calculations \cite{NVF02,LNV05}. Moreover, a survey of different sources
of information on $r_n-r_p$ suggests \cite{JTL04} that $\gamma$=0.9-1.0 fm
is the proper value. Therefore the pionic atoms result 
of $\gamma \geq $1.6 fm is at odds with all we know about neutron density
distributions in nuclei.

\begin{figure}
\includegraphics[scale=0.7,angle=0]{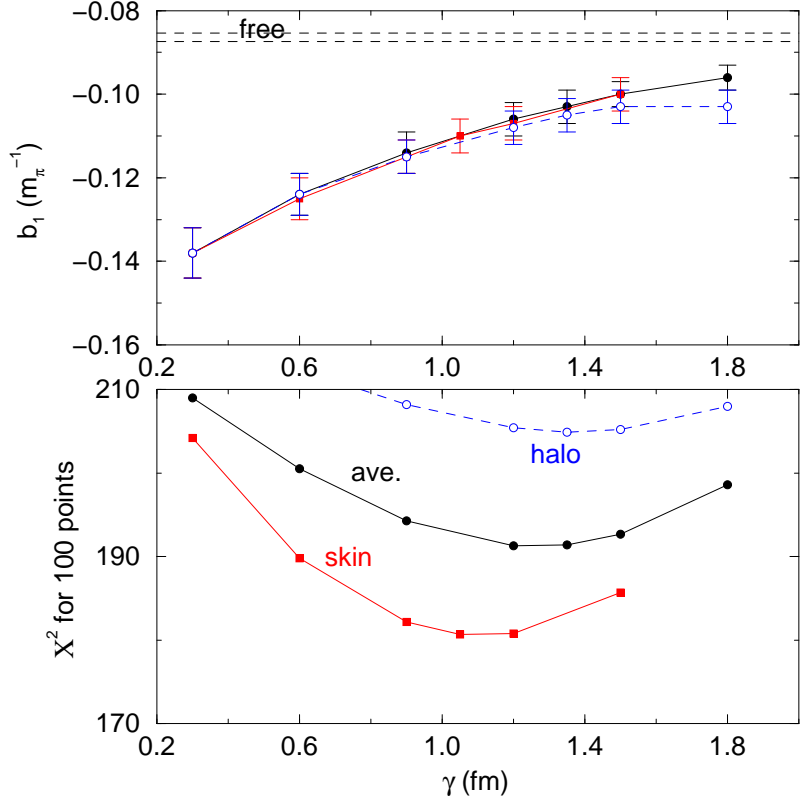}
\caption{Same as Fig.~\ref{fig:piZRC} but with finite range
folding applied to the $p$-wave potential, see text.}
\label{fig:piYC}
\end{figure}

\begin{figure}
\includegraphics[scale=0.7,angle=0]{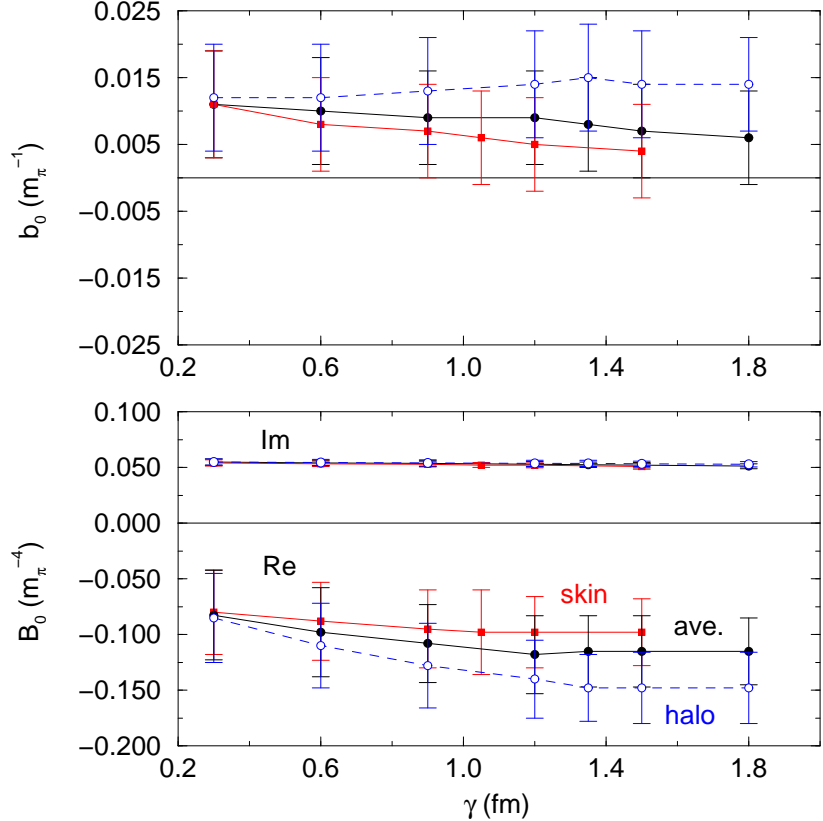}
\caption{Same as Fig.~\ref{fig:piZRC2} but with finite range 
folding applied to the $p$-wave potential, see text.}
\label{fig:piYC2}
\end{figure}

An obvious next step is to try some finite-range (FR) modification of
the otherwise zero range pion-nucleus potential. This can be achieved
by folding the densities with e.g. a Gaussian such that each density
$\rho $ in the potential is replaced by a folded one $\rho ^F$, see 
Eq.~(\ref{eq:fold}). It is found that when folded densities are used the 
$\chi ^2$ becomes a monotonic increasing function of the range parameter 
$\beta$. However, when folding is introduced separately into the $s$-wave 
and into the $p$-wave parts of the potential, then a minimum is obtained
for $\chi ^2$ when the rms radius of the finite range function 
is 0.9$\pm$0.1 fm, provided FR is applied {\it only} to the $p$-wave term. 
The same results are obtained  when a Yukawa FR function is used instead 
of the Gaussian. Figure~\ref{fig:piYC} shows results of global fits to 
pionic atom data when a FR folding with rms radius of 0.9 fm is applied 
to the $p$-wave part of the potential. Comparing with Fig.~\ref{fig:piZRC} 
we note that again the `skin' shape yields the best fit but now it is 
obtained for $\gamma$=1.1~fm, which is an acceptable value. 
However, the difference between $b_1$ and its free $\pi N$ value 
is very large, representing a significant increased repulsion 
in the nuclear medium. Figure~\ref{fig:piYC2} is quite similar to 
Fig.~\ref{fig:piZRC2} with non-zero values for Re$B_0$. 
Although there is no acceptable theory for the empirical parameter $B_0$, 
it is believed that the absolute value of its real part should be smaller 
than the imaginary part \cite{CRi79,SHO95}, which is not the case here. 

The present results mean that the in-medium $s$-wave repulsion, as
obtained from global fits to pionic atom data, is significantly
enhanced compared to expectations, partly via the extra repulsion provided 
by $b_1$ and partly by the dispersive Re$B_0$ being more repulsive
than expected. The sum of these two effects is the well-known
pionic atoms `anomaly'. This is the most bias-free way of presenting the
`anomaly', or `anomalous $s$-wave repulsion'.

\subsubsection{The $s$-wave anomaly and the issue of 
chiral symmetry restoration}
\label{sec:anomaly}

The so-called $s$-wave anomaly or the extra repulsion observed in fits
to pionic atom data had been known for a very long time~\cite{BFG97}.
The previous section presented a state-of-the-art summary of this
topic, based on global fits to 100 data points, respecting our
knowledge on the $r_n-r_p$ rms radii difference and otherwise without
additional assumptions. Global approaches always yielded the result
that the extra repulsion observed in the phenomenological potential
is due to two sources, namely, an enhanced $b_1$ coefficient and 
an unexpectedly large repulsion of the dispersive quadratic 
term, albeit with large uncertainty.
An alternative approach, mainly due to Yamazaki and co-workers, has
been to handle restricted data sets and to reduce
the number of parameters in the potential by making some assumptions.
Basically they used the approach of Seki and Masutani~\cite{SMa83}
where due to the correlations between $b_0$ and Re$B_0$ and the assumption
of an average or an effective density, the two terms plus the isoscalar 
double-scattering contribution of Eq.~(\ref{eq:b0b}) are lumped together, 
resulting in a single effective isoscalar real part linear in the density. 
Using the deeply bound 1$s$ states in Sn isotopes 
(see Table~\ref{tab:piat_s}) together with 1$s$ states in $^{16}$O,
$^{20}$Ne and $^{28}$Si, a total of 12 points, they obtain
$b_1=-0.1149\pm0.0074m_\pi ^{-1}$ (see Ref.~\cite{KYa04} for full details.)
This may be compared with the value of $b_1$ that corresponds to the
minimum of $\chi ^2$ in Fig.~\ref{fig:piYC} for the `skin' shape of 
$\rho _n$, namely $b_1=-0.109\pm0.005m_\pi ^{-1}$. 
The agreement between the two results is very good, and a significant 
discrepancy is thus established beyond any doubt with respect to the 
free $\pi N$ value $b_1^f=-0.0864\pm0.0010m_\pi ^{-1}$ derived from 
the preliminary results of the PSI measurement of the $\pi^-$H 
$1s$ level shift and width \cite{Mar06}. Note, however, the {\it conceptual} 
differences between the two methods: 

\begin{itemize} 

\item The Yamazaki et al. approach makes particular assumptions on the 
neutron densities of only few nuclides for $1s$ atomic states. 
In our global approach, where it is found that the major effect of the 
unknown neutron densities is through $r_n-r_p$, the $(N-Z)/A$ dependence 
of $r_n-r_p$ is an average over 36 nuclides spanning a full range of atomic 
states (from $1s$ to $4f$). 

\item Due to the use of `effective density' as mentioned above, and $N=Z$ 
nuclei dominated by the isoscalar $\pi N$ interaction, it is somewhat 
ambiguous in the Yamazaki et al. approach to determine $b_0$ and Re$B_0$ 
independently of each other. 
Thus, assuming $b_0=0$, they obtain Re$B_0 = -0.033 \pm 0.012m_\pi ^{-4}$ 
[Eq.~(77) of Ref.~\cite{KYa04}]. Our data base includes 100 rather than 12 
points, where all are significant in determining parameters of the $s$-wave 
part of the potential since this part contributes large fractions of the 
strong interaction effects also for states with $l>0$. 
Moreover, no assumptions are made regarding the terms nonlinear in density.  
In addition figure~\ref{fig:piYC2} shows that our deduced value of $b_0$
for the `skin' shape for $\rho _n$ is essentially in agreement with the 
free $\pi N$ value $b_0^f=+0.0068\pm 0.0031m_\pi ^{-1}$, deduced from the 
preliminary PSI results for the $\pi^-$H $1s$ level shift and width 
\cite{Mar06}. 
If this value were assumed in Ref.~\cite{KYa04}, then they would have 
derived a value of Re$B_0 = -0.064 \pm 0.012m_\pi ^{-4}$ agreeing within 
error bars with our values as depicted in Fig.~\ref{fig:piYC2}. 

\end{itemize} 

A systematic study of the uncertainties in parameters of the potential
and their dependence on the size of the data base and on assumptions made 
and constraints imposed in the analysis can be found in Ref.~\cite{FGa03}.
In what follows we focus attention on the in-medium values of $b_1$.
It is shown below that eventually the problem with Re$B_0$ is also solved.

The renewed interest in recent years in the `anomalous' $s$-wave repulsion 
in the pion-nucleus interaction at threshold, as found in phenomenological 
analyses of strong interaction effects in pionic atoms, is partly due to the 
suggestion by Weise~\cite{Wei00,Wei00a} that such enhancement could  be 
expected, at least in the isovector channel, to result from a chirally 
motivated approach where the pion decay constant becomes effectively 
density dependent in the nuclear medium. Since $b_{1}$ in free-space is well 
approximated in lowest chiral-expansion order by the Tomozawa-Weinberg (TW) 
expression~\cite{Tom66,Wei66}
\begin{equation} 
\label{eq:b1TW} 
b_{1}=-\frac{\mu_{\pi N}}{8 \pi f^{2}_{\pi}}=-0.08m^{-1}_{\pi} \,,
\end{equation}
with $\mu_{\pi N}$ the pion-nucleon reduced mass, then it may be argued that 
$b_{1}$ will be modified in pionic atoms if the free-space pion decay constant 
$f_\pi = 92.4$ MeV is modified in the medium. QCD coupled with PCAC relates 
$f_\pi$ to the quark condensate $<\bar q q>$: 
\begin{equation} 
\label{eq:GOR} 
\frac{f_\pi^{*2}(\rho)}{f_\pi^2} = \frac{<\bar q q>_{\rho}}{<\bar q q>_0} 
\simeq {1 - {{\sigma \rho} \over {m_{\pi}^2 f_{\pi}^2}}} \,, 
\end{equation} 
where $\sigma$ is the $\pi N$ sigma term and where the last step 
provides the leading term in a density expansion of the quark condensate 
\cite{DLe91} assuming that the charge-averaged pion mass does not change 
in the medium.\footnote{This holds for the temporal version of $f_\pi$, 
corresponding to the vacuum-to-pion matrix element of the time component 
of the axial current, c.f. Refs.~\cite{TWi95,MOW02}.} Thus $f_\pi$ is expected 
to decrease in the nuclear medium by about 20$\%$ at nuclear-matter density 
$\rho_0 \simeq 0.16$~fm$^{-3}$, using $\sigma \simeq 50$~MeV~\cite{GLS91}. 
The form of Eq.~(\ref{eq:b1TW}) suggests then a density-dependent 
isovector amplitude such that $b_1$ becomes 
\begin{equation} 
\label{eq:ddb1}
b_1(\rho) = \frac{b_1(0)}{1 - {\sigma \rho} / {m_{\pi}^2 f_{\pi}^2}} 
= \frac{b_1(0)}{1-2.3\rho} 
\end{equation}
for $\sigma $=50 MeV and with $\rho$ in units of fm$^{-3}$, 
resulting in an increase of $b_1$ in the nuclear medium by about 60$\%$
at nuclear-matter density $\rho_0 \simeq 0.16$ fm$^{-3}$. This ansatz 
\cite{Wei00,Wei00a} was applied at almost the same time in 
two different analyses 
of pionic atom data. Kienle and Yamazaki \cite{KYa01} outlined a method for 
extracting $b_1$ from analyses of very limited data sets, which is 
essentially the method described above, but using a fixed average value of 
the density-dependent $b_1(\rho)$ of Eq.~(\ref{eq:ddb1}). 
Friedman \cite{Fri02} presented results of global analyses, as outlined above 
but with 60 data points compared to the present 100 points, using explicitly 
Eq.~(\ref{eq:ddb1}) for the density dependent $b_1(\rho)$. It was shown 
that indeed most of the difference between the derived $b_1$ and its free 
$\pi N$ value disappeared when the above density dependence was included. 

\begin{figure}
\includegraphics[scale=0.7,angle=0]{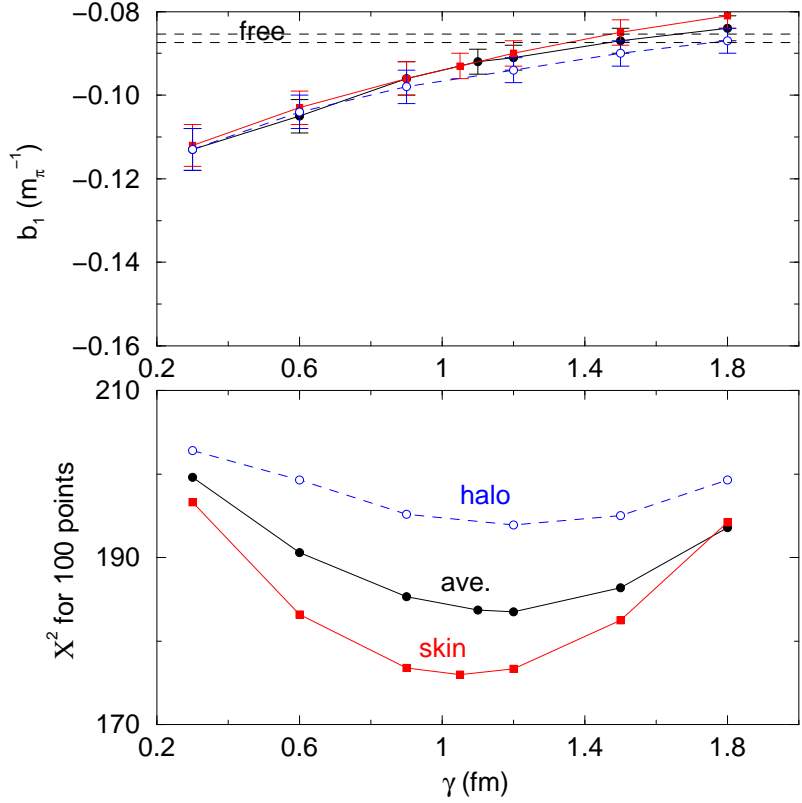} 
\caption{Results of global fits for different values of the neutron 
radius parameter $\gamma $ (Eq.~(\ref{eq:RMF})) with $b_1(\rho)$ given 
by Eq.~(\ref{eq:ddb1}). Lower part - values of $\chi ^2$ for 100 data points 
for three shapes of the neutron density $\rho _n$. 
Upper part - the corresponding isovector parameter 
$b_1$ in comparison to its free $\pi N$ value (marked `free').}
\label{fig:piYW} 
\end{figure} 

\begin{figure}
\includegraphics[scale=0.7,angle=0]{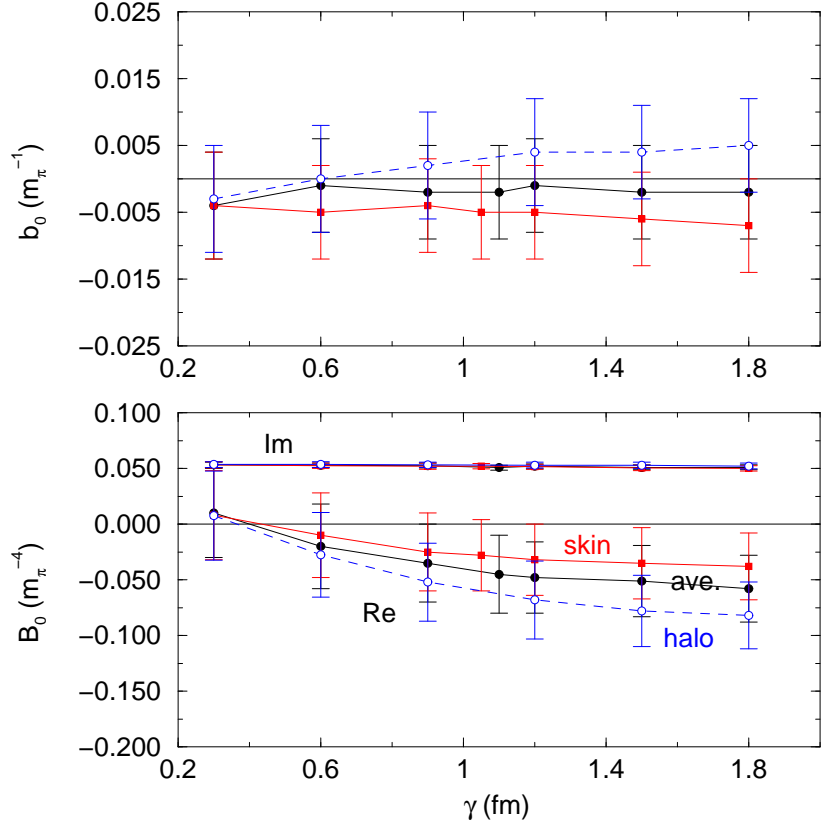}
\caption{Results of global fits for different values of the neutron 
radius parameter $\gamma $ (Eq.~(\ref{eq:RMF})) with $b_1(\rho)$ given 
by Eq.~(\ref{eq:ddb1}). Lower part - values of the complex parameter $B_0$.
Upper part - the resulting isoscalar parameter $b_0$.} 
\label{fig:piYW2}
\end{figure}

Figure~\ref{fig:piYW} shows results of global fits to the 100 data 
points for pionic atoms, with $b_1(\rho)$ given by Eq.~(\ref{eq:ddb1}). 
The parameter $b_1$ stands here for $b_1(\rho=0)$. It is seen that 
for the lowest minimum of $\chi ^2$, i.e. the minimum of the curve 
obtained for the `skin' shape for the neutron density, $b_1$ is 
much closer to the free $\pi N$ value than it was in Fig.~\ref{fig:piYC} 
where a fixed $b_1$ was assumed, and they are almost in agreement. 
Figure~\ref{fig:piYW2} shows that $b_0$ now is consistent with zero and 
that Re$B_0$ is much less repulsive than before and is acceptable 
being almost consistent with zero. 

The renormalization of $f_\pi$ in dense matter, as given by
Eq.~(\ref{eq:GOR}) and leading to the related renormalization of the
isovector amplitude $b_1$ in Eq.~(\ref{eq:ddb1}), may also be derived
under simplifying assumptions by solving the KG equation in infinite
nuclear matter of protons and neutrons for a pion chiral $s$-wave
polarization operator $\Pi_s(E)$ near threshold~\cite{KKW03,KKW03a}.
In chiral-perturbation theory, in the limit of vanishing $m_{\pi}/M$ and
for zero momentum pions, {\bf q} = 0, the leading terms to order
$E/(4\pi f_\pi)$ of the $\pi N$ $s$-wave amplitudes give rise to the
following form of the $s$-wave pion polarization operator:
\begin{equation}
\Pi_s(E) = 2 E V^{(s)}_{\rm opt}(E) \approx
\tau_z{E\over 2f_\pi^2}(\rho_n - \rho_p)
- {(\sigma - \beta E^2)\over f_\pi^2}
(\rho_p + \rho_n) ~,
\label{eq:Vchiral}
\end{equation}
where $E$ is the pion energy including its rest mass $m_\pi$ and
$\tau_z = +1,0,-1$ for $\pi^-,\pi^0,\pi^+$, respectively, in the
notation of Eq.~(\ref{eq:Vopt}). The first, TW isovector term is
the dominant one near threshold, providing repulsion for $\pi ^-$
for all nuclei with neutron excess. The second, isoscalar term is
nearly zero at threshold, so it is reasonable to fit $\beta$ by
requiring $\beta = \sigma / m_{\pi}^2$. The polarization
operator $\Pi_s(E)$ satisfies a KG equation (for {\bf q} = 0):
\begin{equation}
E^2 - m_{\pi}^2 - \Pi_s(E) = 0
\label{eq:KGpol}
\end{equation}
which by inserting Eq.~(\ref{eq:Vchiral}) becomes
\begin{equation}
(1 - {\sigma\rho \over m_{\pi}^2 f_{\pi}^2}) (E^2 - m_{\pi}^2)
- \tau_z {E \over 2f_\pi^2}(\rho_n - \rho_p) = 0 \,,
\label{eq:disp}
\end{equation}
with $\rho = \rho_p + \rho_n$. When recast into the form
\begin{equation}
\Pi_s(E) = E^2 - m_{\pi}^2 = \tau_z \frac{E}{2 f_{\pi}^{*2}}
(\rho_n - \rho_p) \,,
\label{eq:renorm}
\end{equation}
with the effective density-dependent pion decay constant
$f_{\pi}^*$ defined by
\begin{equation}
f_{\pi}^{*2} = f_{\pi}^2 \left(1 - \frac{\sigma \rho}
{m_{\pi}^2 f_{\pi}^2}\right) \,,
\label{eq:fpi*}
\end{equation}
it is seen to be equivalent, at threshold,
to Eqs.~(\ref{eq:b1TW}-\ref{eq:ddb1})
with the renormalized effective pion decay constant $f_{\pi}^*$.
Equation~(\ref{eq:renorm}) strictly speaking is satisfied for one specific
value of energy (`self energy') near threshold, $E \gtrsim m_{\pi}$.
Approximating then $\Pi_s(E)$ by the right-hand side at $E = m_{\pi}$,
the appearance of $f_{\pi}^*$ instead of $f_{\pi}$ may be attributed
to a wave-function renormalization effect \cite{KKW03,KKW03a}.

Switching from infinite nuclear matter considerations to actual pionic-atom
calculations, the finite-size Coulomb potential $V_c$ needs to be introduced
properly into a KG equation in which $\Pi_s(E)$ serves as a {\it given} input,
not as an entity to solve Eq.~(\ref{eq:KGpol}) for.
Schematically, Eq.~(\ref{eq:KGpol}) is replaced by
\begin{equation}
\left[ (E-V_c)^2 - m_{\pi}^2  - \Pi_s(E-V_c) \right] \psi = 0 \,,
\label{eq:KGcoul}
\end{equation}
where the Coulomb potential $V_c$ enters via the minimal substitution
requirement~\cite{ETa82}, $\Pi_s(E) \to \Pi_s(E - V_c)$.
When the chiral version of the $s$-wave pion polarization operator
Eq.~(\ref{eq:Vchiral}) was used in a KG equation of the type of
Eq.~(\ref{eq:KGcoul}), it was found in the global fits to pionic atom
data reported in Ref.~\cite{FGa04} that a large over-correction of $b_1$
occurred, to a value $b_1 = -0.068 \pm 0.004 m_\pi ^{-1}$, significantly
less repulsive than $b_1^f$. In this case, it is the combined effect of
the isoscalar and the isovector amplitudes that is responsible for
over-shooting $b_1^f$. Therefore, although the effect of including the
energy dependence of the {\bf q} = 0 chiral amplitudes goes in the
desired direction, it does not provide a quantitative resolution of
the $s$-wave anomaly problem. Different results, and conclusions,
are obtained when fits are made to partial data sets which may not
carry sufficient statistical significance to decide one way or another
on this issue \cite{KKW03,KKW03a}.

\begin{figure}
\includegraphics[scale=0.7,angle=0]{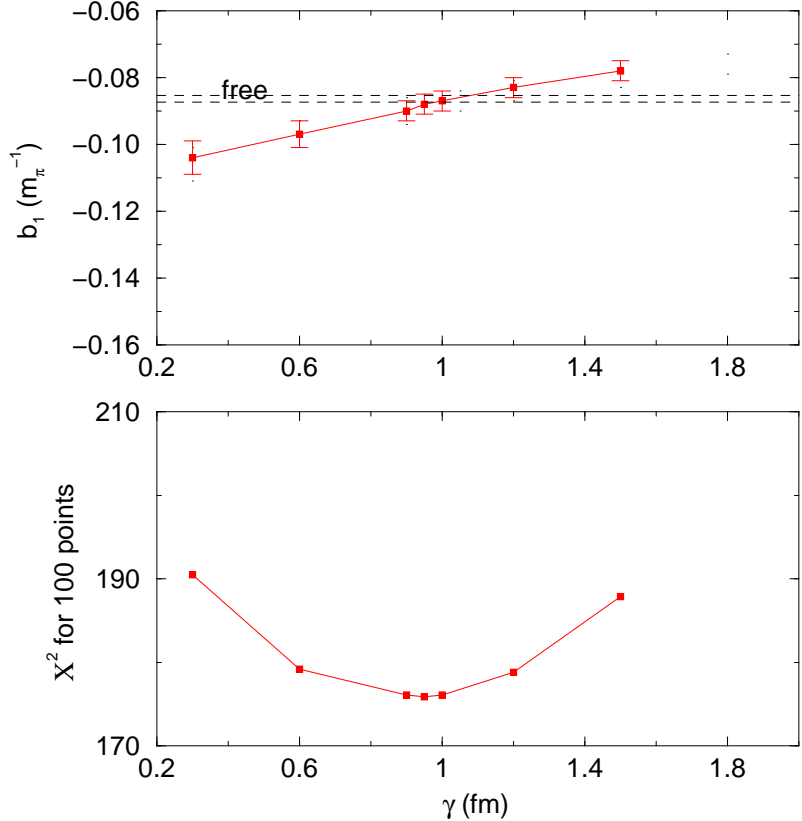}
\caption{Results of global fits for different values of the neutron
radius parameter $\gamma $ (Eq.~(\ref{eq:RMF})) with $b_1(\rho)$ given
by Eq.~(\ref{eq:ddb1}) and the empirical energy dependence of the free
$b_0(E)$ included. Lower part - values of $\chi ^2$ for 100 data points
for the `skin' shape of the neutron density $\rho _n$. Upper part - the
corresponding isovector parameter $b_1$ in comparison to its free $\pi N$
value (marked `free').}
\label{fig:piYEW}
\end{figure}

Friedman and Gal~\cite{FGa04} discussed also an alternative procedure
in which the empirical {\it on-shell} energy dependent $\pi N$ amplitudes
are used for implementing the minimal substitution requirement
$E \to E - V_c$. As pointed out by Ericson~\cite{Eri94} the on-shell
approximation follows naturally for strongly repulsive short-range $NN$
correlations from the Agassi-Gal theorem~\cite{AGa73} for scattering off
non-overlapping nucleons. The energy dependence of the `empirical'
amplitudes is weaker than that of the `chiral' ones. For $b_1(E)$ there
is hardly any energy dependence, in contrast to the energy dependence
of the `chiral' $b_1(E)$ which by itself would suffice to recover
$b_1^f$, the free-space value of $b_1$, in the pionic-atom global fits.
For $b_0(E)$ the empirical energy dependence is only about 60\% of the
`chiral' effect. When this weaker empirical energy dependence of $b_0$
was used, the resulting $b_1$ was still too repulsive in comparison with
$b_1^f$, but closer to $b_1^f$ than the $b_1$ resulting in the conventional
fixed $b_1$ model.

Figure~\ref{fig:piYEW} is similar to Fig.~\ref{fig:piYW} but with the
energy dependence of the free $\pi N$ $b_0$ included {\it in addition}
to applying the density-dependent renormalization of Eq.~(\ref{eq:ddb1})
for $b_1$.
Comparing with Fig.~\ref{fig:piYW} the values of $b_1$ have shifted
now and at the minimum of $\chi ^2$ the agreement with the free
$\pi N$ value is perfect. The resulting $b_0$ (not shown)
of $b_0=-0.009\pm0.007 m_\pi ^{-1}$ is close to the free value and
Re$B_0$ (not shown) is essentially zero at $-0.005\pm0.035 m_\pi ^{-4}$.
The chiral-motivated isovector $b_1(\rho)$ Eq.~(\ref{eq:ddb1})
together with the empirical energy dependence therefore provides all the
required extra repulsion, with the isoscalar $b_0$ being essentially zero
and with no significant dispersive Re$B_0$ term needed.

\subsubsection{Radial sensitivity of pionic atoms}
\label{sec:piradsen}

The chiral-motivated dependence of the isovector term $b_1$, as given
by Eq.~(\ref{eq:ddb1}), almost completely removed the `anomaly' observed
when a fixed value was used for $b_1$. Not only with Eq.~(\ref{eq:ddb1})
does now one get for $b_1(0)$ the free $\pi N$ value, but also the fit
value of $b_0$ is very close to its free value and Re$B_0$ is zero.
This is obtained without the need to introduce `effective' density
or to make any assumptions regarding Re$B_0$. It is concluded, therefore,
that Eq.~(\ref{eq:ddb1}) is a fair representation of the medium-modification
of the isovector term of $q(r)$, Eq.~(\ref{eq:EE1s}). It is instructive to
examine further the radial sensitivity of this term.

The radial sensitivity of exotic atom data was addressed before
\cite{BFG97} with the help of a `notch test', introducing a local
perturbation into the potential and studying the changes in the
fit to the data as function of position of the perturbation. The
results gave at least a semi-quantitative information on what are
the radial regions which are being probed by the various types of
exotic atoms. However, the radial extent of the perturbation
was somewhat arbitrary and only very recently that approach was
extended \cite{BFr07} into a mathematically well-defined limit.

In order to study the radial sensitivity of {\it global}
fits to exotic atom data, it is necessary to define the radial position 
parameter globally using as reference e.g. the known charge distribution 
for each nuclear species in the data base. The radial position $r$
is then defined as $r=R_c+\eta a_c$, where $R_c$ and $a_c$ are the radius 
and diffuseness parameters, respectively, of a 2pF charge 
distribution \cite{FBH95}. In that way $\eta$ becomes the relevant radial 
parameter when handling together data for several nuclear species along 
the periodic table. The value of $\chi ^2$ is regarded now as a functional
of a global optical potential $V(\eta)$, i.e. $\chi ^2=\chi^2[V(\eta)]$,
where the parameter $\eta$ is a {\it continuous} variable.
It leads to \cite{BFr07,FD_wikipedia}
\begin{equation} 
\label{eq:dchi2} 
d\chi^2 = \int d\eta \frac{\delta \chi^2}{\delta V(\eta)} \delta V(\eta) \;,
\end{equation}
where
\begin{equation}
\label{eq:FD}
\frac{\delta \chi^2[V(\eta)]}{\delta V(\eta')}
= \lim_{\sigma \rightarrow 0}\lim_{\epsilon _V \rightarrow 0 }
\frac{\chi^2[V(\eta)+\epsilon _V\delta_{\sigma}(\eta-\eta')]-\chi^2[V(\eta)]}
     {\epsilon _V}\;
\end{equation}
is the functional derivative (FD) of $\chi^2[V]$.
The notation $\delta_{\sigma}(\eta-\eta')$ stands for an approximated
$\delta$-function and $\epsilon _V $ is a change in the potential.
From Eq.~(\ref{eq:dchi2}) it is seen that the FD determines 
the effect of a local
change in the optical potential on $\chi^2$. Conversely it can be said that
the optical potential sensitivity to the experimental data is determined by
the magnitude of the FD.
Calculation of the FD may be carried out by multiplying the
best fit potential by a factor
\begin{equation}
\label{eq:FDfac}
 f=1+\epsilon \delta_{\sigma}(\eta-\eta')
\end{equation}
using a normalized Gaussian with a range parameter $\sigma$ for the
smeared $\delta$-function,
\begin{equation}
\label{eq:Gauss}
\delta_{\sigma}(\eta-\eta')=\frac{1}{\sqrt{2\pi}\sigma}e^{-(\eta-\eta')^2/2\sigma^2}.
\end{equation}
For finite values of $\epsilon$ and $\sigma$ the FD can 
then be approximated by
\begin{equation}
\label{eq:FDnum}
\frac{\delta \chi^2[V(\eta)]}{\delta V(\eta')}
\approx \frac{1}{V(\eta')}
\frac{\chi^2[V(\eta)(1+\epsilon\delta_{\sigma}(\eta-\eta'))]
      -\chi^2[V(\eta)]}
     {\epsilon}\;.
\end{equation}
The parameter $\epsilon $ is used for a {\it fractional}
change in the potential
and the limit $\epsilon \to 0$ is obtained numerically for several
values of $\sigma $ and then extrapolated to $\sigma =0$.

\begin{figure}
\includegraphics[scale=0.7,angle=0]{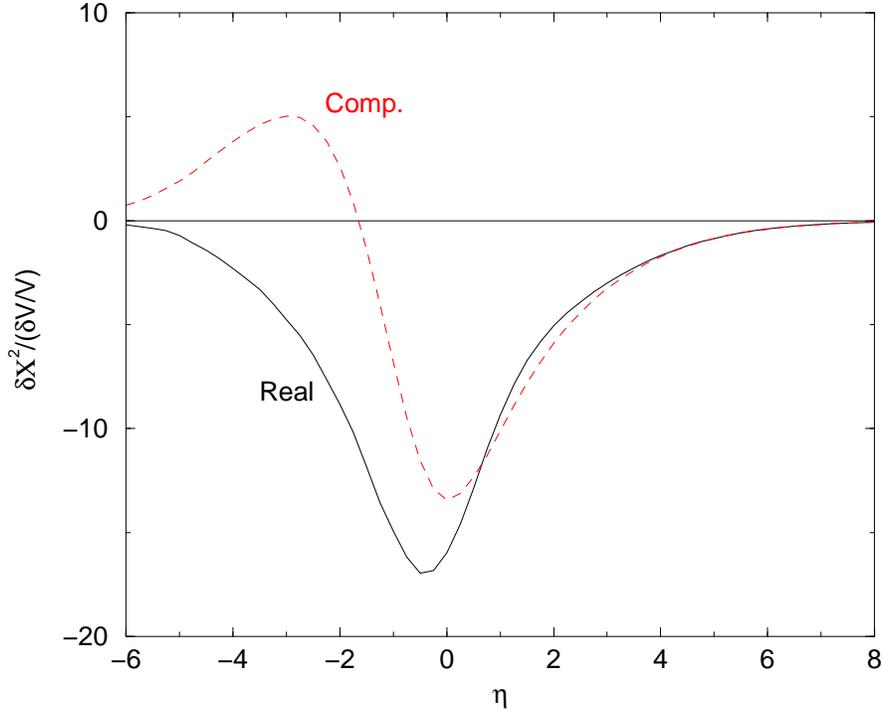}
\caption{Functional derivatives of the best fit $\chi ^2$
with respect to relative changes in the real part (solid curve) and 
with respect to relative changes in
the full complex (dashed) $s$-wave
pion nucleus potential Eq.~(\ref{eq:EE1s}), with $b_1(\rho)$ of
Eq.~(\ref{eq:ddb1})}
\label{fig:piFDW}
\end{figure}

Figure \ref{fig:piFDW} shows the FDs with respect to 
relative changes in the real part 
and with respect to relative changes in the full complex 
$s$-wave part of the best-fit pion-nucleus potential  
 where the chiral $b_1(\rho)$ is assumed.
Calculations of the FD with respect to the imaginary 
part of the potential show
additivity of the FDs, hence the difference between the FD for the
full complex potential and for the real part is the FD 
with respect to relative changes in the 
imaginary part of  the $s$-wave part of the potential.
It is immediately clear from the figure that both parts of the potential
have quite similar importance in determining strong interaction
effects in pionic atoms. Turning to radial  sensitivity, recall that
$\eta =-2.2$ corresponds to 90\% of the central density of the nuclear
charge and $\eta =2.2$ corresponds to 10\% of that density. It is 
therefore clear that pionic atom data are sensitive to the $s$-wave part 
of the potential over a region where the densities change between free
space to the full central density. It is no wonder that
using a fixed value for $b_1$ in fits to data led to `abnormal' values 
for this parameter. With the success of the chiral $b_1(\rho)$
there is no reason for employing approximations such as linearization, 
by using an effective density.

\subsection{Pion elastic scattering}
\label{sec:piscat}

\begin{figure}
\includegraphics[scale=0.7,angle=0]{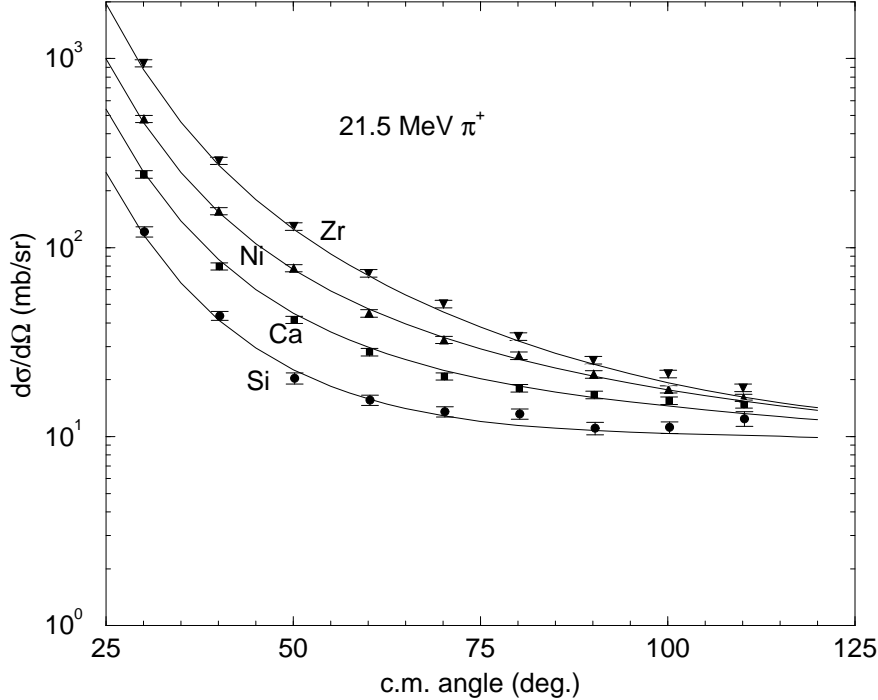}
\caption{Experimental results for elastic scattering of $\pi ^+$
compared with calculations.}
\label{fig:piscatt+}
\end{figure}

The applicability of the Kisslinger potential \cite{Kis55} 
and its Ericson-Ericson extension \cite{EEr66} 
(Eqs.~(\ref{eq:EE1s}-\ref{eq:angtr}))
to elastic scattering of pions by nuclei had been demonstrated early
in the days of the `pion factories' \cite{SMC79,SMa83}, mostly at energies
higher than 40-50~MeV. Following the success with the `chiral' density
dependence of $b_1$ in the subthreshold regime of pionic atoms, it is 
interesting to study the behavior of the pion-nucleus potential across
threshold into the scattering regime and to examine
whether the pionic atom `anomaly' is observed also above threshold
when using a fixed value for the parameter $b_1$. Of particular interest is
the question of whether the density dependence of the isovector amplitude 
as given by Eq.~(\ref{eq:ddb1}) is required 
by the scattering data. In the scattering scenario, unlike in the atomic
case, one can study both charge states of the pion, thus increasing
sensitivities to isovector effects and to the energy dependent effects
due to the Coulomb interaction.
Looking for earlier suitable data for elastic scattering, 
it was somewhat surprising to realize that at
kinetic energies well below~50 MeV there seemed to be only one set of
high quality data available for both charge states of the pion obtained in
the same experiment,
namely, the data of Wright et al.~\cite{Wri88} for 19.5 MeV pions on 
calcium, with predominantly the $N=Z$ isotope $^{40}$Ca.
For that reason precision measurements of elastic scattering
of 21.5 MeV $\pi ^+$ and $\pi ^-$ by several nuclei were performed very
recently at PSI \cite{FBB04,FBB05} with the aim of analyzing 
the results in terms of the
same effects as found
in pionic atoms.
The experiment was dedicated to studying the elastic scattering of {\it both}
pion charge states and special emphasis was placed on the absolute
normalization of the cross sections, which was based on the parallel
measurements of Coulomb scattering of muons.

\begin{figure}
\includegraphics[scale=0.7,angle=0]{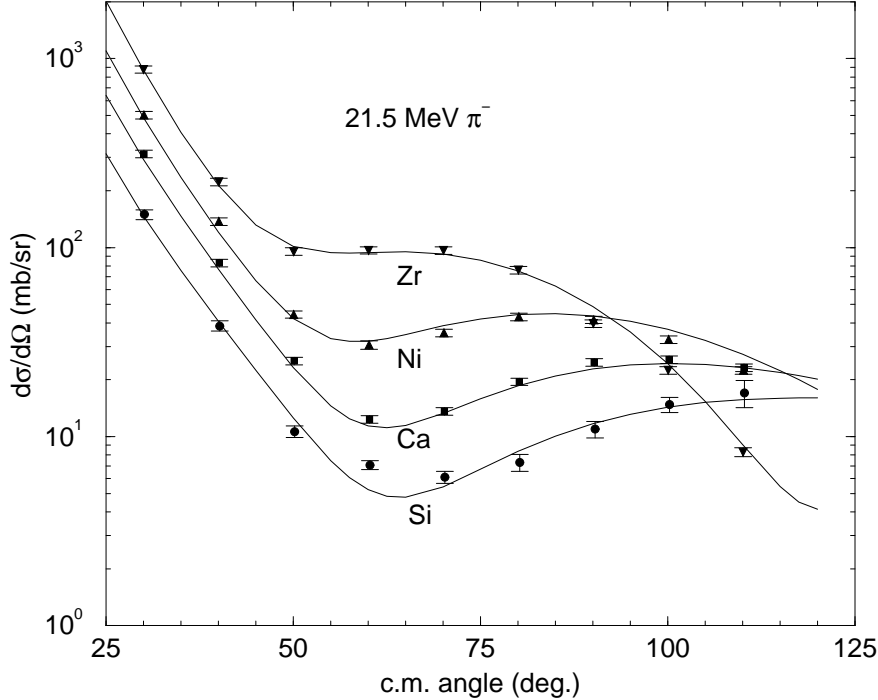}
\caption{Experimental results for elastic scattering of $\pi ^-$
compared with calculations.}
\label{fig:piscatt-}
\end{figure}

Figures \ref{fig:piscatt+} and \ref{fig:piscatt-} show comparisons
between experimental differential cross sections for the elastic
scattering of 21.5~MeV $\pi^+$ and $\pi^-$ by Si, Ca, Ni and Zr
and calculations made with the best-fit EW potential, defined below. 
The fit to the data was made to all eight angular distributions put 
together, see Ref.~\cite{FBB05} for details. 
Table~\ref{tab:scatt} summarizes values of $b_1$ obtained for the various
models, with C indicating a conventional (fixed $b_1$) potential,
W indicating the use of the chiral-motivated $b_1(\rho)$ of
Eq.~(\ref{eq:ddb1}) and EW stands for using $b_1(\rho)$ and including also 
the empirical energy dependence of $b_0$. 
It is evident that with the C potential using a fixed value for $b_1$ 
the fit to elastic scattering results yields too repulsive a value for 
$b_1$ in comparison with the corresponding value for the free $\pi N$ 
interaction, much the same as is the case with pionic atoms. When 
the chiral motivated $b_1(\rho)$ of Eq.~(\ref{eq:ddb1}) is used
$\chi ^2$ is reduced significantly and the
resulting $b_1$ is essentially in agreement with the free value. 
Adding the empirical energy dependence of $b_0$
brings the two values into full agreement,
much the same as is the case with pionic atoms.
As with pionic atoms, the chiral energy dependence in 
Eq.~(\ref{eq:Vchiral}) caused large over-correction
of the resulting $b_1$ which turns out to be significantly less
repulsive than its free $\pi N$ value.

\begin{table}
\caption{Values of $b_1$ from fits to elastic scattering
of 21.5 MeV $\pi ^\pm$ by Si, Ca, Ni and Zr. \newline
The free $\pi N$ value is  $b_1^f=-0.0864\pm0.0010m_\pi ^{-1}$.}
\label{tab:scatt}
\begin{ruledtabular}
\begin{tabular}{lccc}
 model    &  C & W & EW \\ \hline
$b_1 (m_\pi ^{-1}$)& $-0.114\pm0.006$ &
 $-0.081\pm0.005$ & $-0.083\pm0.005$\\
$\chi ^2$ for 72 points & 134 & 88  & 88 \\
\end{tabular}
\end{ruledtabular}
\end{table}

It is therefore concluded that the extra-repulsion or $s$-wave
anomaly is observed also in the scattering regime and
that the medium-modification of the isovector term
of the local part of the pion-nucleus potential is of the same
nature both below and just above threshold.

\subsection{Conclusions}
\label{sec:piconc}

Pionic atoms form the oldest type of exotic atom of a strongly-interacting
particle and both experiment and theory had achieved maturity already
in the 1990s, with the nagging problem of anomalously enhanced repulsion
in the  $s$-wave part of the potential essentially unresolved.
The discrepancy between the in-medium $b_1$ and its free $\pi N$ value
became clearer with the ever improving accuracy of the experimental
results on pionic hydrogen. In contrast, it seemed that 
experimental X-ray spectroscopy
of medium-weight and heavy pionic atoms reached a dead end without
new experiments in the last 20 years. 
The predicted  existence of well-defined 
1$s$ and $2p$ states in heavy pionic atoms, due to saturation of
widths caused by the $s$-wave repulsion, prompted the pioneering (d,$^3$He)
experiments which supplied strong interaction data unreached otherwise.
Although the overall picture of pionic atom potentials has not changed
due to these new data, they gave added impetus to the study of medium
effects in the pion-nucleus interaction at threshold. The 
issue of the enhanced repulsion
in the  $s$-wave part of the potential appears to have been resolved with
the chiral-motivated $b_1(\rho)$ of Eq.~(\ref{eq:ddb1}) which was shown
to apply also at 21 MeV thanks to a recent dedicated experiment on 
elastic scattering of $\pi ^\pm$ by nuclei.
It may be concluded that the dominant effect of the nuclear environment
on the real part of the pion-nucleus potentials close to threshold is
given by Eq.~(\ref{eq:ddb1}).
Note, however, that the absorption terms quadratic in density 
remain largely phenomenological.

 
\section{${\bf \bar K}$ nuclear physics} 
\label{sec:kbars} 

\subsection{Preview} 
\label{sec:Kbarprev} 

\begin{figure}[t] 
\centerline{\includegraphics[scale=0.7,angle=0]{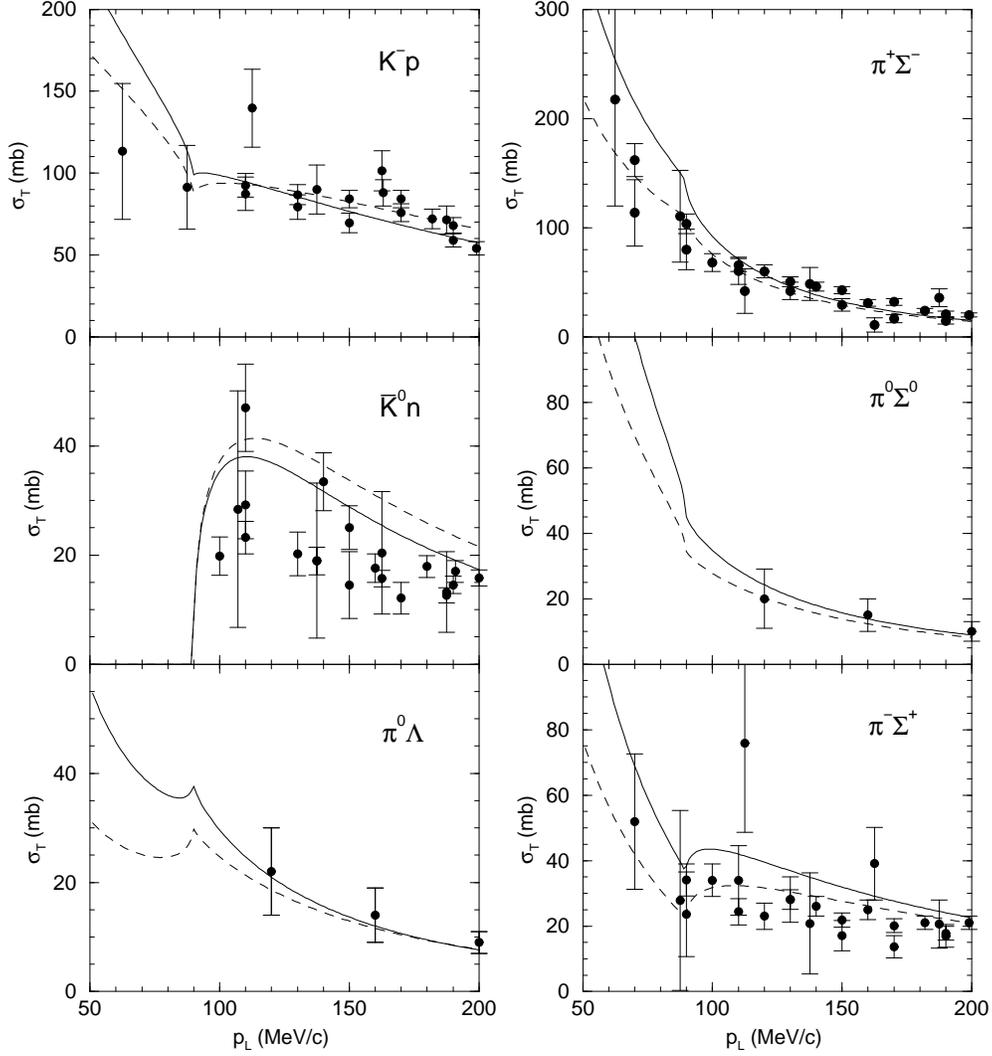}} 
\caption{Calculations from Ref.~{\protect \cite{CFG01}} of cross sections 
for $K^{-}p$ scattering and reactions. The dashed lines show free-space 
chiral-model coupled-channel calculations using amplitudes from 
Ref.~{\protect \cite{WKW96}}. The solid lines show chiral-model 
coupled-channel calculations using slightly varied parameters in order 
to fit also the $K^-$-atom data for a (shallow) optical potential 
calculated self consistently.}  
\label{fig:kminuspdata} 
\end{figure} 

The $\bar K$-nucleus interaction near threshold is strongly attractive and 
absorptive as suggested by fits to the strong-interaction shifts and widths 
of $K^-$-atom levels~\cite{BFG97}. Global fits yield `deep' density dependent 
optical potentials with nuclear-matter depth 
Re$V_{\bar K}(\rho_0)\sim-$(150-200) MeV~\cite{FGB93,FGB94,FGM99,MFG06,BFr07}, 
whereas in the other extreme case several studies that fit the low-energy 
$K^-p$ reaction data, including the $I=0$ quasibound state $\Lambda(1405)$ 
as input for constructing self consistently density dependent optical 
potentials, obtain relatively `shallow' potentials with 
Re$V_{\bar K}(\rho_0) \sim -$(40-60) MeV~\cite{SKE00,ROs00,CFG01,TRP01}. 
For a recent update of these early calculations, see Ref.~\cite{TRO06}. 
An example of a chirally motivated coupled-channel fit to the low-energy 
$K^-p$ cross sections is shown in Fig.~\ref{fig:kminuspdata} from 
Ref.~\cite{CFG01}. This calculation is based on the free-space chiral 
coupled-channel amplitudes used in the in-medium calculations of Waas et 
al.~\cite{WKW96} following the earlier work of Kaiser et al.~\cite{KSW95}. 
By imposing self consistency in its nuclear part (solid lines) the calculation 
\cite{CFG01} leads to a weakly density dependent shallow $\bar K$-nucleus 
potential in terms of the effective scattering length $a_{\rm eff}(\rho)$ of 
Fig.~\ref{fig:arho} in Sec.~\ref{sec:int}. 
As is shown below, `shallow' potentials are 
substantially inferior to `deep' ones in comprehensive fits to $K^-$-atom 
data. The issue of depth of Re$V_{\bar K}$ is reviewed below and the 
implications of a `deep' potential for the existence and properties of 
$\bar K$-nucleus quasibound states are discussed. Since the two-body 
$\bar K N$ interaction provides a starting point in many theoretical works 
for constructing the $\bar K$ nuclear optical potential $V_{\bar K}$, 
we start with a brief review of the $\bar K N$ data available near the 
$K^- p$ threshold.

\subsection{The $K^- p$ interaction near threshold} 
\label{sec:Kpnearth} 

The $K^- p$ data at low energies provide a good experimental base upon 
which models for the strong interactions of the $\bar K N$ system have 
been developed. Near threshold the coupling to the open $\pi \Sigma$ and 
$\pi \Lambda$ channels is extremely important, as may be judged from the 
size of the $K^- p$ reaction cross sections, particularly 
$K^- p \to \pi^+ \Sigma^-$, with respect to the $K^- p$ elastic 
cross sections shown in Fig.~\ref{fig:kminuspdata}. Theoretical models often 
include also the closed $\eta \Lambda,~\eta \Sigma,~K \Xi$ 
channels~\cite{ORa98}. Other threshold constraints are provided by the 
accurately determined threshold branching ratios~\cite{TDS71,NAD78} 
\begin{equation}
\label{eq:gammaK}
\gamma = \frac{\Gamma(K^- p \to \pi^+ \Sigma^-)}{\Gamma(K^- p \to
\pi^- \Sigma^+)} = 2.36 \pm 0.04 \,,
\end{equation}
\begin{equation} 
\label{eq:Rc} 
R_c = \frac{\Gamma(K^- p \to \pi^+ \Sigma^-,~\pi^- \Sigma^+)}
{\Gamma(K^- p \to {\rm all~inelastic~channels})} = 0.664 \pm 0.011 \,,
\end{equation}
\begin{equation} 
\label{eq:Rn} 
R_n  = \frac{\Gamma(K^- p \to \pi^0 \Lambda)}{\Gamma(K^- p \to 
\pi^0 \Lambda,~\pi^0 \Sigma^0)} = 0.189 \pm 0.015 \,. 
\end{equation} 
Additional sources of experimental information are the $\pi \Sigma$ invariant 
mass spectrum in the $I=0$ channel from various reactions, and the $K^- p$ 
scattering length deduced from the recent measurements at KEK~\cite{IHI97} 
and at Frascati~\cite{BBC05} (the DEAR collaboration) using a Deser-based 
formula~\cite{MRR04}: 
\begin{equation} 
\label{eq:Deser} 
\epsilon_{1s} - {\rm i} \frac{\Gamma_{1s}}{2} \approx 
-2\alpha^3\mu_{K^-p}^2a_{K^-p} 
(1-2\alpha\mu_{K^-p}({\ln}~\alpha - 1)a_{K^-p}) \,, 
\end{equation} 
where $\alpha$ is the fine-structure constant. The value of 
$a_{K^-p}$ derived from the DEAR measurement using this expression,  
$a_{K^-p} = (-0.45 \pm 0.09) + {\rm i}(0.27 \pm 0.12)~{\rm fm}$, 
appears inconsistent with most comprehensive fits to the low-energy 
$K^- p$ scattering and reaction data, as discussed by Borasoy 
et al.~\cite{BNW05a,BNW05b,BNW06,BMN06}. A dissenting view, however, 
is reviewed recently by Oller et al.~\cite{OPV07}. 

\begin{figure}
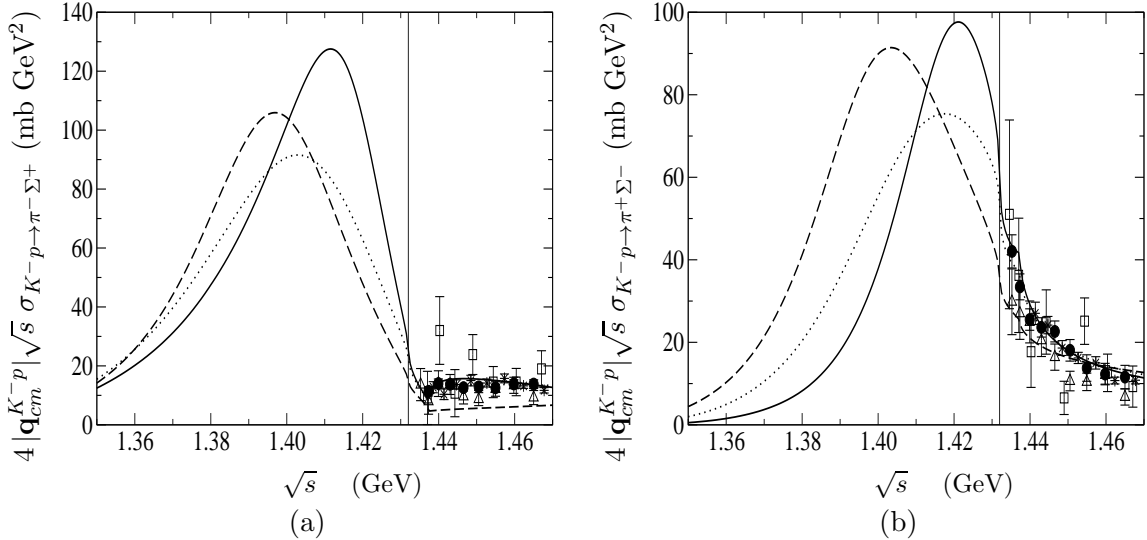

\centering
\begin{tabular}{p{0.4cm}cp{1.0cm}c}
 &
\begin{overpic}[height=6cm,width=0.40\textwidth,clip]{fg07fig18a.eps}
  \put(-11,-4){\rotatebox{90}{{\scalebox{1.0}{ 
    $4\,|\mathbf{q}_{cm}^{K^- p}| \sqrt{s} \ \sigma_{K^- p \to \pi^- \Sigma^+}$
(mb GeV$^2$)}}}} 
\put(45,-8){\scalebox{0.9}{$\sqrt{s}$ \quad (GeV)}}
\end{overpic}
& &
\begin{overpic}[height=6cm,width=0.40\textwidth,clip]{fg07fig18b.eps}
  \put(-11,-4){\rotatebox{90}{{\scalebox{1.0}{
$4\,|\mathbf{q}_{cm}^{K^- p}| \sqrt{s} \ \sigma_{K^- p \to \pi^+ \Sigma^-}$
(mb GeV$^2$)}}}} 
  \put(45,-8){\scalebox{0.9}{$\sqrt{s}$ \quad (GeV)}}
\end{overpic}
\\[3.5ex]
& (a) & & (b)
\end{tabular} 
\caption{Calculated cross sections for $K^- p \to \pi^{\mp} \Sigma^{\pm}$ 
multiplied by $4 |\mathbf{q}_{cm}^{K^- p}| \sqrt{s}$ and continued below the 
$K^- p$ threshold (vertical line), for three chiral coupled-channel fits to 
the $K^- p$ low-energy data. The fit shown by the solid (dashed) lines 
excludes (includes) the DEAR value for $a_{K^- p}$. 
Figure taken from Ref.~{\protect \cite{BNW05b}}.} 
\label{fig:weise8} 
\end{figure} 

\begin{figure}[t] 
\centerline{\includegraphics[height=7cm,width=9cm]{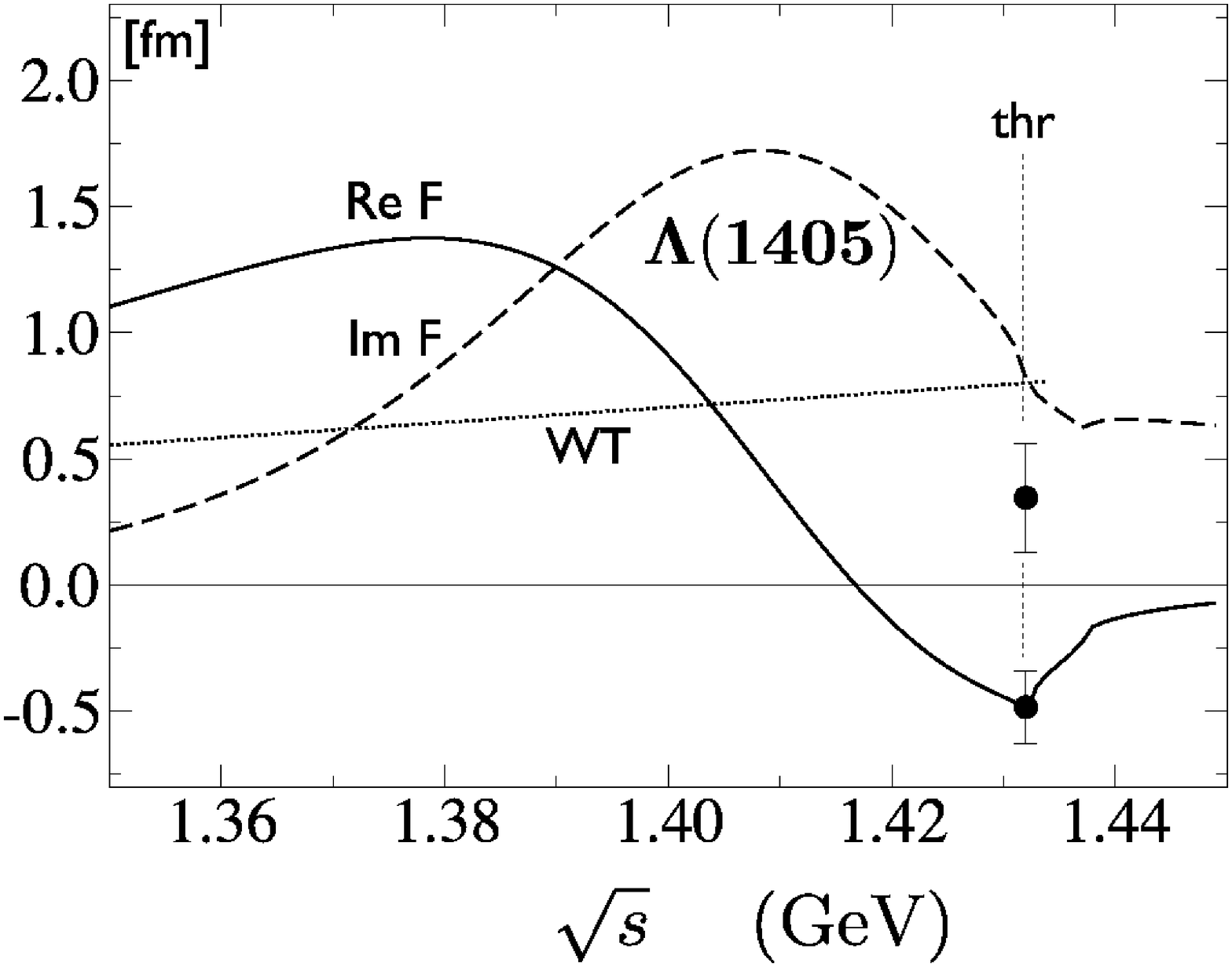}} 
\caption{Real and imaginary parts of the $K^- p$ forward elastic 
scattering amplitude, fitted within a NLO chiral SU(3) coupled-channel 
approach to $K^- p$ scattering and reaction data. The line denoted WT 
is the (real) LO Tomozawa-Weinberg $K^- p$ driving-term amplitude. 
The DEAR measurement~{\protect \cite{BBC05}} value for $a_{K^- p}$ is 
shown with error bars. Figure taken from Ref.~{\protect \cite{Wei07}}, 
based on the work of Ref.~{\protect \cite{BNW05b}}.} 
\label{fig:weisehyp06} 
\end{figure}

\begin{figure}[t]
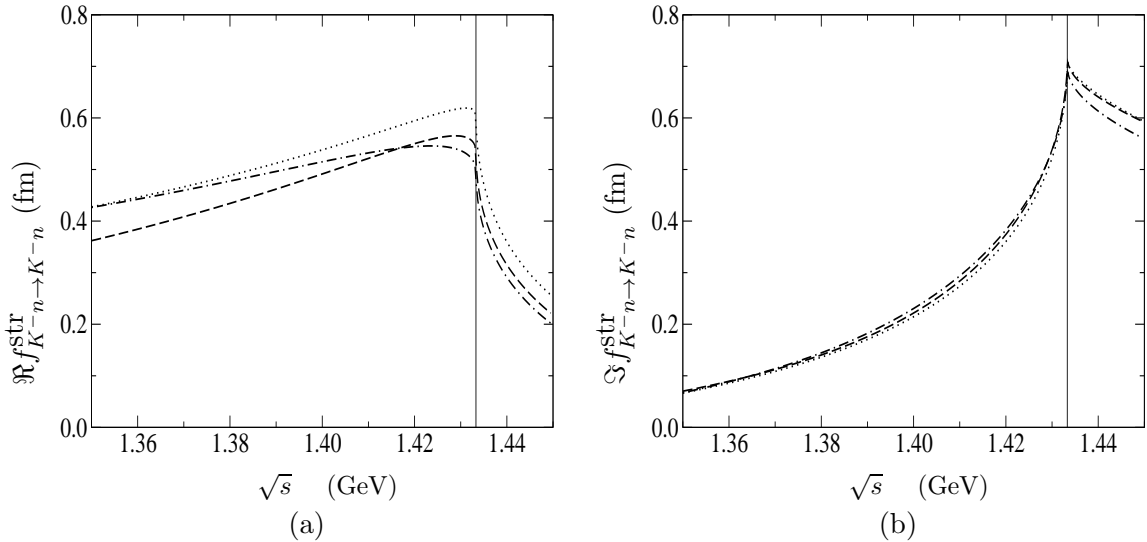
 
\centering 
\begin{tabular}{cp{1cm}c}
\begin{overpic}[height=6cm,width=0.40\textwidth,clip]{fg07fig20a.eps}
  \put(-11,14){\rotatebox{90}{{\scalebox{1.0}{{$\Re{f^{\textrm{str}}_{K^- n 
  \to K^- n}}$ (fm)}}}}}
  \put(40,-8){\scalebox{0.9}{$\sqrt{s}$ \quad (GeV)}}
\end{overpic}
& &
\begin{overpic}[height=6cm,width=0.40\textwidth,clip]{fg07fig20b.eps}
  \put(-11,14){\rotatebox{90}{{\scalebox{1.0}{{$\Im{f^{\textrm{str}}_{K^- n 
  \to K^- n}}$ (fm)}}}}}
  \put(40,-8){\scalebox{0.9}{$\sqrt{s}$ \quad (GeV)}}
\end{overpic}
\\[3.5ex]
(a) & & (b)
\end{tabular}
\caption{Real (left) and imaginary part (right) of the $K^- n$ forward elastic 
scattering amplitude, continued below the $K^- n$ threshold 
(vertical line). The various lines correspond to different interaction 
models, the dashed line standing for the Tomozawa-Weinberg interaction. 
Figure taken from Ref.~{\protect \cite{BNW05b}}.} 
\label{fig:weise11} 
\end{figure}

The $K^- p$ elastic, charge-exchange and reaction cross sections shown in 
Fig.~\ref{fig:kminuspdata} refer to energies above the $K^- p$ threshold 
at $\sqrt{s}=1432$~MeV. However, by developing potential models, or limiting 
the discussion to phenomenological K-matrix analyses, $\bar K N$ amplitudes 
are obtained that allow for analytic continuation into the nonphysical region 
below threshold. Using a K-matrix analysis, this was the way Dalitz and Tuan 
predicted the existence of the $\Lambda(1405)$ $\pi \Sigma$, $I=0$ resonance 
in 1959~\cite{DTu59}. Recent examples from a coupled-channel potential model 
calculation~\cite{BNW05b} based on low-energy chiral expansion are shown 
in the next three figures. Figure~\ref{fig:weise8} depicts the quantity 
$4 |\mathbf{q}_{cm}^{K^- p}| \sqrt{s} \sigma_{K^-p\to\pi^{\mp}\Sigma^{\pm}}(s)$ 
obtained by continuing the $K^-p\to\pi\Sigma$ amplitudes below the $K^- p$ 
threshold. One sees clearly the resonant behavior of the extrapolated 
cross sections due to the $\Lambda(1405)$. This resonance is also 
seen in Fig.~\ref{fig:weisehyp06} where the real and imaginary parts of the 
$K^- p$ elastic scattering amplitude, continued analytically below the $K^- p$ 
threshold, are plotted. The discrepancy with Im$a_{K^-p}$ deduced from the 
DEAR measurement~\cite{BBC05}, as given above, is clearly seen. 
In contrast to amplitudes which allow for the $I=0$ $\bar K N$ channel and 
thus exhibit a resonance effect due to the $\Lambda(1405)$, the purely $I=1$ 
$K^- n$ amplitude does not show such effects below threshold, as seen in 
Fig.~\ref{fig:weise11} where the real and imaginary parts of the $K^- n$ 
elastic scattering amplitude, within the same coupled-channel model, are 
shown. The model dependence of the $K^- n$ elastic scattering amplitude, 
as given by the three different curves, is considerably weaker than the 
model dependence of amplitudes in which the $\Lambda(1405)$ resonance enters, 
e.g. the $K^- p \to \pi^{\mp} \Sigma^{\pm}$ amplitudes related to the cross 
sections shown in Fig.~\ref{fig:weise8}.  

\subsection{$\bar K$-nucleus potentials} 
\label{sec:kbarpot} 

The gross features of low-energy $\bar K N$ physics, as demonstrated 
in the previous section by chiral coupled-channel fits to the low-energy 
$K^-p$ scattering and reaction data, are encapsulated in the leading-order 
Tomozawa-Weinberg (TW) vector term of the chiral effective 
Lagrangian~\cite{WRW97}. The Born approximation for the $\bar K$-{\it nuclear} 
optical potential $V_{\bar K}$ due to the TW interaction term yields then 
a sizable attraction: 
\begin{equation} 
\label{eq:chiral} 
V_{\bar K}=-\frac{3}{8f_{\pi}^2}~\rho\sim -55~\frac{\rho}{\rho_0}~~{\rm MeV} 
\end{equation} 
for $\rho _0 = 0.16$ fm$^{-3}$, where $f_{\pi} \sim 93$ MeV is the 
pseudoscalar meson decay constant. Iterating the TW term plus 
next-to-leading-order terms, 
within an {\it in-medium} coupled-channel approach constrained 
by the $\bar K N - \pi \Sigma - \pi \Lambda$ data near the 
$\bar K N$ threshold, roughly doubles this $\bar K$-nucleus attraction. 
It is found (e.g. Ref. \cite{WKW96}) that the $\Lambda(1405)$ quickly 
dissolves in the nuclear medium at low density, so that 
the repulsive free-space scattering length $a_{K^-p}$, as function of 
$\rho$, becomes {\it attractive} well below $\rho _0$. Since the purely 
$I=1$ attractive scattering length $a_{K^-n}$ is only weakly density 
dependent, the resulting in-medium $\bar K N$ isoscalar scattering length 
$b_0(\rho)={\frac{1}{2}}(a_{K^-p}(\rho)+a_{K^-n}(\rho)$) translates into 
a strongly attractive $V_{\bar K}$: 
\begin{equation} 
\label{eq:trho} 
V_{\bar K}(r) = -{\frac{2\pi}{\mu_{KN}}}~b_0(\rho)~\rho(r)~, 
~~~~{\rm Re}V_{\bar K}(\rho_0) \sim -110~{\rm MeV}\,. 
\end{equation} 
This in-medium $\bar K N$ isoscalar scattering length $b_0(\rho)$ is the same 
one as the effective scattering length $a_{\rm eff}$ plotted as function of 
the density $\rho$ in Fig.~\ref{fig:arho} in Sec.~\ref{sec:inmedium}. 
It is useful to record that $b_0(\rho_0) \approx 1$~fm corresponds to 
$V_{\bar K}(\rho_0) \approx -100$~MeV. 
However, when $V_{\bar K}$ is calculated {\it self consistently}, 
namely by including $V_{\bar K}$ in the propagator $G_0$ used in the 
Lippmann-Schwinger equation determining $b_0(\rho)$, as demonstrated by 
Eq.~(\ref{eq:sc}) in Sec.~\ref{sec:inmedium}, one obtains 
Re$V_{\bar K}(\rho_0)\sim -$(40-60) MeV~\cite{SKE00,ROs00,TRP01,CFG01}. 
The main reason for this weakening of $V_{\bar K}$, 
approximately going back to that calculated using Eq.~(\ref{eq:chiral}), 
is the strong absorptive effect which $V_{\bar K}$ exerts within $G_0$ to 
suppress the higher Born terms of the $\bar K N$ TW potential. 

Additional considerations for estimating  $V_{\bar K}$ are listed below.  
 
\begin{itemize} 

\item QCD sum-rule estimates~\cite{Dru06} for vector (v) and scalar (s) 
self-energies: 
\begin{eqnarray} 
\label{eq:QCDv} 
\Sigma_v(\bar K) &\sim & -\frac{1}{2}~\Sigma_v(N)~\sim~
-\frac{1}{2}~(200)~{\rm MeV}~ =~-100~{\rm MeV}\,,\\ 
\Sigma_s(\bar K) &\sim & \frac{m_s}{M_N}~\Sigma_s(N)~\sim~
\frac{1}{10}~(-300)~{\rm MeV}~ =~ -30~{\rm MeV}\, ,
\label{eq:QCDs}
\end{eqnarray}
where $m_s$ is the strange-quark (current) mass. The factor 1/2 in 
Eq.~(\ref{eq:QCDv}) is due to the one nonstrange antiquark $\bar q$ in the 
$\bar K$ meson out of two possible, and the minus sign is due to G-parity going 
from $q$ to $\bar q$. This rough estimate gives then 
$V_{\bar K}(\rho_0) \sim -130$~MeV. 

\item The QCD sum-rule approach essentially 
refines the mean-field argument~\cite{SGM94,BRh96} 
\begin{equation} 
\label{eq:MF} 
V_{\bar K}(\rho_0)~\sim~\frac{1}{3}~(\Sigma_s(N)-\Sigma_v(N))~\sim~
-170~{\rm MeV}\,,
\end{equation} 
where the factor 1/3 is again due to the one nonstrange antiquark in the 
$\bar K$ meson, but here with respect to the three nonstrange quarks of 
the nucleon. 

\item The ratio of $K^-/K^+$ production cross sections in nucleus-nucleus and
proton-nucleus collisions near threshold, measured by the Kaon Spectrometer 
(KaoS) collaboration~\cite{SBD06} at SIS, GSI, yields an estimate 
$V_{\bar K}(\rho_0) \sim -80$~MeV by relying on Boltzmann-Uehling-Uhlenbeck 
(BUU) transport calculations 
normalized to the value $V_K(\rho_0) \sim +25$~MeV. 
Since $\bar K NN \to YN$ absorption processes 
apparently were disregarded in these calculations, a deeper $V_{\bar K}$ may 
follow once nonmesonic absorption processes are included. 

\item Capture rates of $(K^{-}_{\rm stop},\pi)$ reactions to specific 
$\Lambda$ hypernuclear states provide a sensitive measure for the strength 
of the $K^-$ optical potential at threshold. A very strong potential, 
as discussed in Sec.~\ref{sec:Kdeep}, generates $\bar K$-{\it nuclear} 
quasibound states which due to orthogonality with the $K^-$ atomic states 
force the wavefunctions of the latter to oscillate and become suppressed 
within the nuclear volume, and thus to reduce their effectiveness in the 
calculation of the $(K^{-}_{\rm stop},\pi)$ transition matrix element. 
The results of DWIA calculations~\cite{CFG01} for capture on $^{12}$C into 
the $1^-$ ground state configurations of $_{\Lambda}^{12}{\rm C}$ and 
$_{\Lambda}^{12}{\rm B}$ are shown in Table~\ref{tab:Krates}, using several 
fitted $K^-$ optical potentials which are ordered according to their depth.  
It is clearly seen that the deeper the $K^-$ optical potential is, 
the lower the calculated rate becomes. Unfortunately, all the calculated 
rates shown in Table~\ref{tab:Krates} are much lower than the measured 
values $R_{\rm exp}$~\cite{THO94,ACE03}, making it difficult to draw 
definitive conclusions. Furthermore, the experimental capture rates depart 
strongly from the ratio $2:1$ expected from charge independence for the ratio 
$R(_{\Lambda}^{12}{\rm C})/R(_{\Lambda}^{12}{\rm B})$ for the rates 
$R_{\rm exp}$. 

\begin{table} 
\caption{Calculated $(K^{-}_{\rm stop},\pi)$ rates on $^{12}$C per stopped 
$K^-$ (in units of $10^{-3}$) for $1p_{N} \rightarrow 1s_{\Lambda}$ capture 
into the $1^{-}$ ground-state configurations in $^{12}_{\Lambda}$C 
and $^{12}_{\Lambda}$B, for various fitted optical potentials ordered 
according to their depth. Table taken from Ref.~{\protect \cite{CFG01}}.} 
\label{tab:Krates} 
\begin{ruledtabular} 
\begin{tabular}{cccccl} 
final $_{\Lambda}^{\rm A}\rm Z\rule{0em}{1.4em}$ 
 & \makebox[2cm]{chiral} & \makebox[2cm]{effective} 
 & \makebox[2cm]{fixed} & \makebox[2cm]{DD} & 
\makebox[2cm]{$R_{\rm exp} \times 10^3$} \\ \hline 
$_{\Lambda}^{12}\rm C$ & $0.231$ & $0.169$ & $0.089$ & $0.063$ 
& $0.98 \pm 0.12$~\cite{THO94} \\ 
$_{\Lambda}^{12}\rm B$ & $0.119$ & $0.087$ & $0.046$ & $0.032$ 
& $0.28 \pm 0.08$~\cite{ACE03} \\ 
\end{tabular} 
\end{ruledtabular} 
\end{table} 

\end{itemize}

\subsection{Fits to $K^-$ -atom data}
\label{sec:Kat}

The $K^-$-atom data used in global fits~\cite{BFG97} span a range of nuclei
from $^7$Li to $^{238}$U, with 65 level-shifts, widths and transition yields 
data points. It was shown already in the mid 1990s~\cite{BFG97} that 
although a reasonably good fit to the data is obtained 
with the generic $t\rho $ potential of Eq.~(\ref{eq:Vopt}) with an effective
complex parameter $b_0$ corresponding to attraction, greatly improved fits 
are obtained with a density-dependent potential, where the fixed $b_0$ 
is replaced by $b_0+B_0 [\rho(r) /\rho_0 ]^\alpha $, with $b_0, B_0$ and 
$\alpha \geq 0$ determined by fits to the data. Fitted potentials of this 
kind are marked DD. This parameterization offers the advantage of fixing 
$b_0$ at its (repulsive) free-space value in order to 
respect the low-density limit, while relegating the expected in-medium 
attraction to the $B_0$ term which goes with a higher power of the density. 
As mentioned in Sec.~\ref{sec:Kbarprev}, the $t\rho$ best-fit potentials 
have real parts which are less than 100 MeV deep for medium-weight and heavy 
nuclei. The corresponding density-dependent potentials are more attractive, 
150-200 MeV deep, hence the `shallow' {\it vs} `deep' terminology. 
Chirally inspired approaches that fit the low-energy $K^-p$ reaction data 
predict attractive potentials of depths $\sim$(110-120) MeV~\cite{WKW96} 
and, additionally, by requiring self consistency in the construction of 
the optical potential, lead to yet shallower potentials with 
Re$V_{\bar K}(\rho_0) \sim -$(40-60) MeV~\cite{ROs00,CFG01}. 
Recent experimental reports on candidates for $\bar K$-nuclear deeply
bound states in the range of binding $B_{\bar K} \sim$~100-200 MeV
\cite{SBF04,SBF05,KHA05,ABB05} again highlighted the question of how much 
attractive the $\bar K$-nucleus interaction is below the $\bar K N$ threshold. 
Therefore the motivation for re-analysis of a comprehensive set of kaonic 
atom data is two fold. First is the question of `deep' {\it vs.} `shallow' 
real $\bar K$-nucleus potential, in light of recent possible experimental 
evidence for the existence of deeply bound kaonic states whose binding 
energies exceed the depth of the shallow type of potential obtained from fits 
to kaonic atom data. However, if the deep variety of potential is confirmed, 
then the dependence of the $\bar K N$ interaction on the nuclear density 
becomes of prime concern. This density dependence is the second point which 
motivated the re-analysis of kaonic atoms data \cite{MFG06}
although there have not been any new experimental results on strong
interaction effects in kaonic atoms since the early 1990s.

\begin{figure}[t]
\centering
\includegraphics[height=8cm,width=7.5cm]{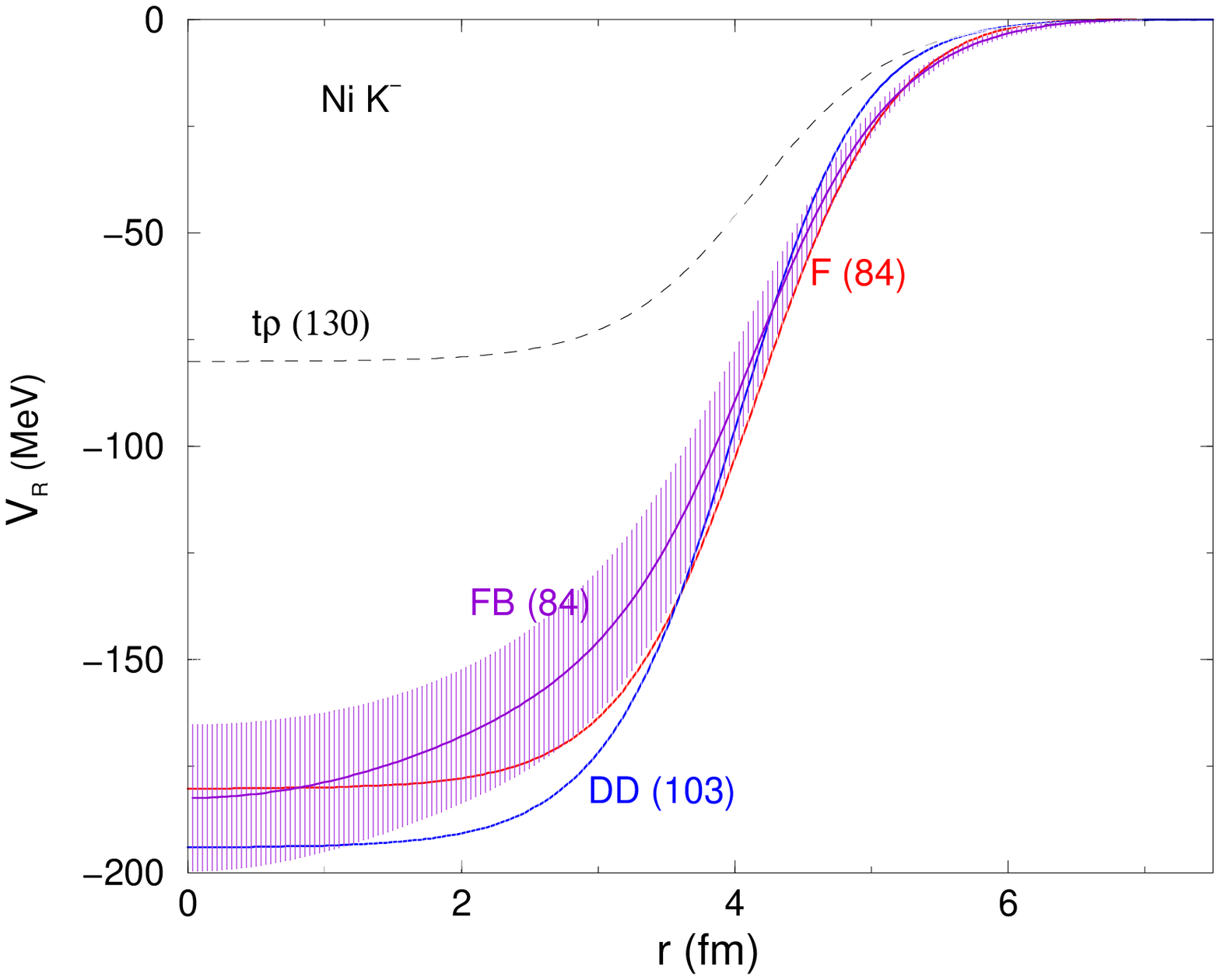}
\hspace*{3mm}
\includegraphics[height=8cm,width=7.5cm]{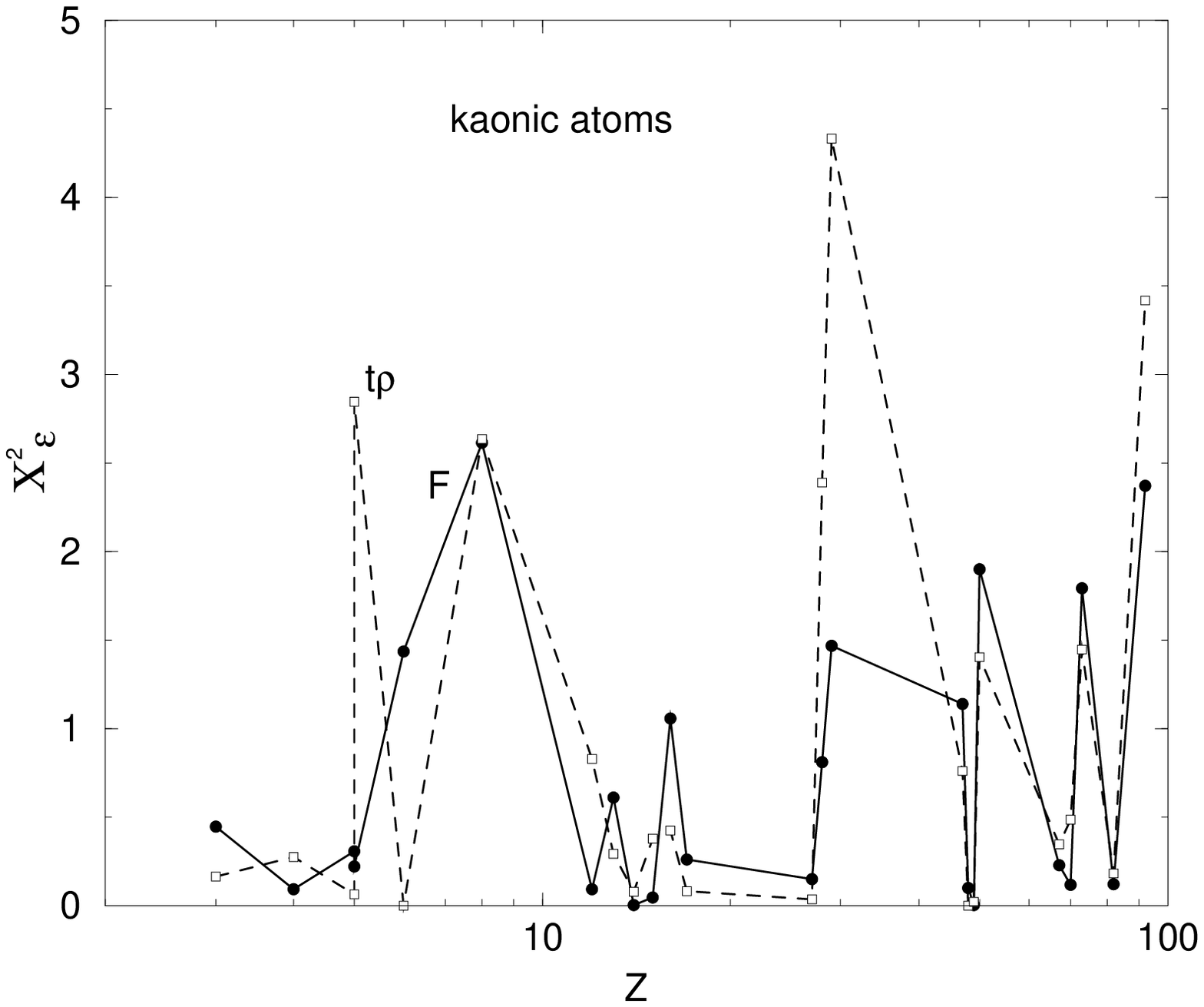}
\caption{Left: real part of the ${\bar K}$-${^{58}}$Ni potential
obtained in a global fit to $K^-$-atom data using the model-independent
FB technique~{\protect \cite{BFr07}}, in comparison with
other best-fit potentials and  $\chi^2$ values in parentheses.
Right: contributions to the $\chi^2$ of $K^-$ atomic shifts for the {\it deep}
density-dependent potential F from Ref.~{\protect \cite{MFG06}} and for the
{\it shallow} $t \rho $ potential.}
\label{fig:kbarNiVR}
\end{figure}

The departure of the optical potential from the fixed-$t$ $t \rho $
approach was introduced \cite{MFG06} with the help of a geometrical model, 
where one {\it loosely} defines in coordinate space an `internal' region 
and an `external' region by using the multiplicative functions $F(r)$ in
the former and $[1-F(r)]$ in the latter, where $F(r)$ is defined as
\begin{equation}
\label{eq:F}
F(r)~=~\frac{1}{e^x +1}
\end{equation}
with $x~=~(r-R_x)/a_x$. Then clearly $F(r)~\rightarrow~1$ 
for $(R_x - r) >> a_x$, which defines the internal region. 
Likewise $[1~-~F(r)]~\rightarrow~1$ for $(r - R_x) >> a_x$, which defines the 
external region. 
Thus $R_x$ forms an approximate border between the internal and the external
regions, and {\it if} $R_x$ is close to the nuclear surface then
the two regions will correspond to the high density and low density
regions of nuclear matter, respectively.
The fixed $b_0$
in the $t\rho$ potential was replaced by
\begin{equation}
\label{eq:DDF}
b_0~\rightarrow ~B_0~F(r)~+~b_0~[1~-~F(r)] 
\end{equation}
where the parameter $b_0$ represents the low-density interaction 
and the parameter $B_0$ represents
the interaction
inside the nucleus. This division into regions
of high and low densities is meaningful
{\it provided} $R_x$ is close to the radius of the nucleus
and $a_x$ is of the order of 0.5~fm. This  is indeed the case,
as found  in global
fits to kaonic atom data \cite{MFG06}.
We note that unlike with pionic and antiprotonic atoms, the dependence
of kaonic atom fits on the rms radius of the neutron distribution 
is marginal.

\begin{figure}[t] 
\centering
\includegraphics[height=8cm,width=8.0cm]{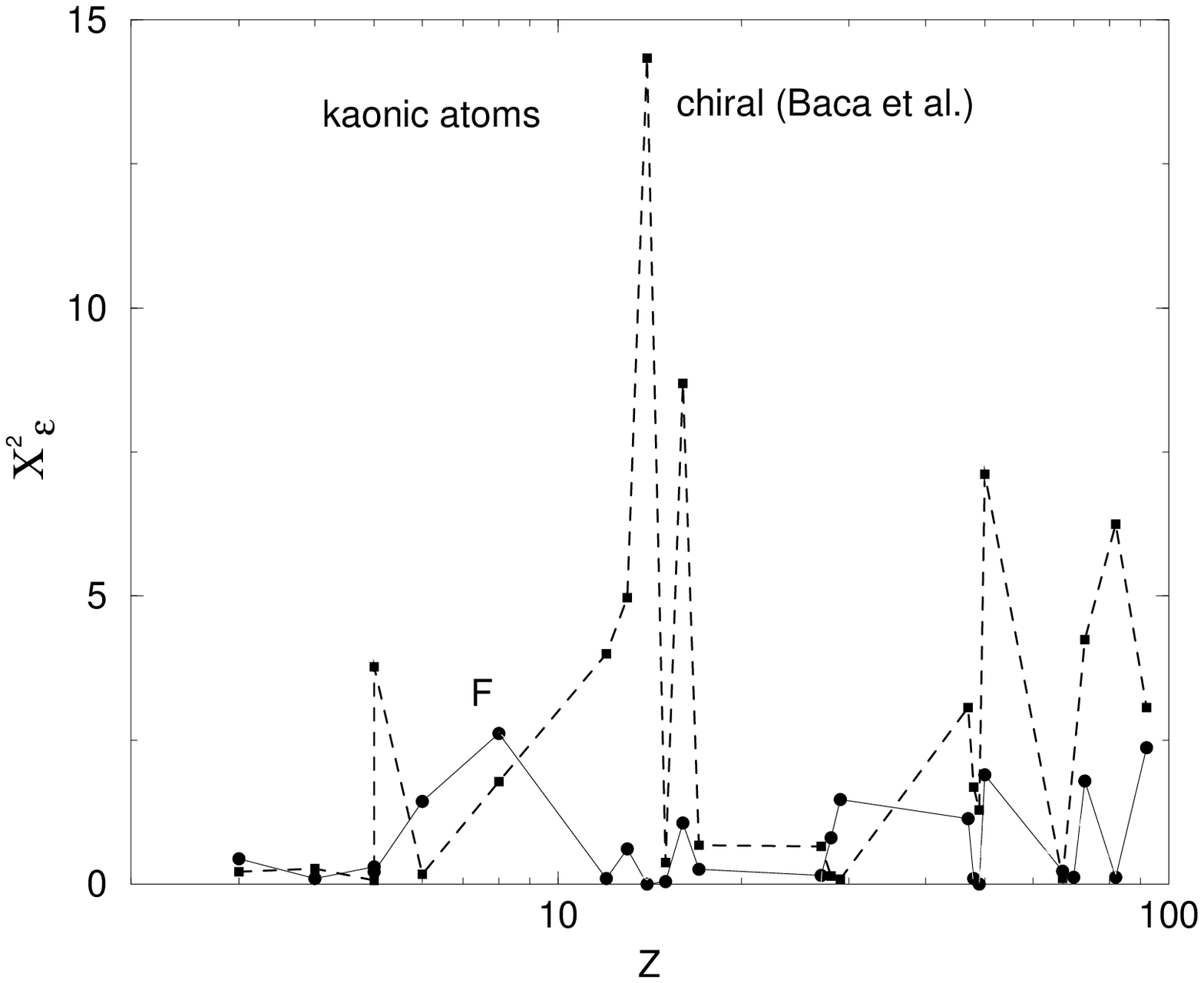} 
\hspace*{3mm}
\includegraphics[height=8cm,width=7.8cm]{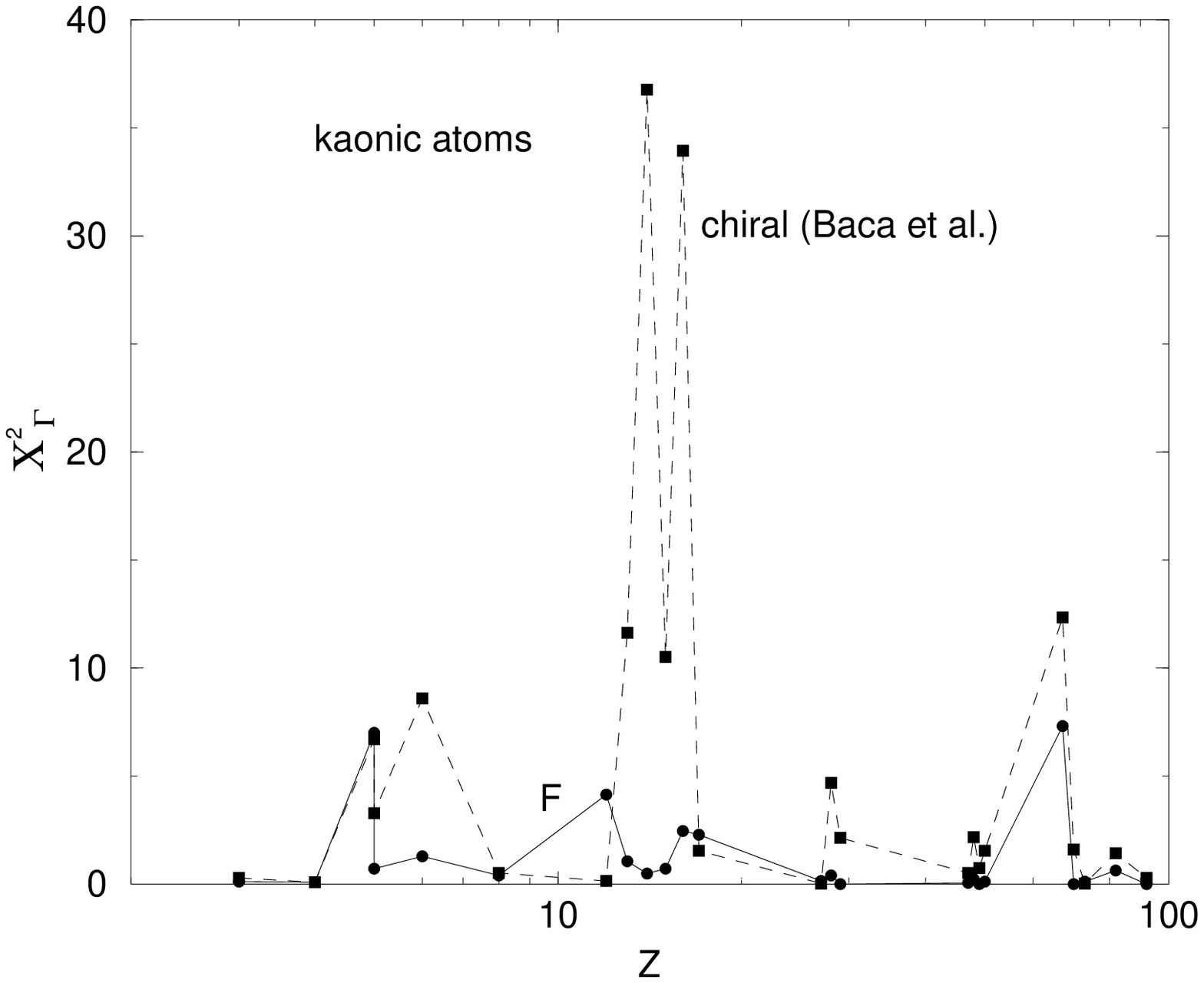} 
\caption{Contributions to the $\chi^2$ of $K^-$ atomic shifts (left)
and widths (right) for the {\it deep} density-dependent potential F from 
Ref.~{\protect \cite{MFG06}} and for the {\it shallow} chirally-based 
potential from Ref.~{\protect \cite{BGN00}}.} 
\label{fig:kbarBaca}
\end{figure} 

Figure \ref{fig:kbarNiVR} (left) shows, as an example, the real part
of the best-fit potential for $^{58}$Ni obtained with the various
models discussed above, i.e. the simple $t \rho $ model and its 
DD extension, and the geometrical model F, with the corresponding values of
$\chi ^2$ for 65 data points in parentheses. Also shown, with an error
band, is a Fourier-Bessel (FB) fit \cite{BFr07} that is discussed below.
We note that, although the two density-dependent
potentials marked DD and F have very different parameterizations, the
resulting potentials are quite similar. In particular, the shape of potential
F departs appreciably from  $\rho (r)$ for $\rho (r)/\rho_0 \leq 0.2$, where
the physics of the $\Lambda(1405)$ is expected to play a role.
The density dependence of the potential F provides by far the best fit ever 
reported for any global $K^-$-atom data fit, and the lowest $\chi ^2$ value 
as reached by the model-independent FB method.
On the right-hand side of the figure are shown the individual contributions
to $\chi ^2$ of the shifts for the deep F potential and the shallow $t \rho $ 
potential. Figure \ref{fig:kbarBaca} shows comparisons between $\chi ^2$ 
values of the shifts and of the widths for the F potential and the yet 
shallower chirally-based potential of Baca et al. \cite{BGN00}. 
It is self evident that the agreement between calculation and experiment 
is substantially better for the deep F potential than for the shallow
chiral potentials. 

The question of how well the real part of the
$K^-$-nucleus potential is determined was discussed in Ref.~\cite{BFr07}.
Estimating the uncertainties of hadron-nucleus potentials as function
of position is not a simple task. For example, in the `$t\rho $'
approach the shape of the potential is determined by the nuclear
density distribution and the uncertainty in the strength parameter,
as obtained from $\chi ^2$ fits to the data,
implies a fixed {\it relative} uncertainty at all radii, which is, of course,
unfounded. Details vary when more elaborate
forms such as `DD' or `F' are used, but one is left essentially
with {\it analytical continuation} into the nuclear interior of potentials
that might be well-determined only close to the nuclear surface.
`Model-independent' methods have been used in analyses of elastic scattering
data for various projectiles \cite{BFG89}
to alleviate this problem. However, applying e.g. the Fourier-Bessel (FB) 
method in global analyses of kaonic atom data, one ends up
with too few terms in the series, thus making the uncertainties unrealistic
in their dependence on position.
This is illustrated in Fig. \ref{fig:kbarNiVR} by the `FB' curve, 
obtained by adding a Fourier-Bessel series to a `$t\rho $' potential. 
Only three terms in the series are needed to achieve a $\chi ^2$ of 84
and the potential becomes deep, in agreement with the other two `deep'
solutions. The error band obtained from the FB method \cite{BFG89} is, 
nevertheless, unrealistic because only three FB terms are used. However, 
an increase in the number of terms is found to be unjustified numerically.

\begin{figure}[t]
\centerline{\includegraphics[height=9cm]{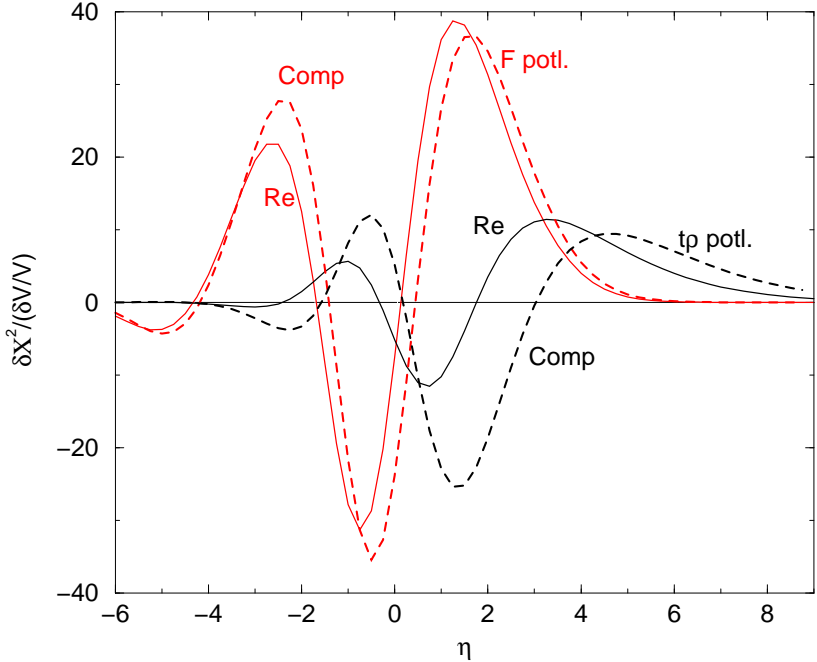}}
\caption{Functional derivatives of kaonic atoms $\chi ^2$ with respect to 
the fully complex (Comp, dashed lines) and real (Re, solid lines) potential 
as function of $\eta$, where $r=R_c+\eta a_c$, with $R_c$ and $a_c$ the 
radius and diffuseness parameters, respectively, of the charge distribution.
Results are shown for the $t\rho $ potential and for the $t(\rho )$ `F' 
potential of Ref. \cite{MFG06} obtained from global fits to kaonic atom data.}
\label{fig:katFD}
\end{figure}

The functional derivative (FD) method for identifying the radial regions
to which exotic atom data are sensitive is described in detail in
Sec.~\ref{sec:piradsen}. This method was applied 
in Ref.~\cite{BFr07} to the F and $t \rho $ kaonic atom
potentials and results are shown in Fig.~\ref{fig:katFD}
where $\eta$ is a global parameter defined by
$r=R_c+\eta a_c$, with $R_c$ and $a_c$ the radius and diffuseness
parameters, respectively, of the charge distribution.
From the figure it can be inferred that the sensitive region
for the real $t\rho $ potential is between $\eta =-1.5$ and $\eta =6$
whereas for the F potential it is between $\eta =-3.5$ and $\eta =4$.
Recall that $\eta =-2.2$ corresponds to 90\% of the central charge density
and $\eta =2.2$ corresponds to 10\% of that density. It therefore
becomes clear that within the $t\rho $ potential there is no sensitivity 
to the interior of the nucleus whereas with the $t(\rho )$ `F' potential,
which yields greatly improved fit to the data, there is sensitivity
to regions within the full nuclear density.
The different sensitivities result from
the potentials themselves: for the $t \rho $ 
potential the interior of the nucleus
is masked essentially by the strength of the imaginary potential.
In contrast, for the F potential not only is its imaginary part significantly
smaller than the imaginary part of the $t \rho $ potential 
\cite{MFG06} but also
the additional attraction provided
by the deeper potential enhances the {\it atomic} wavefunctions within
the nucleus \cite{BFG97} thus creating the sensitivity at smaller radii.
As seen in the figure, the functional derivative for the complex 
F potential is well approximated by that for its real part. 

It is concluded that optical potentials derived from the observed
strong-interaction effects in kaonic atoms are sufficiently deep to
support strongly-bound antikaon states, but it does not necessarily
imply that such states are sufficiently narrow to be resolved unambiguously 
from experimental spectra. Moreover, the discrepancy between
the very shallow chirally motivated potentials \cite{ROs00,CFG01}, the 
intermediate potentials of depth around 100 MeV~\cite{WKW96} and
the deep phenomenological potentials of type `F' remains an open problem. 
It should also be kept in mind that these depths relate to $\bar K$ 
potentials at {\it threshold}, whereas the information required for 
$\bar K$-nuclear quasibound states is at energies of order 100 MeV below 
threshold.

\subsection{Deeply bound $K^-$ atomic states}
\label{sec:Kdeep}

\begin{figure}[t]
\centerline{\includegraphics[height=9cm]{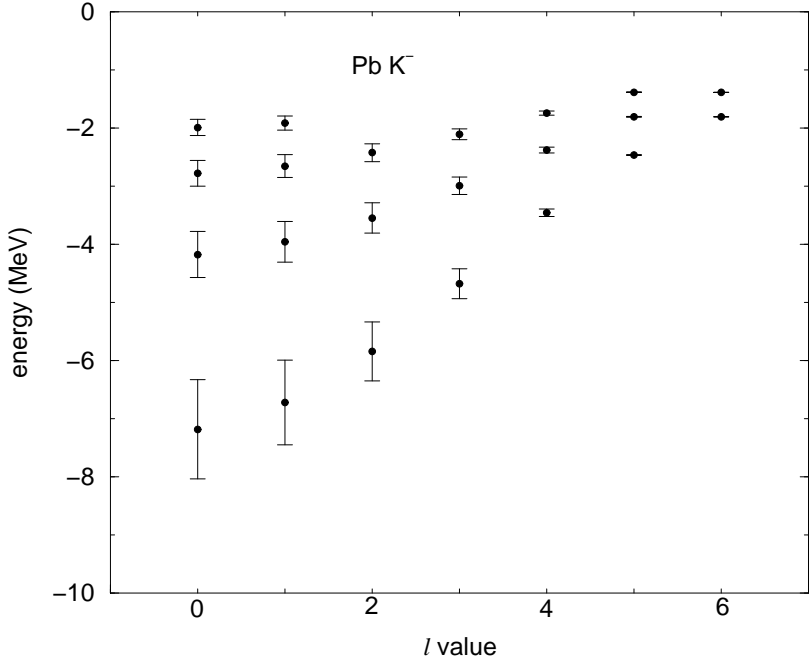}}
\caption{Calculated energies of $K^-$ atomic states in $^{208}$Pb.
The lowest energy for each $l$ value corresponds to $n=l+1$.
The bars represent the widths of the states.}
\label{fig:Kspect}
\end{figure}

Somewhat paradoxically, due to the strong absorptive imaginary part
of the $K^-$-nucleus potential, relatively narrow deeply-bound
atomic states are expected to exist
which are quite independent of the real potential. 
Such states are indeed found in
numerical calculations as can be seen in  Fig.~\ref{fig:Kspect}
where calculated binding energies and widths of atomic 
states of $K^-$ in $^{208}$Pb are shown for several $l$-values, down
to states which are inaccessible via the X-ray cascade. For $^{208}$Pb, 
the last observed $K^-$ atomic circular state is the $7i$, corresponding to 
$l=6$. The general physics behind this phenomenon is similar to that
responsible for the deeply-bound pionic atom states, although there
are differences in the mechanism. 
For a Schr\"odinger equation
the width of a state is given {\it exactly} by Eq.~(\ref{eq:gamma})
and if a  normalized atomic wavefunction is expelled from the nucleus, then
small widths are expected due to the reduced overlap between the
atomic wavefunction and the imaginary potential. (For the KG equation
there are small changes in the expression Eq.~(\ref{eq:gamma}), see 
Refs.~\cite{FGa99a,FGa99b}.)
The mechanism behind the pionic atom deeply bound states is simply
the repulsive real part of the $s$-wave potential. 
In contrast, phenomenological
kaonic atom potentials are attractive, but the strengths of the imaginary
part of the potential are such that the decay of the wavefunction 
as it enters the nucleus is equivalent to repulsion, resulting in
narrow atomic states due to the reduced overlap as discussed above.
It is seen from Fig.~\ref{fig:Kspect} that there is a saturation
phenomenon where widths hardly increase 
for $l \leq 2$, 
contrary to intuitive expectation.
The repulsive effect of sufficiently strong absorption
is responsible for the general property of saturation of widths of
atomic states and saturation of reaction cross sections above threshold,
observed experimentally for antiprotons (see Sec. \ref{sec:pbarsdeep}.)

\begin{figure}[t]
\centering
\includegraphics[height=8.0cm,width=7.5cm]{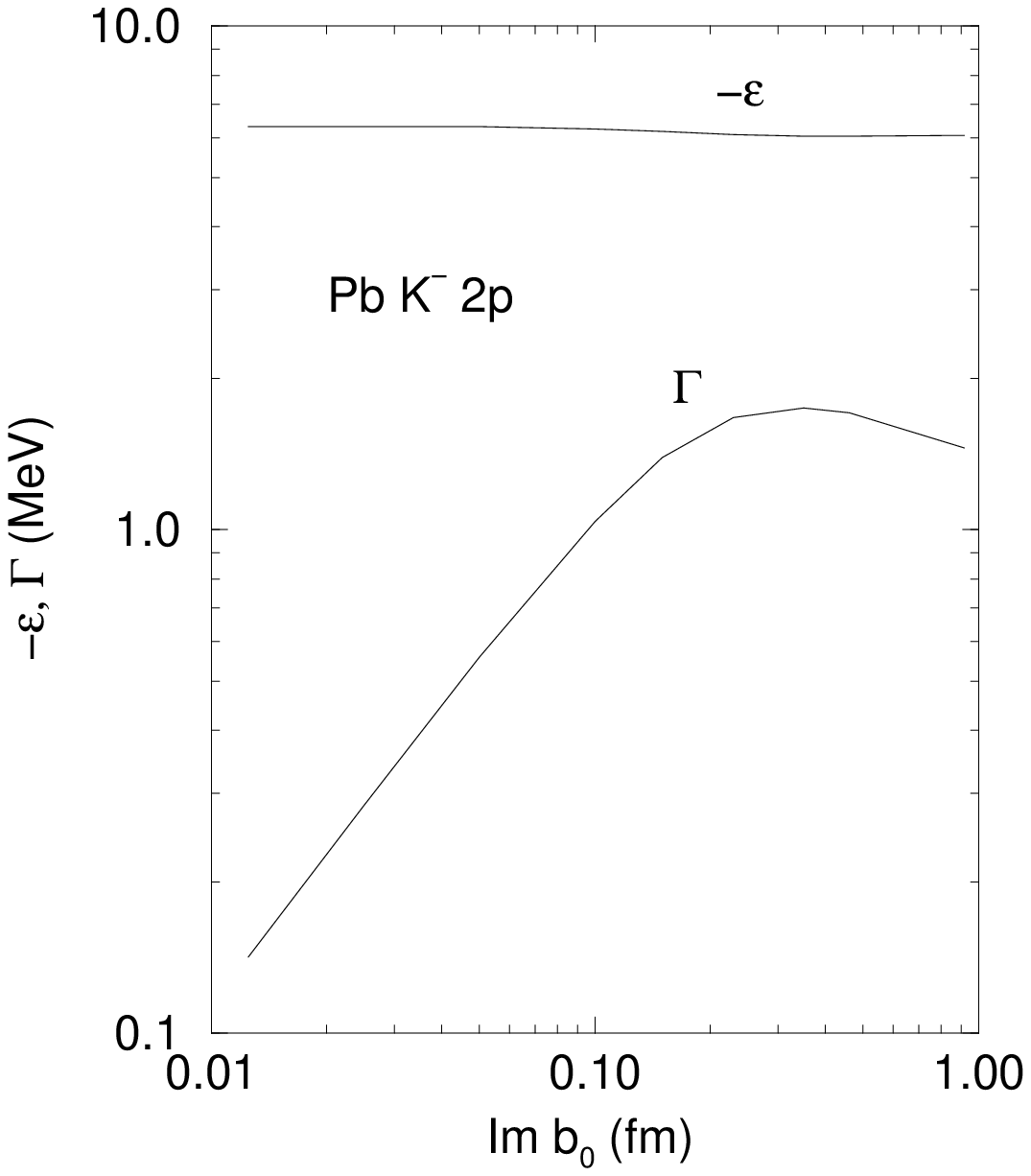}
\hspace*{3mm}
\includegraphics[height=8.0cm,width=7.5cm]{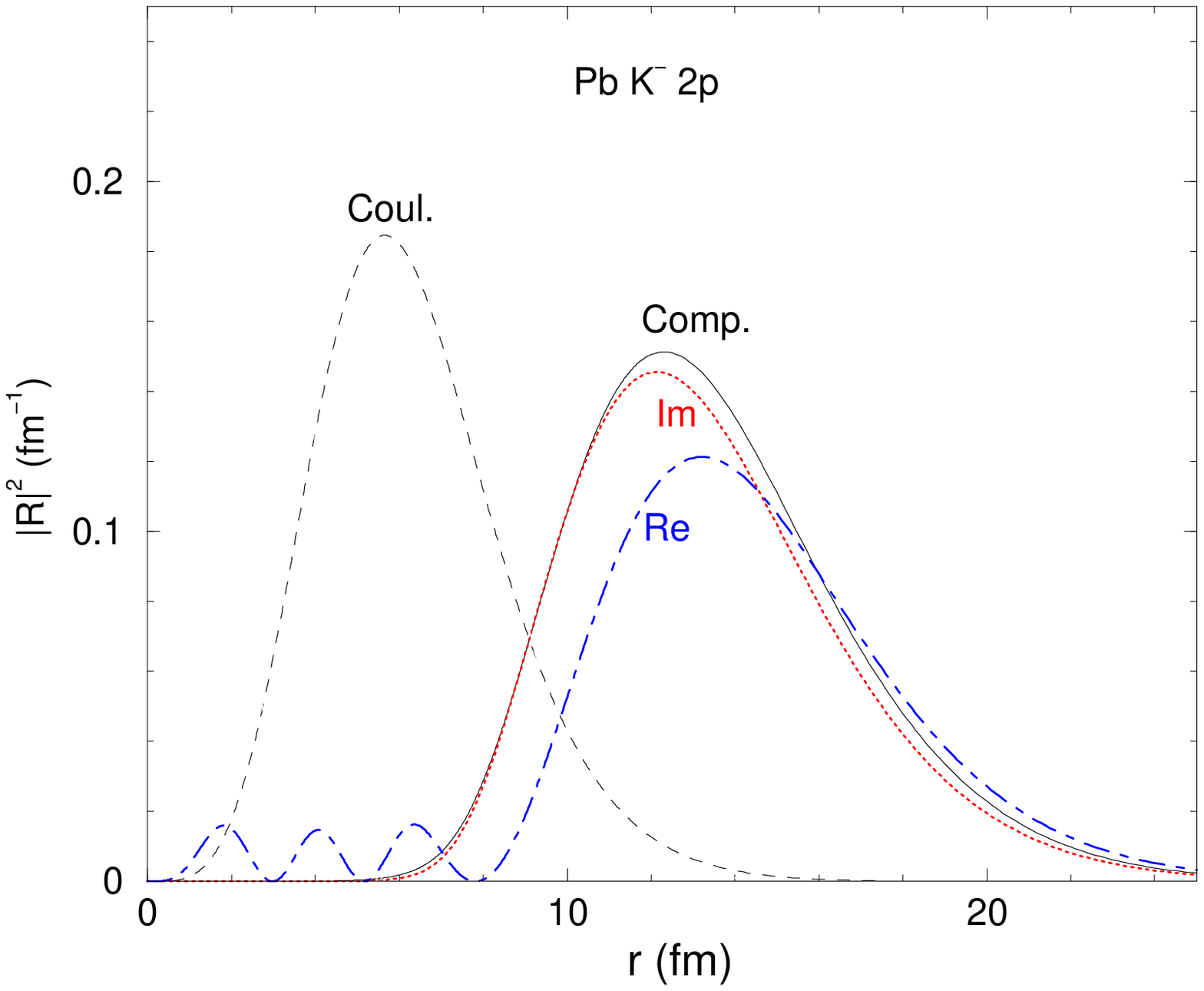}
\caption{Saturation of width $\Gamma$ for the $2p$ `deeply bound' $K^-$
atomic state in $^{208}$Pb as function of absorptivity, Im$b_0$, for
Re$b_0 = 0.62$~fm (left) and wavefunctions for this state (right).}
\label{fig:sat}
\end{figure}

The left-hand side of Fig.~\ref{fig:sat} shows the saturation of widths
as function of Im$b_0$ for the 2$p$ state of kaonic atoms of $^{208}$Pb.
For small values of Im$b_0$ the calculated width increases linearly
but already at 20\% of the best-fit value of 0.9~fm saturation sets in
and eventually the width goes down with further increase of the absorption.
Note that the real part of the binding energy, represented here by the
strong interaction level shift $\epsilon $, is hardly changing with Im$b_0$.
The right-hand side of Fig.~\ref{fig:sat} shows radial wavefunctions 
for the 2$p$  atomic $K^-$ state in $^{208}$Pb for several combinations 
of potentials. The dashed curve marked `Coul' is for the Coulomb potential 
only, and with a half-density radius for $^{208}$Pb 
of 6.7 fm it clearly overlaps strongly with the nucleus. Adding the full
complex optical potential the solid curve marked `Comp' shows that the 
wavefunction is expelled from the nucleus, and the dotted curve
marked `Im' shows that this repulsion is effected by the imaginary
part of the potential. Clearly the overlap of the wavefunction with the 
nucleus is dramatically reduced compared to the Coulomb-only situation.
An interesting phenomenon is displayed by the
dot-dashed curve marked `Re'. It shows the wavefunction when the 
real potential is added to the Coulomb potential, demonstrating
significant {\it repulsion} of the wavefunction by the added 
{\it attractive} potential.
The explanation for this bizarre result 
is provided by the three small peaks inside the nucleus
which are due to the orthogonality of the {\it atomic} wavefunction and
strongly-bound $K^-$ 
{\it nuclear} wavefunctions having the same $l$-values. This extra structure 
of the wavefunction in the interior effectively disappears when the 
imaginary potential is included.

\subsection{Deeply bound $K^-$ nuclear states in light nuclei}
\label{sec:nuclear db}

No saturation mechanism holds for the width of $\bar K$-nuclear states which 
retain very good overlap with the potential. Hence, the questions to ask 
are (i) whether it is possible at all to bind {\it strongly} $\bar K$ mesons 
in nuclei, and (ii) are such quasibound states sufficiently narrow to allow 
observation and identification? 
The first question was answered affirmatively by Nogami~\cite{Nog63} as early 
as 1963 arguing that the $K^-pp$ system could acquire about 10 MeV 
binding in its $I=1/2,~L=S=0$ state. Yamazaki and Akaishi, using a complex 
energy-independent $\bar K N$ potential within a single-channel 
$K^-pp$ calculation~\cite{YAk02}, reported a binding energy 
$B \sim 50$~MeV and width $\Gamma \sim 60$~MeV. Preliminary results of 
revised Antisymmetrized Molecular Dynamics (AMD) calculations by Dot\'e and 
Weise~\cite{Wei07,DWe06} which implicitly account for $\bar K N - \pi \Sigma$ 
coupling agree on $B$ but estimate $\Gamma \sim 100$~MeV. Coupled-channel 
$\bar K NN - \pi \Sigma N$ Faddeev calculations~\cite{SGM06,SGM07,ISa06,ISa07} 
of $K^-pp$ have confirmed this order of magnitude of binding, 
$B~\sim$~55-75~MeV, differing on the width; the calculations by Shevchenko 
et~al.~\cite{SGM06,SGM07} give large values, $\Gamma \sim 100$~MeV, for the 
mesonic width. These Faddeev calculations overlook the $\bar K NN \to YN$ 
coupling to nonmesonic channels which are estimated to add, conservatively, 
20 MeV to the overall width. Altogether, the widths calculated for the 
$K^-pp$ quasibound state are likely to be so large as to make it difficult 
to identify it experimentally~\cite{KHa07}.  

The current experimental and 
theoretical interest in $\bar K$-nuclear bound states was triggered back in 
1999 by the suggestion of Kishimoto~\cite{Kis99} to look for such states 
in the nuclear reaction $(K^{-},p)$ in flight, and by Akaishi and 
Yamazaki~\cite{AYa99,AYa02} who suggested to look for a $\bar K NNN$ $I=0$ 
state bound by over 100 MeV for which the main $\bar K N \to \pi \Sigma$ 
decay channel would be kinematically closed. In fact, Wycech had conjectured 
that the width of such states could be as small as 20 MeV~\cite{Wyc86}. 
Some controversial evidence for relatively narrow states was presented 
initially in $(K^{-}_{\rm stop},p)$ and $(K^{-}_{\rm stop},n)$ reactions on 
$^4$He (KEK-PS E471)~\cite{SBF04,SBF05} but has recently been 
withdrawn (KEK-PS E549/570)~\cite{Iwa06}. $\bar K$-nuclear states were also 
invoked to explain few statistically-weak irregularities in the neutron 
spectrum of the $(K^{-},n)$ in-flight reaction on $^{16}$O (BNL-AGS, parasite 
E930)~\cite{KHA05}, but subsequent $(K^{-},n)$ and $(K^{-},p)$ reactions on 
$^{12}$C at $p_{\rm lab}=1$~GeV/c (KEK-PS E548)~\cite{Kis06} have not 
disclosed any peaks beyond the appreciable strength observed below the 
$\bar K$-nucleus threshold. Ongoing experiments by the FINUDA spectrometer 
collaboration at DA$\Phi$NE, Frascati, already claimed evidence for 
a relatively broad $K^- pp$ deeply bound state ($B \sim 115$~MeV) in 
$K^{-}_{\rm stop}$ reactions on Li and $^{12}$C, by observing back-to-back 
$\Lambda p$ pairs from the decay $K^-pp\to\Lambda p$~\cite{ABB05}, but these 
pairs could naturally arise from conventional absorption processes at 
rest when final-state interaction is taken into account~\cite{MOR06}. 
Indeed, the $K^-_{\rm stop}pn\to \Sigma^- p$ reaction on $^6$Li has also 
been recently observed~\cite{ABB06}. Another recent 
claim of a very narrow and deep $K^- pp$ state ($B \sim 160$~MeV, 
$\Gamma \sim 30$~MeV) is also based on observing decay $\Lambda p$ pairs, 
using $\bar p$ annihilation data on $^4$He from the OBELIX spectrometer 
at LEAR, CERN~\cite{Bre06}. The large value of $B_{K^- pp}$ over 100 MeV 
conjectured by these experiments is at odds with {\it all} the few-body 
calculations of the $K^- pp$ system listed above. One cannot rule out that 
the $\Lambda p$ pairs assigned in the above analyses to $K^-pp$ decay in fact 
result from nonmesonic decays of different clusters, say the $\bar K NNN$ $I=0$ 
quasibound state. A definitive study of the $K^- pp$ quasibound 
state (or more generally $\{\bar K[NN]_{I=1}\}_{I=1/2}$) 
could be reached through fully exclusive formation reactions, such as: 
\begin{equation} 
\label{eq:nag}
K^-+^3{\rm He}~ \to ~ n + \{\bar K[NN]_{I=1}\}_{I=1/2,I_z=+1/2},~~~ 
p + \{\bar K[NN]_{I=1}\}_{I=1/2,I_z=-1/2} \, , 
\end{equation} 
the first of which is scheduled for day-one experiment in J-PARC~\cite{Nag06}. 
Finally, preliminary evidence for a $\bar K NNN$ $I=0$ state with 
$B = 58 \pm 6$~MeV, $\Gamma = 37 \pm 14$~MeV has been recently presented 
by the FINUDA collaboration on $^6$Li by observing 
back-to-back $\Lambda d$ pairs~\cite{ABB07}. 
It is clear that the issue of $\bar K$ nuclear 
states is far yet from being experimentally resolved 
and more dedicated, systematic searches are necessary.

\subsection{RMF dynamical calculations of $\bar K$ quasibound nuclear states} 
\label{sec:RMF} 

\begin{figure}[t]
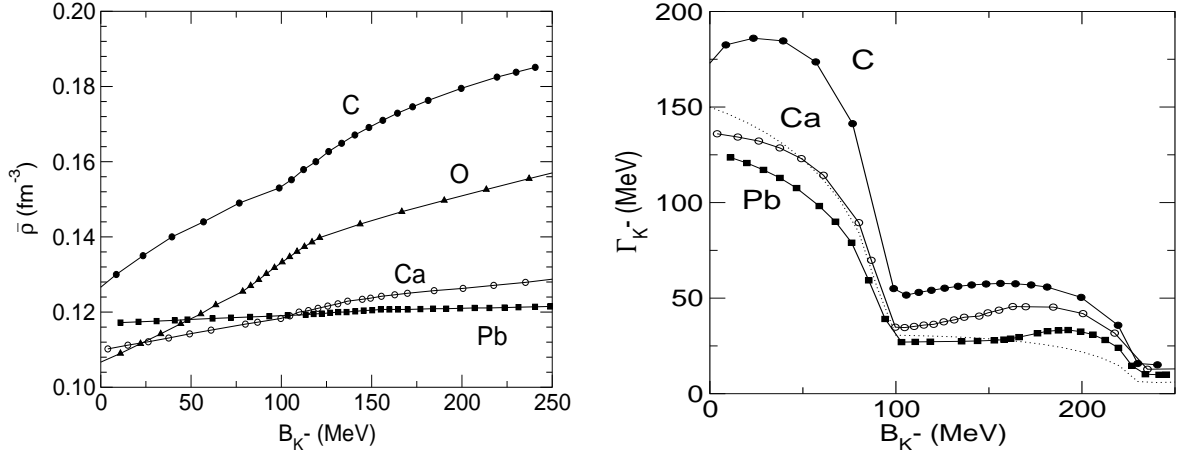
 
\centering 
\includegraphics[height=6cm,width=7.5cm]{fg07fig26a.eps} 
\hspace*{3mm} 
\includegraphics[height=6cm,width=7.5cm]{fg07fig26b.eps} 
\caption{Dynamically calculated average nuclear density $\bar \rho$ (left) 
and widths $\Gamma_{K^-}$ (right) of $1s$ $K^-$-nuclear states in the nuclei 
denoted, as function of the $1s$ $K^-$ binding energy~\cite{MFG06}. }
\label{fig:rho} 
\end{figure}

In this model, spelled out in Refs.~\cite{MFG05,MFG06,GFG07}, the (anti)kaon 
interaction with the nuclear medium is incorporated by adding to ${\cal L}_N$ 
the Lagrangian density ${\cal L}_K$: 
\begin{equation}
\label{eq:Lk}
{\cal L}_{K} = {\cal D}_{\mu}^*{\bar K}{\cal D}^{\mu}K -
m^2_K {\bar K}K
- g_{\sigma K}m_K\sigma {\bar K}K\; .
\end{equation} 
The covariant derivative
${\cal D_\mu}=\partial_\mu + ig_{\omega K}{\omega}_{\mu}$ describes
the coupling of the (anti)kaon to the vector meson $\omega$.
The(anti)kaon coupling to the isovector $\rho$ meson was found to have 
negligible effects. 
The $\bar K$ meson induces additional source terms in the equations of motion 
for the meson fields $\sigma$ and $\omega_0$. It thus affects the scalar 
$S = g_{\sigma N}\sigma$ and the vector $V = g_{\omega N}\omega_0$ potentials 
which enter the Dirac equation for nucleons, and this leads to rearrangement 
or polarization of the nuclear core, as shown on the left-hand side of 
Fig.~\ref{fig:rho} for the calculated average nuclear density 
$\bar \rho = \frac{1}{A}\int\rho^2d{\bf r}$ 
as a function of $B_{K^-}$ for $K^-$ nuclear $1s$ states across the periodic 
table. It is seen that in the light $K^-$ nuclei, $\bar \rho$ increases 
substantially with $B_{K^-}$ to values about 50\% higher than without the 
$\bar K$. The increase of the central nuclear densities is bigger, 
up to 100\%, and is nonnegligible even in the heavier $K^-$ nuclei where it 
is confined to a small region of order 1~fm. 
Furthermore, in the Klein-Gordon 
equation satisfied by the $\bar K$, the scalar $S = g_{\sigma K}\sigma$ and 
the vector $V = -g_{\omega K}\omega_0$ potentials become 
{\it state dependent} through the {\it dynamical} density dependence of the 
mean-field potentials $S$ and $V$, as expected in a RMF calculation. 
An imaginary ${\rm Im}V_{\bar K} \sim t\rho$ was added, fitted to the 
$K^-$ atomic data~\cite{FGM99}. It was then suppressed by an energy-dependent 
factor $f(B_{\bar K})$, considering the reduced phase-space for the initial 
decaying state and assuming two-body final-state kinematics for the decay 
products in the $\bar K N \to \pi Y$ mesonic modes ($80\%$) and in the 
$\bar K NN \to Y N$ nonmesonic modes ($20\%$).  

The RMF coupled equations were solved self-consistently. For a rough idea, 
whereas the static calculation gave $B_{K^-}^{1s}~=~132$~MeV
for the $K^-$ $1s$ state in $^{12}$C, using the values 
$g^{\rm atom}_{\omega K},~g^{\rm atom}_{\sigma K}$ corresponding to the 
$K^-$-atom fit, the dynamical calculation gave $B_{K^-}^{1s}~=~172$~MeV. 
In order to scan a range of values for $B_{K^-}^{1s}$, the coupling 
constants $g_{\sigma K}$ and $g_{\omega K}$ were varied in given intervals 
of physical interest. 

Beginning approximately with $^{12}$C, the following conclusions may be drawn: 

\begin{itemize} 

\item For given values of $g_{\sigma K},g_{\omega K}$, the $\bar K$ binding 
energy $B_{\bar K}$ saturates as function of $A$, except for a small increase 
due to the Coulomb energy (for $K^-$). 

\item The difference between the binding energies calculated dynamically and 
statically, $B_{\bar K}^{\rm dyn} - B_{\bar K}^{\rm stat}$, is substantial 
in light nuclei, increasing with $B_{\bar K}$ for a given value of $A$, and 
decreasing monotonically with $A$ for a given value of $B_{\bar K}$. 
It may be neglected only for very heavy nuclei. The same holds for the 
nuclear rearrangement energy $B_{\bar K}^{\rm s.p.} - B_{\bar K}$ which is 
a fraction of $B_{\bar K}^{\rm dyn} - B_{\bar K}^{\rm stat}$. 
 
\item The width $\Gamma_{\bar K}(B_{\bar K})$ decreases monotonically 
with $A$, according to the right-hand side of Fig.~\ref{fig:rho} 
\begin{figure}[t] 
\centerline{\includegraphics[scale=0.5,angle=-90]{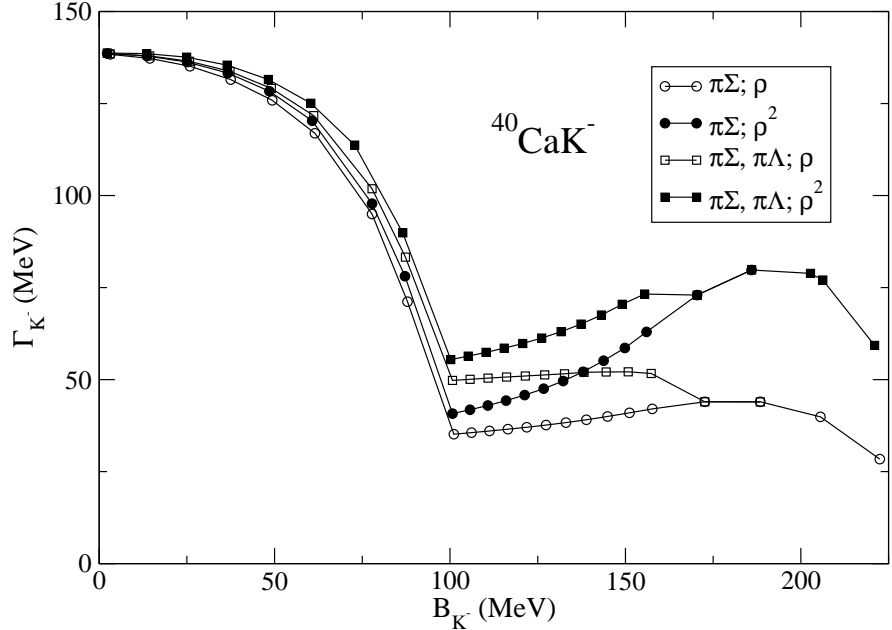}} 
\caption{Dynamically calculated widths of the $1s$ $K^-$-nuclear 
state in $^{~~40}_{K^-}$Ca for $\pi \Sigma + \pi \Lambda$ compared to 
$\pi \Sigma$ final mesonic absorption channels, and for $\rho^2$ compared 
to $\rho$ dependence of the final nonmesonic absorption channels, 
as function of the $K^-$ binding energy, from Gazda et al.~\cite{GFG07}.}
\label{fig:Gamma}  
\end{figure} 
which shows calculated 
widths $\Gamma_{K^-}$ as function of the binding energy $B_{K^-}$ for 
$1s$ states in $^{~~12}_{K^-}$C, $^{~~40}_{K^-}$Ca and $^{~~208}_{K^-}$Pb. 
The dotted line shows the static `nuclear-matter' limit 
corresponding to the $K^-$-atom fitted value ${\rm Im}b_0=0.62$ fm and for 
$\rho(r)=\rho_0=0.16$ fm$^{-3}$, using the same phase-space suppression factor 
as in the `dynamical' calculations. It is clearly seen that the functional 
dependence $\Gamma_{K^-}(B_{K^-})$ follows the shape of the dotted line. 
This dependence is due primarily to the binding-energy dependence of the 
suppression factor $f(B_{K^-})$ which falls off rapidly until
$B_{K^-} \sim 100$~MeV, where the dominant
$\bar K N \rightarrow \pi \Sigma$ gets switched off, and then stays
rather flat in the range $B_{K^-} \sim$~100-200~MeV where the width is 
dominated by the $\bar K NN \to YN$ absorption modes. The widths
calculated in this range are considerably larger than given by the dotted 
line (except for Pb in the range $B_{K^-} \sim$~100-150~MeV) due 
to the dynamical nature of the RMF calculation, whereby the nuclear density 
is increased by the polarization effect of the $K^-$. 
Adding perturbatively the residual width neglected in this calculation, 
partly due to the $\bar K N \to \pi \Lambda$ secondary mesonic decay channel, 
a representative value for a lower limit $\Gamma_{\bar K} \geq 50 \pm 10$~MeV 
holds in the binding energy range $B_{K^-} \sim$~100-200~MeV.
Fig.~\ref{fig:Gamma} shows the effect of splitting the $80\%$ 
mesonic decay width, previously assigned all to $\pi \Sigma$ absorption 
channels, between $\pi \Sigma$ ($70\%$) and $\pi \Lambda$ ($10\%$), and also 
of simulating the $20\%$ nonmesonic absorption channels by a $\rho^2$ 
dependence compared to ${\rm Im}V_{\bar K} \sim t\rho$ used by Mare\v{s} 
et al.~\cite{MFG06}. These added contributions ~\cite{GFG07} make the lower 
limit $\Gamma_{\bar K} \geq 50 \pm 10$~MeV a rather conservative one. 

\end{itemize}


\section{$K^+$ mesons}
\label{sec:Kplus}

\subsection{Overview of the $K^+$-nucleus interaction}
\label{sec:kplusexp}

There are obviously no $K^+$ exotic atoms to provide information on the
$K^+$-nucleus interaction below threshold. Nevertheless, it has been found that
for pions and antiprotons the optical potentials cross smoothly from the
atomic into the scattering regime, and therefore studies of the $K^+$-nucleus
interaction at low energies above threshold are relevant to the general topic
of medium modifications of the interaction. In fact, $K^+$ mesons provide 
a clear example of such modifications.

The $K^+N$ interaction below the pion-production threshold is fairly weak and
featureless and this merit has motivated past suggestions to probe 
nuclear in-medium effects by studying scattering and reaction processes 
with $K^+$ beams below 800~MeV/c; see Ref.~\cite{DWa82} for an early review. 
The insufficiency of the impulse-approximation $t_{\rm free}\rho$ form of 
the $K^+$ - nucleus optical potential, where $t_{\rm free}$ is the 
free-space $K^+N$ $t$ matrix, was somewhat of a surprise already in the 1980s. 
Limited total cross-section data \cite{BGK68} on carbon, and elastic
and inelastic differential cross section data \cite{MBC82} on carbon and
calcium showed problems with the $t\rho$ potential,  particularly
with respect to its reaction content (`reactivity' below).
In order to account for the increased reactivity in $K^+$ - nucleus 
interactions, Siegel et al.~\cite{SKG85} and later Peterson~\cite{Pet99} 
suggested that nucleons `swell' in the
nuclear medium, primarily by increasing the dominant hard-core $S_{11}$
phase shift. Brown et al.~\cite{BDS88} suggested that the extra
reactivity was due to the reduced in-medium masses of exchanged vector
mesons, and this was subsequently worked out in detail in Ref.~\cite{CLa96}.
Another source for increased reactivity in $K^+$ - nucleus interactions
was discussed in the 1990s and is due to meson exchange-current 
effects~\cite{JKo92,GNO95}.

\begin{figure} 
\includegraphics[scale=0.7]{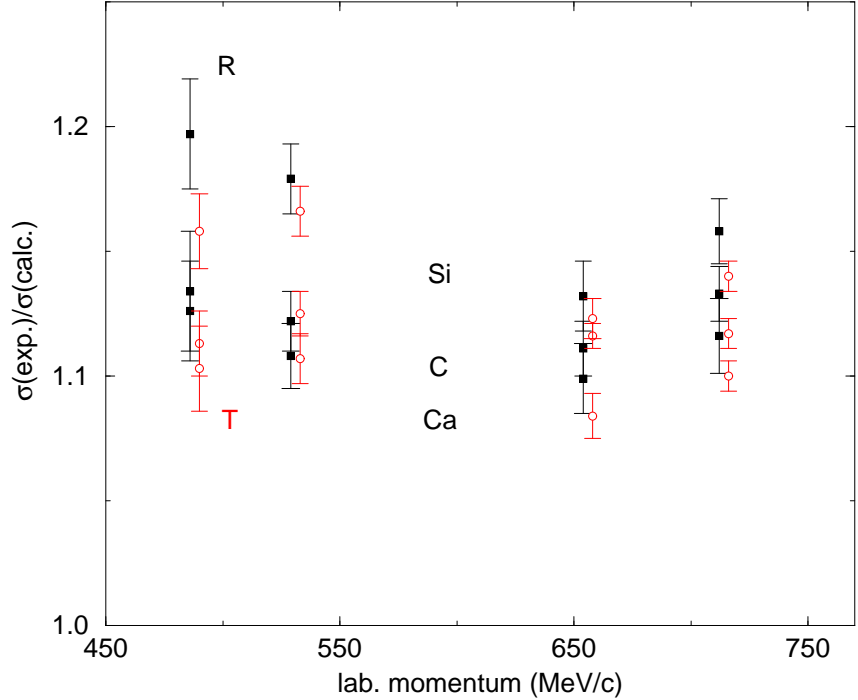} 
\caption{Ratios between experimental and calculated cross sections 
for calculations based on $V_{\rm opt}^{(K^+)}=t\rho$ fits to $^6$Li. 
Solid squares represent $\sigma_R$, circles represent $\sigma_T$.} 
\label{fig:kratios} 
\end{figure}

Some further experimental progress was made during the early 1990s,
consisting mostly of measuring attenuation cross sections in $K^+$
transmission experiments at the BNL-AGS on deuterium and several other
nuclear targets in the momentum range $p_{\rm lab} =$~450-740~MeV/c 
\cite{KAA92,SWA93,WAA94} and extracting $K^+$-nucleus total cross sections. 
The same transmission data were then reanalyzed to extract in addition 
total reaction cross sections~\cite{FGW97}, which are less dependent 
on the potential used in converting measured attenuation into cross 
sections and eventually self-consistent final values of $K^+$ integral 
cross sections (reaction and total) on $^6$Li, $^{12}$C, $^{28}$Si and 
$^{40}$Ca were published in Refs.~\cite{FGM97a,FGM97b}. 
These integral cross-section data gave clear evidence for the density 
dependence of the increased reactivity suggested by the earlier data, 
as demonstrated in Fig.~\ref{fig:kratios} from Ref.~\cite{FGM97a}. 
Plotted in the figure are ratios of experimental to calculated integral 
cross sections for $^{12}$C, $^{28}$Si and $^{40}$Ca, where the calculated 
cross sections use a $t\rho$ potential fitted to the $^6$Li data. 
These ratios, for the denser nuclei, deviate considerably from the value 
of one in a way which is largely independent of the beam momentum. 

Other measurements during the 1990s include $K^+$ quasielastic scattering 
on several targets at 705 MeV/c~\cite{KPS95} and new 
measurements of $K^+$ elastic and inelastic differential cross sections on 
$^6$Li and $^{12}$C at 715 MeV/c~\cite{MBB96}, further analyzed 
in Ref.~\cite{CSP97}. These data and analyses lent support to the substantial 
medium modifications demonstrated above on the basis of studying integral 
cross sections. By the late 1990s, experimentation in
$K^+$ - nuclear physics had subsided, and with it died out also theoretical
interest although the problems with medium modifications of the interaction
remained, as is shown below. Theoretical interest in $K^+$ - nuclear physics 
to some extent has been revived recently \cite{GFr05,GFr06,TCR06,AVG05}, 
particularly in connection with possible contributions due to $\Theta^+$ 
pentaquark degrees of freedom, as is also discussed below.

\subsection{Kaon-nucleus optical potential}
\label{sec:kpluspotl}

The starting form adopted for the kaon-nucleus optical potential 
$V_{\rm opt}$, following Eq.~(\ref{eq:KGs}) of Sec.~\ref{sec:pot}, 
is the simplest possible $t\rho$ form:
\begin{equation}
\label{eq:kplusVopt}
2 \varepsilon^{(A)}_{\rm red} V_{\rm opt}(r) = -4\pi F_A b_0 \rho(r) ~~,
\end{equation}
where $\varepsilon^{(A)}_{\rm red}$ is the center-of-mass (c.m.) 
reduced energy, 
\begin{equation} 
\label{eq:kin1}  
(\varepsilon^{(A)}_{\rm red})^{-1}=E_p^{-1}+E_A^{-1} 
\end{equation} 
in terms of the c.m. total energies for the projectile and target
respectively, and  
\begin{equation}
\label{eq:kin2}
F_A = \frac {M_A \sqrt{s}}{M(E_A+E_p)}
\end{equation}
is a kinematical factor resulting from the transformation
of amplitudes between the $KN$ and the $K^+$ - nucleus c.m. systems,
with $M$ the free nucleon mass, $M_A$ the mass of the target nucleus
and $\sqrt s$ the total projectile-nucleon energy in their c.m. system.
The parameter $b_0$ in Eq.~(\ref{eq:kplusVopt}) reduces in the impulse
approximation to the (complex) isospin-averaged $KN$ scattering amplitude 
in the forward direction. For $^6$Li and for $^{12}$C the modified 
harmonic oscillator (MHO) form was used for the nuclear densities whereas 
for $^{28}$Si and for $^{40}$Ca the two-parameter Fermi (2pF) form was 
used and minor changes were made to parameters of the neutron density 
to check sensitivities to $\rho _n$. The data base for the analysis were 
the 32 integral cross sections for $K^+$ on $^6$Li, $^{12}$C, $^{28}$Si 
and $^{40}$Ca from Ref.~\cite{FGM97b}.

\begin{table}
\caption{Fits to the eight $K^+$ - nuclear integral cross sections
\protect\cite{FGM97b} at each of the four laboratory momenta $p_{\rm lab}$
(in MeV/c), using different potentials.}
\label{tab:FGa04}
\begin{tabular}{ccccccc}
\hline \hline
$p_{\rm lab}$~~&~~$V_{\rm opt}$~~&~~Re$b_0$(fm)~~&~~Im$b_0$(fm)~~&
~Re$B$(fm$^4$)~&~Im$B$(fm$^4$)~&~$\chi^2/N$  \\  \hline
488&$t\rho$ &$-$0.203(26)&0.172(7)& & & 16.3 \\
   &$t_{\rm free}\rho$&$-$0.178&0.153 & & &  \\
   &Eq.(\ref{eq:kpDD1}) &$-$0.178&0.122(5)&0.52(20)&0.88(8) &1.18 \\
   &Eq.(\ref{eq:kpDD2}) &$-$0.178&0.129(4)&0.17(11)&0.62(6) &0.27 \\
 & & & & & & \\
531&$t\rho$ &$-$0.196(39)&0.202(9)& & & 56.3 \\
   &$t_{\rm free}\rho$&$-$0.172&0.170 & & &  \\
   &Eq.(\ref{eq:kpDD1}) &$-$0.172&0.155(14)&1.79(46)&0.72(27) &7.01 \\
   &Eq.(\ref{eq:kpDD2}) &$-$0.172&0.146(5) &0.46(21)&0.78(7) &3.94 \\
 & & & & & & \\
656&$t\rho$ &$-$0.220(50)&0.262(12)& & & 54.9 \\
   &$t_{\rm free}\rho$&$-$0.165&0.213 & & &  \\
   &Eq.(\ref{eq:kpDD1}) &$-$0.165&0.203(18)&1.66(80)&0.89(36) &7.24 \\
   &Eq.(\ref{eq:kpDD2}) &$-$0.165&0.204(5) &2.07(19)&0.77(7) &0.32 \\
 & & & & & & \\
714&$t\rho$ &$-$0.242(53)&0.285(15)& & & 67.7 \\
   &$t_{\rm free}\rho$&$-$0.161&0.228 & & &  \\
   &Eq.(\ref{eq:kpDD1}) &$-$0.161&0.218(24)&1.40(95)&1.10(48) &9.3 \\
   &Eq.(\ref{eq:kpDD2}) &$-$0.161&0.218(6)&1.51(43)&0.97(9)& 1.24 \\
\hline \hline
\end{tabular}
\end{table}

Fits to the integral cross sections were made 
\cite{GFr05,GFr06} separately at each of
the four momenta, varying the complex parameter $b_0$, and the results are
summarized in Table \ref{tab:FGa04}, marked as $t\rho$ for each momentum.
From the values of $\chi ^2$ per point it is seen that the fits are 
unacceptably poor and the resulting Re$b_0$ and Im$b_0$ disagree with
the corresponding free $K^+ N$  values (marked as $t_{\rm free}\rho$ 
and derived from the $KN$ phase shifts as given by SAID~\cite{SAID}). 
The discrepancies are particularly noticeable for Im$b_0$, which are 
determined to good accuracy. Evidently the experimental results indicate 
significant increase in reactivity, as mentioned in Sec.~\ref{sec:kplusexp}.

The obvious next step is to effectively make $b_0$ density dependent by
introducing higher powers of the density, such as

\begin{equation}
\label{eq:kpDD1}
b_0~ \rho (r) \rightarrow b_0~ \rho (r)~+~B~ \rho^2 (r)~,
\end{equation}  
where both parameters $b_0$ and $B$ are to be determined from fits to
the data. The results \cite{GFr05,GFr06} 
are also shown in Table \ref{tab:FGa04}, marked
as Eq.~(\ref{eq:kpDD1}), where Re$b_0$ was kept fixed at its respective
free $KN$ value. The improvement in the fits to the data is evident
from the reduction of the values of $\chi ^2$, but except for the
lowest incoming momentum the quality of the fits suggests that something 
is still missing. Guided by earlier analyses \cite{FGM97a,FGM97b} that 
achieved much improved fits by introducing the {\it average} nuclear
density ${\bar \rho}$ for each target nuclide, 

\begin{equation}
\label{eq:rhoave}
{\bar \rho}=\frac{1}{A}\int\rho^2d{\bf r} \,,
\end{equation}
Eq.~(\ref{eq:kpDD1}) was replaced by the following ansatz~\cite{GFr05,GFr06}: 
\begin{equation}
\label{eq:kpDD2}
b_0~ \rho (r) \rightarrow b_0~ \rho (r)~+~B~ {\bar \rho}~\rho (r)~.
\end{equation}
The results of this prescription are also shown in Table \ref{tab:FGa04},
marked as Eq.~(\ref{eq:kpDD2}) and it is clear that the fits to the data
are very good. This use of the average nuclear density singles out the $^6$Li 
target from the other three targets, due to its average density ${\bar \rho}$ 
being close to 50\% of the corresponding values for the other targets. 
This is a purely phenomenological observation without (as yet) any theory 
behind it.

\begin{figure}
\includegraphics[scale=0.8,angle=0]{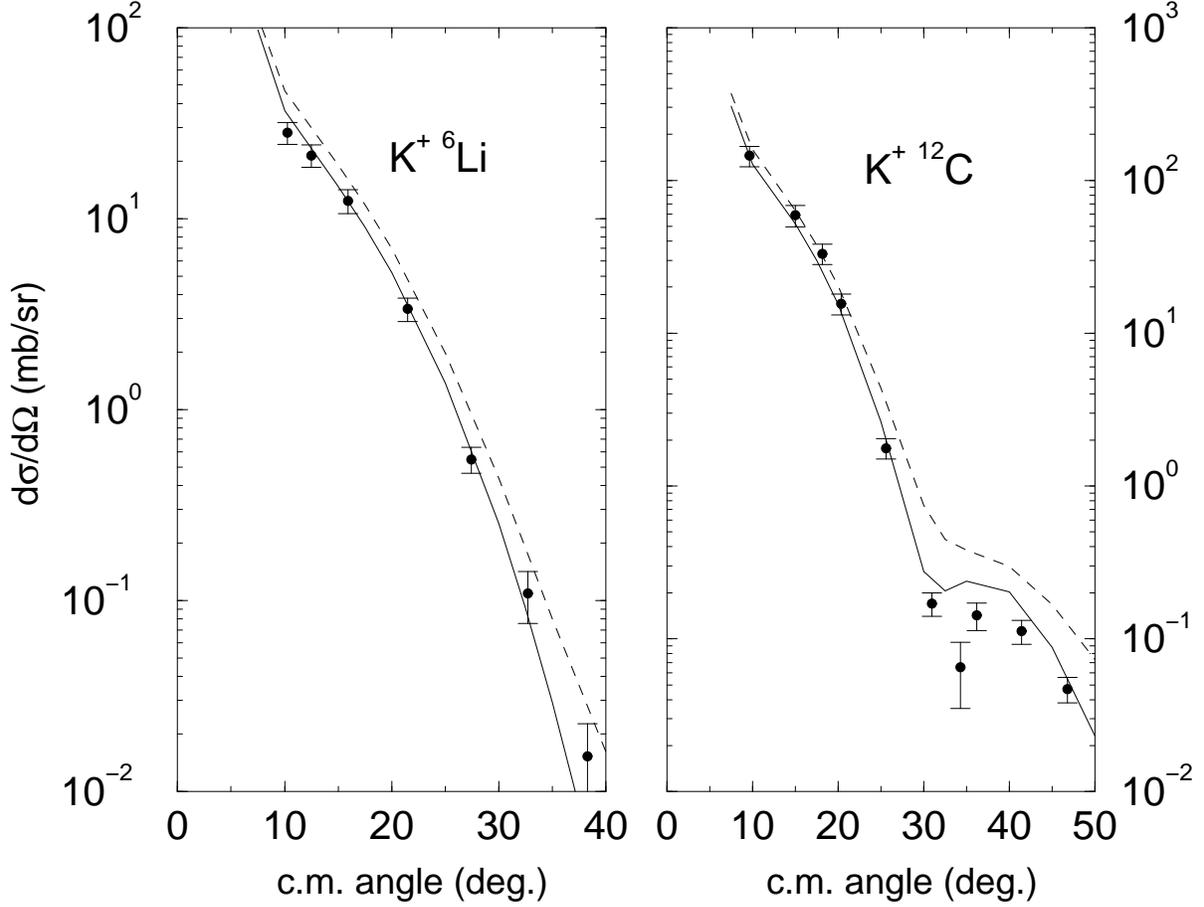}
\caption{Comparison between measured differential cross sections for
$K^+$ elastic scattering at $p_{\rm lab}=715$ MeV/c on $^6$Li and
$^{12}$C \protect{\cite{CSP97}} and best-fit calculations using
Eq.~(\ref{eq:kpDD1}) (dashed lines) and Eq.~(\ref{eq:kpDD2}) (solid lines).}
\label{fig:kplusfig2}
\end{figure}

In order to further test the picture that emerges from the analysis
of the integral cross sections for the $K^+$ - nucleus interaction,
the analysis was repeated~\cite{GFr06}
including also differential cross sections for the
elastic scattering of $K^+$ by some of the  target nuclei.
Fits were made to the combined integral and differential cross sections
at 714 MeV/c consisting of the eight integral cross sections and the 17
differential cross sections from Ref.~\cite{CSP97}, using the potentials
of either Eq.~(\ref{eq:kpDD1}) or Eq.~(\ref{eq:kpDD2}).
Figure \ref{fig:kplusfig2} shows that again the ${\bar \rho}\rho$
version of the potential Eq.~(\ref{eq:kpDD2}) is preferred and the fits to
the differential cross sections are good. The potential parameters 
obtained from the fits to the combined integral and differential
cross sections agree, within uncertainties, with the corresponding 
values in Table~\ref{tab:FGa04}.

Prior to discussing in the next section the reactive content of the above 
forms of density-dependent $K^+$ - nucleus optical potentials, it is worth 
noting that the splitting of Im$V_{\rm opt}$ in Table \ref{tab:FGa04} 
into its two reactive components Im$b_0$ and Im$B$ appears well 
determined by the data at all energies, and perhaps is even model 
independent, particularly for the ${\bar \rho}\rho$ version 
Eq.~(\ref{eq:kpDD2}) of the optical potential for which very accurate 
values of Im$b_0$ are derived. These values of Im$b_0$ are close to, 
but somewhat below the corresponding free-space values, a feature which 
is observed in calculations which replace the $t\rho$ form by $g\rho$ 
where nuclear phase space effects are considered explicitly~\cite{TCR06}. 
The values derived for Im$B$ are roughly independent of the form of the 
piece added to $t\rho$, $\Delta V_{\rm opt}$, whether Eq.~(\ref{eq:kpDD1}) 
or Eq.~(\ref{eq:kpDD2}) are used to derive these values from the data. 
In contrast, the two components of Re$V_{\rm opt}$ 
are correlated strongly when Re$b_0$ is also varied, largely cancelling 
each other into a resulting poorly determined Re$V_{\rm opt}$. 
For this reason, it will appear difficult to offer any conclusive model for 
the physics underlying the real part of $\Delta V_{\rm opt}$.

\subsection{$K^+$ absorption cross sections} 
\label{sec:kabs} 

\begin{figure}
\includegraphics[scale=0.7,angle=0]{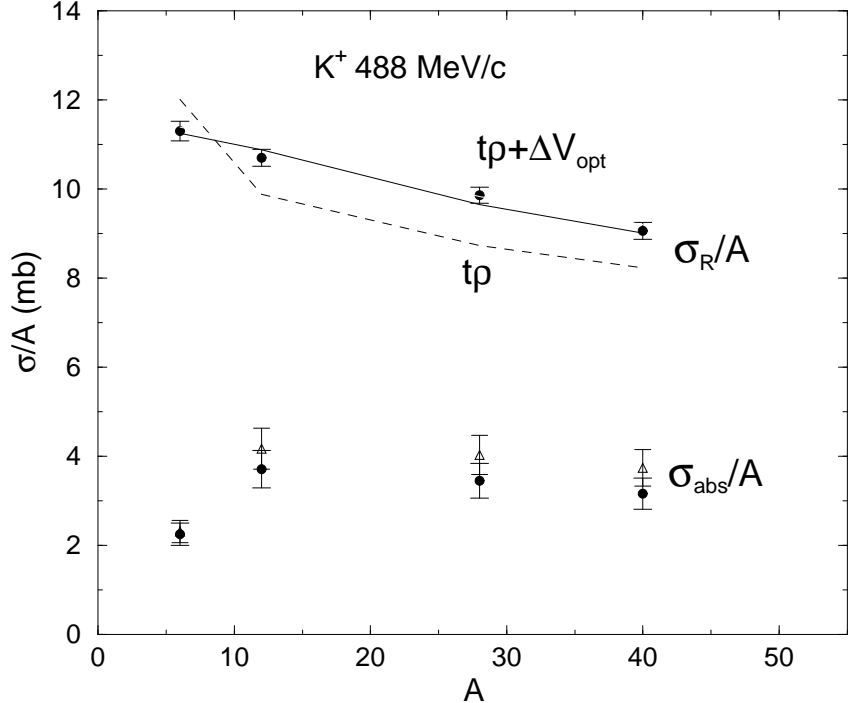}
\caption{Data and calculations~\protect\cite{GFr05} for $K^+$ reaction cross
sections per nucleon ($\sigma_R/A$) at $p_{\rm lab}=488$ MeV/c are shown 
in the upper part. Calculated $K^+$ absorption cross sections per nucleon
($\sigma_{\rm abs}/A$) are shown in the lower part, see text.}
\label{fig:kplusfig1}
\end{figure} 

The effect of $\Delta V_{\rm opt}$, within the improved fits to the 
$K^+$-nuclear integral cross sections, is demonstrated in the upper part of 
Fig.~\ref{fig:kplusfig1} for the reaction cross sections per nucleon 
$\sigma_R/A$ at 488 MeV/c, where the calculated cross sections using 
a best-fit $t\rho$ optical potential (dashed line) are compared with the 
experimental values listed in Ref.~\cite{FGM97b}. The $t\rho$ fit fails 
to reconcile the $^6$Li data with the data on the other, denser nuclei. 
If $^6$Li is removed from the data base, then it becomes 
possible to fit reasonably well the data for the rest of the nuclei, 
but the rise in Im$b_0$ with respect to its free-space value is then 
substantially higher than that for the $t\rho$ potential when $^6$Li is 
included. At the higher energies, $t\rho$ fits which exclude $^6$Li are 
less successful than at 488 MeV/c, while also requiring a substantial 
rise in Im$b_0$, which means increased values of the in-medium $KN$ 
total cross sections with respect to the corresponding free-space values. 
This has been observed also in a $K^+$ - nucleus quasifree-scattering 
analysis~\cite{Pet04}, for $K^+$ mesons incident on C, Ca, Pb at 
$p_{\rm lab}=705$ MeV/c~\cite{KPS95}. Also shown in the upper part 
of Fig.~\ref{fig:kplusfig1}, by the solid line marked 
$t\rho + \Delta V_{\rm opt}$, are calculated reaction cross sections 
at 488 MeV/c using Eq.~(\ref{eq:kpDD2}). This is a very good fit. 
Recently, Tolos et al.~\cite{TCR06} have demonstrated that a similarly 
substantial improvement in the reproduction of reaction cross sections 
could be achieved microscopically by coupling in pentaquark degrees 
of freedom. It is tempting to assume that the effects of absorbing $K^+$ 
mesons into a pentaquark configuration are given by the difference between 
the solid curve and the dashed curve in Fig.~\ref{fig:kplusfig1} for nuclear 
targets heavier than $^6$Li. However, for a quantitative estimate of the $K^+$ 
{\it absorption} cross sections one needs to do a more explicit calculation. 

In close analogy to analyses of pionic atoms and of low-energy pion-nuclear 
scattering reactions in which the parameter $B_0$ (cf. Eq.~(\ref{eq:EE1s})) 
is related to $\pi^-$ nuclear absorption processes on two and on more 
nucleons, the additional piece $\Delta V_{\rm opt}$ due to the nonzero value 
of the parameter $B$ in Eqs.~(\ref{eq:kpDD1},\ref{eq:kpDD2}) represents $K^+$ 
nuclear absorption into $\Theta^+$ - nuclear final states. Here $\Theta^+$, 
with mass $M_{\Theta^+} \approx 1540$~MeV, is an hypothetical $S=+1$ `exotic' 
pentaquark baryon, searches for which have not provided 
conclusive evidence (for a recent review see Ref.~\cite{Sch06}). 
The abnormally small upper limit $\Gamma_{\Theta^+} < 1$~MeV for the width 
of the $\Theta^+$ pentaquark deduced from some of these searches 
implies a negligible $\Theta^+ \to KN$ coupling, but this does not limit the  
coupling $\Theta^+N \to KNN$ which is related to virtual decays such as  
$\Theta^+ \to \pi K N$~\cite{LOM05}. Gal and Friedman~\cite{GFr05,GFr06} 
estimated the nuclear absorption cross sections of $K^+$ mesons by using two 
slightly different versions of the distorted-wave Born approximation: 
\begin{equation}
\label{eq:abs1}
\sigma_{\rm abs}^{(K^+)} \sim  -~ {\frac{2}{\hbar v}} 
\int {\rm Im} (\Delta V_{\rm opt}(r))~
|\Psi_{(\Delta V_{\rm opt}=0)}^{(+)}({\bf r})|^2~d{\bf r} ~~,
\end{equation} 
and 
\begin{equation}
\label{eq:abs2}
\sigma_{\rm abs}^{(K^+)} \sim  -~ {\frac{2}{\hbar v}} 
\int {\rm Im} (\Delta V_{\rm opt}(r))~
|\Psi^{(+)}({\bf r})|^2~d{\bf r} ~~, 
\end{equation} 
where the distorted waves $\Psi_{(\Delta V_{\rm opt}=0)}^{(+)}$ are
calculated discarding $\Delta V_{\rm opt}$.  

Calculated absorption cross sections {\it per target nucleon} at
$p_{\rm lab}=488$ MeV/c are shown in the lower part of Fig.~\ref{fig:kplusfig1}
for the fit using Eq.~(\ref{eq:kpDD2}) for $V_{\rm opt}$ in 
Table~\ref{tab:FGa04}. The triangles are for expression (\ref{eq:abs1}) 
and the solid circles are for expression (\ref{eq:abs2}). The error bars 
plotted are due to the uncertainty in the parameter Im$B$. It is seen that 
these calculated absorption cross sections, for the relatively dense targets 
of $^{12}$C, $^{28}$Si and $^{40}$Ca, are proportional to the mass number 
$A$, and the cross section per target nucleon due to Im$B \neq 0$ is 
estimated as close to 3.5 mb. Although the less successful 
Eq.~(\ref{eq:kpDD1}) gives cross sections larger by $40\%$ at this particular 
incident momentum, this value should be regarded an upper limit, since the 
best-fit density-dependent potentials of Refs.~\cite{FGM97a,FGM97b} yield 
values smaller than 3.5 mb by a similar amount. 
The experience gained from studying $\pi$-nuclear absorption \cite{ASc86} 
leads to the conclusion that $\sigma_{\rm abs}(K^+NN)$ is smaller than the 
extrapolation of $\sigma_{\rm abs}^{(K^+)}/A$ in Fig.~\ref{fig:kplusfig1} to
$A=1$, and since the $KN$ interaction is weaker than the $\pi N$ interaction
one expects a reduction of roughly $50\%$, so that 
$\sigma_{\rm abs}(K^+NN)~\sim$~1-2~mb. 

In Fig.~\ref{fig:kplusfig1}, the considerably smaller absorption cross 
section per nucleon calculated for the relatively low-density $^6$Li 
nucleus suggests a cross section of order fraction of millibarn, 
in a possible missing-mass search based on observing the final proton 
in the two-body reaction $K^+ d \rightarrow \Theta^+ p$. 
This cross section is not expected to exhibit marked resonance behavior near 
$p_{\rm lab} \sim 440$ MeV/c, which corresponds to the $\Theta^+$(1540) 
resonance assumed rest mass, even if $\Theta^+$ is very narrow. 
For the heavier nuclear targets too, the assignment of the excess reactivity 
observed in $K^+$-nuclear cross sections as due to $S=+1$ pentaquark degrees 
of freedom does not require the existence of a {\it narrow} $KN$ resonance. 
It only assumes that pentaquark degrees of freedom are spread over this 
energy range with sufficient spectral strength. For nuclear targets other 
than deuterium, given the magnitude of the $K^+$ nuclear absorption cross 
sections as reviewed here, ($K^+,p$) experiments could prove useful. This 
reaction which has a `magic momentum' about $p_{\rm lab} \sim 600$ MeV/c, 
where the $\Theta^+$(1540) is produced at rest, is particularly suited to 
study bound or continuum states in {\it hyponuclei} ($S=+1$ nuclei 
according to the terminology suggested by Alfred Goldhaber~\cite{Gol82}). 

In conclusion, the available $K^+$ nuclear cross section data at 
$p_{\rm lab} \sim$~450-800~MeV/c reveal substantial reactivity beyond 
that produced by the impulse approximation, or for that purpose by any 
effective $t\rho$ form of the $K^+$ optical potential. It was shown that 
this extra reactivity may be explained by adding a two-nucleon absorption 
channel $K^+ nN \rightarrow \Theta^+ N$, where the $\Theta^+$ degrees of 
freedom need not materialize within a narrow energy bin. This provides 
a density-dependent mechanism that couples in $S=+1$ pentaquark degrees of 
freedom in a way which is insensitive to the width of their spectral 
distribution~\cite{GFr05}. 
While there is no firm support at present for this conjecture from other 
phenomenological sources, a robust experimental program of measuring 
low-energy $K^+d$ and $K^+$ - nuclear scattering and reaction cross sections 
in the range $p_{\rm lab} \sim$~300-800~MeV/c, and particularly about 
400 Mev/c, would be extremely useful to decide whether or not $S=+1$ 
pentaquark degrees of freedom are involved in $K^+$ - nuclear dynamics.



\section{Antiprotons}
\label{sec:pbars}

\subsection{Overview of the ${\bar p}$-nucleus potential}
\label{sec:pbarspotl} 

In line with the other types of exotic atoms, the interaction of 
antiprotons with nuclei at threshold is described in terms of an optical 
potential, which in the simplest $t \rho$ form is given by 
\begin{equation}
\label{eq:pbarspotl}
2\mu V_{{\rm opt}}(r) = -4\pi(1+\frac{\mu}{M}
\frac{A-1}{A})[b_0(\rho_n+\rho_p)
  +b_1(\rho_n-\rho_p)]~~,
\end{equation}
where $\mu$ is the reduced mass of the $\bar p$,
 $\rho_n$ and $\rho_p$ are the neutron and proton density
distributions normalized to the number of neutrons $N$ and number
of protons $Z$, respectively, $A=N+Z$, and $M$ is the mass of the
nucleon. The factor $(A-1)/A$ above, which was omitted 
from the potential for pions, 
is included here due to the larger mass of the $\bar p$.
Because of the large cross section for annihilation of 
$\bar p$ on a single nucleon, the interaction is expected
 to be dominated by the imaginary part of the potential and
the absorption of $\bar p$ is expected to take place
at the extreme surface regions of the nucleus.
As a result it is unlikely that $\bar p$ atoms will provide
 information on the potential deep into the nucleus and
 the above simplest $t \rho$ form is a useful starting
point for analyzing antiprotonic atom data.
Previous attempts
to add to the potential 
a $p$-wave term or non linear terms \cite{BFG95,BFG97} are not
found to be justified in a phenomenological approach 
when the overall picture is considered, respecting also constraints 
satisfied by neutron density distributions, as described below.
More specifically, we find that an imaginary part of a $p$-wave
potential compatible with the Paris potential \cite{CLL94,PLL94,BLL99}
could be accommodated, but then Im$b_0$ is found to be incompatible 
with the Paris potential. 
These remarks apply particularly to analyses based on the 
high-quality data of the PS209 collaboration, which is the basis for
the present analysis. However, an isovector term $b_1(\rho_n-\rho_p)$
is included in Eq.~(\ref{eq:pbarspotl})
because the present data base is rich with groups of isotopes
of the same element.

Proton densities for the above potential are taken, as before, from
the known charge distributions \cite{FBH95} by unfolding the finite
size of the proton. For the neutrons it is, again, a matter of choosing
an adequate model, that will be in line with the bulk of information
on neutron densities \cite{JTL04}. The importance of the {\it shapes}
of neutron density distributions $\rho _n$ was realized long ago, when
single-particle densities (SP) were found \cite{BFG95,BFG97} to produce
better fits to the data of $\bar p$ atoms compared to fits based on the 
two-parameter Fermi (2pF) shape, because of the sensitivity of 
${\bar p}$ atom data to the extreme outer reaches of the nucleus. 
An alternative to SP densities is to use simple parameterizations such as 
the 2pF form for $\rho _n$ and to accommodate different shapes, relative 
to the protons, by taking the `skin' or the `halo' version for 
$\rho _n$ \cite{JTL04}, or their average. In what follows we adopt the 
latter approach for global fits to $\bar p$ data because in any case the 
SP densities are not particularly suitable for nuclei far removed from 
closed shells and because we are interested in average properties. 
We therefore use the approach of Sec.~\ref{sec:neutdens} with the
parameterization of $r_n-r_p$ given by Eq.~(\ref{eq:RMF}).

\subsection{Antiprotonic atom data}
\label{sec:pbarsdata}

As mentioned in Sec.~\ref{sec:exp}, the experimental situation 
with ${\bar p}$ atoms has changed significantly in the last decade 
with the publication by the PS209 collaboration \cite{TJC01} 
of high-quality data for several sequences of isotopes along the periodic
table. This set of X-ray data has greater accuracies compared to older
data used in earlier analyses, and thanks to the full coverage of the 
periodic table by these new data we do not mix in the present analysis 
old data with the new results but use only the PS209 results \cite{TJC01},
including the revised experimental results for Cd, Sn and Te isotopes
\cite{STC03,KWT04} obtained after correcting for E2 resonance effects.
In what follows we address only {\it spin-averaged} quantities for 
antiprotonic atoms within a global approach to the hadron-nucleus interaction.

Strong interaction effects in antiprotonic atoms are 
reported \cite{TJC01} as level shifts
and widths for the lowest levels reached in the X-ray cascade and as 
`upper' level widths deduced from the yields of the transitions, based on 
intensity ratios and calculations
of the atomic cascade process. These yield values had been converted 
in Ref. \cite{TJC01} into
 upper level widths with the help of the calculated radiation widths. 
However, it is easy to see that 
$\chi ^2$ values for the deduced upper level widths
may be different from the corresponding values calculated for the yields.
As the yields are the experimentally determined quantities, we have 
converted the quoted upper level widths back to transition yields,
and used these in the global $\chi ^2$ fits.

The X-ray data used in the present analysis are for the following 
nuclear targets:
$^{16,18}$O, $^{40,42,43,44,48}$Ca, $^{54,56,57,58}$Fe,
$^{58,60,62,64}$Ni, $^{90,96}$Zr, $^{106,116}$Cd, $^{112,116,120,124}$Sn,
$^{122,124,126,128,130}$Te and $^{208}$Pb,
a total of 90 data points \cite{TJC01}.

In addition to the conventional method of studying strong interaction
effects by observation of X-ray emission from exotic atoms, there
is for antiprotons a radiochemical method which is capable of providing
information on the absorption of $\bar p$ by nuclei \cite{Jas93}.
In brief, this method is based on the high probability for annihilation
of $\bar p$ on a single peripheral nucleon which leads to a residual nucleus
containing one neutron or one proton less than the target nucleus $(Z,N)$.
When the two residual nuclei $(Z,N-1)$ and $(Z-1,N)$ are radioactive, 
then measuring their activities can provide the {\it ratio} between the 
probability for $\bar p$ annihilation on a neutron to that on a proton. 
This is based on the reasonable assumption that following such extremely 
peripheral annihilation the resulting pions will not interact with the 
residual nucleus. Considering that the absorption takes place within 
a narrow range of radii in the outer surface region of nuclei, these ratios 
may provide information on the ratios between neutron and proton densities 
at that region, supplementing the information provided by the atomic X-rays.

Experimental ratios of $\bar p$ absorption on neutrons to absorption on
protons were taken from Refs.~\cite{LJT98,SHK99}. Initial calculations
showed that very large contributions to the resulting $\chi ^2$ came
from $^{106}$Cd and $^{112}$Sn and subsequently these two nuclei were
excluded from the data set. Possible explanations for the problem
with these two nuclei in terms of a $\bar p p$ quasi bound state are given
in Ref.~\cite{Wyc01}.
We have therefore used 17 values of absorption ratios
for the following nuclei: $^{48}$Ca, $^{58}$Ni, $^{96}$Zr,
$^{100}$Mo, $^{96,104}$Ru, $^{116}$Cd, $^{124}$Sn, $^{128,130}$Te,
$^{144,154}$Sm, $^{148}$Nd, $^{160}$Gd, $^{176}$Yb, $^{232}$Th and
$^{238}$U.

\subsection{Analyses of antiprotonic atom X-ray data}
\label{sec:pbarsana}

\begin{figure}
\includegraphics[scale=0.7,angle=0]{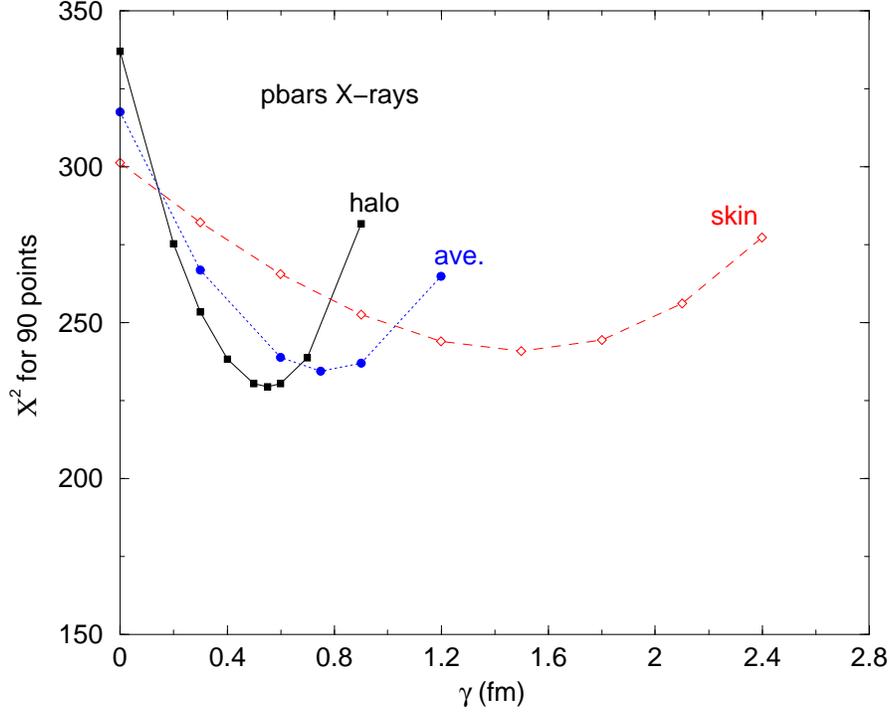}
\caption{Global best-fit $\chi ^2$ values for zero-range $\bar p$-nucleus
potentials as function of the $r_n-r_p$ parameter $\gamma$ of
Eq.(\ref{eq:RMF}) for three shapes of the neutron density $\rho _n$.}
\label{fig:pbarsZRg}
\end{figure}

\begin{figure}
\includegraphics[scale=0.7,angle=0]{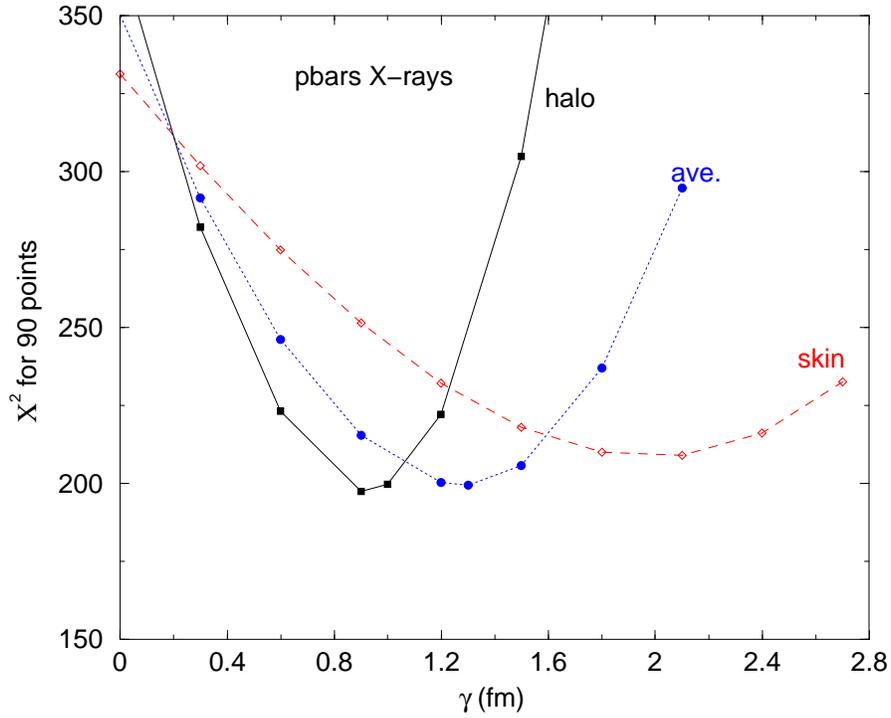}
\caption{Same as Fig.~\ref{fig:pbarsZRg} but for finite-range potentials
with a Gaussian parameter $\beta$=0.9~fm.}
\label{fig:pbarsFRg}
\end{figure}

Detailed analyses of the results of the PS209 collaboration have been
published in a series of papers, dedicated each to a particular 
subset of the data such as neighboring nuclei or isotopes of the same
element. In several cases it is necessary to take into consideration
the effects of possible E2 resonances, when energy of a nuclear E2
transition is very close to the energy of the atomic transition
being studied. In what follows we discuss only  {\it global} fits 
to the entire data set of 90 points as part of a study of 
medium-modification of the $\bar p N$ 
interaction \cite{FGa04,FGM05}. 

\begin{figure}
\includegraphics[scale=0.7,angle=0]{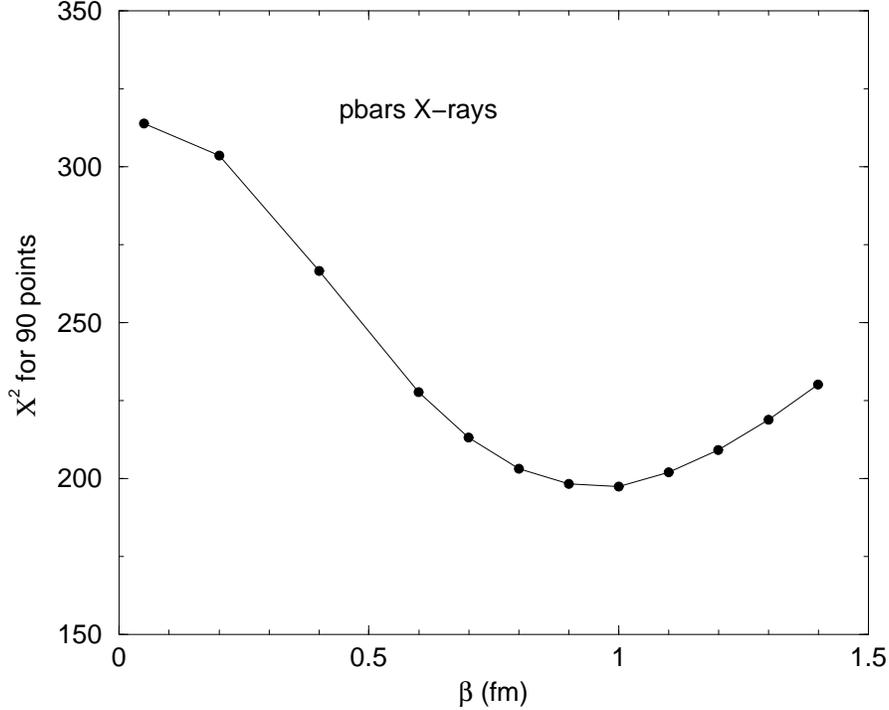}
\caption{Global best-fit $\chi ^2$ values as function of a 
Gaussian range $\beta$
for the halo shape for $\rho _n$ with $\gamma$=1.0~fm.}
\label{fig:pbarsFEag}
\end{figure}

Figure~\ref{fig:pbarsZRg} shows the $\chi ^2$ values for the best-fit
potential of the type Eq.~(\ref{eq:pbarspotl}) obtained with only two
adjustable parameters, the real and imaginary parts of $b_0$. The halo
shape for $\rho _n$ yields the lowest value of $\chi ^2$ but the 
minimum at $\gamma \approx $0.5 fm is unacceptable as representing 
the average dependence of $r_n-r_p$ on the neutron excess, as discussed 
in Sec.~\ref{sec:neutdens}. In Fig.~\ref{fig:pbarsFRg} are shown similar 
results for a finite-range version of the potential, obtained with Gaussian 
folding, as given by Eq.~(\ref{eq:fold}), using a range parameter of 
$\beta$=0.9~fm. The lowest $\chi ^2$ is significanty lower than
the corresponding value in Fig.~\ref{fig:pbarsZRg} and is obtained for
$\gamma \approx $0.9~fm, which is a most acceptable value, see 
Sec.~\ref{sec:neutdens} and Ref.~\cite{JTL04}. The FR parameter 
$\beta$=0.9~fm is chosen because over a range of values of $\gamma$
a minimum of $\chi ^2 $ is obtained for this value of $\beta$=0.9~fm, 
as seen in Fig.~\ref{fig:pbarsFEag}. This minimum means a $\chi ^2$ per
point of 2.2 which is quite good. The parameters of the potential
are Re$b_0$=1.1$\pm$0.1~fm, Im$b_0$=1.8$\pm$0.1~fm for $\gamma$=1.0~fm
and $\delta =-$0.035~fm, see Eq.~(\ref{eq:RMF}).
These parameters are not qualitatively distinct from the parameters 
obtained recently by Wycech et al.~\cite{WHJ07} using somewhat different values 
for $\beta$ and $\gamma$, and also including a $p$-wave absorptive term 
in the $\bar p$ optical potential.

\begin{figure}
\includegraphics[scale=0.7,angle=0]{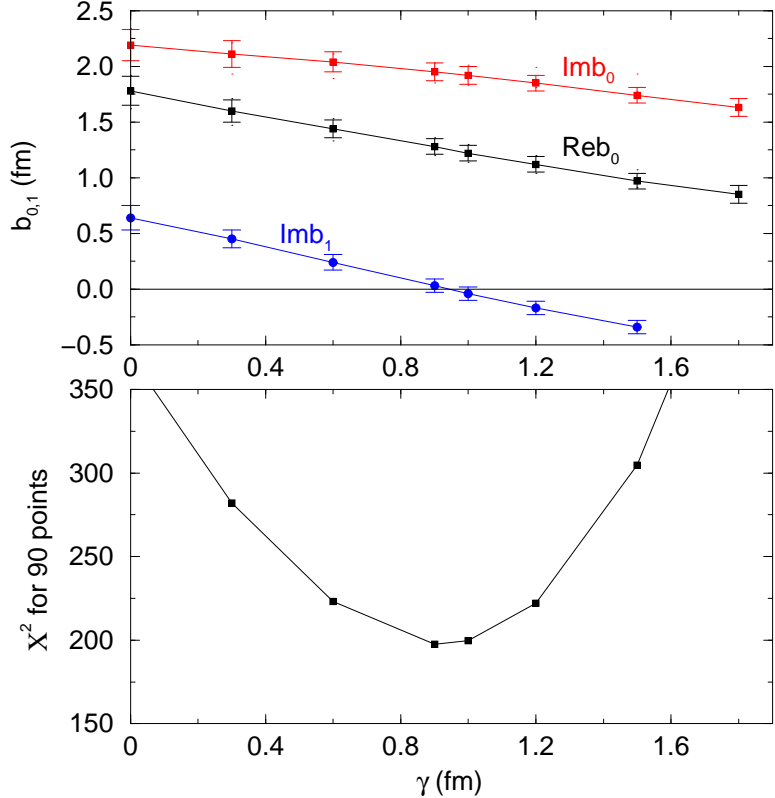}
\caption{Global best-fits  for FR $\bar p$-nucleus
potentials with $\beta$=0.9~fm 
as function of the $r_n-r_p$ parameter $\gamma$ of
Eq.~(\ref{eq:RMF}) for the halo shape of the neutron density $\rho _n$
 including an isovector term Im$b_1$.}
\label{fig:pbarscombo}
\end{figure}

Figure~\ref{fig:pbarscombo} shows results when the isovector parameter
$b_1$ is  varied in the fit in addition to the isoscalar parameter
$b_0$. It is found that Re$b_1$ is always consistent with zero (not shown)
whereas the other three parameters vary monotonically with the 
neutron radius parameter $\gamma$. 
It is seen that the minimum of $\chi ^2$ is obtained for the same
value of $\gamma \approx $1.0 fm as before and the quality of
fit is not improved. Moreover, at the best-fit point Im$b_1$ is
consistent with zero. Note that a non-zero value for this parameter
will be obtained if a very different value of $\gamma$ is used to 
represent neutron densities.

It is interesting to make a few comparisons between the values of
the differences between rms radii of neutron and of 
proton distributions  implied by the global best-fit value
of $\gamma$=1.0~fm, and differences obtained in detailed analyses
of a small group of $\bar p$ atoms. 
For example, for $^{120}$Sn it is found in Ref.~\cite{STC03} that 
$r_n-r_p=0.08^{+0.03}_{-0.04}$~fm whereas our global expression
yields 0.13$\pm$0.02~fm for this difference
if we assign from Fig.~\ref{fig:pbarsFRg} 
an estimated uncertainty of $\pm$0.1~fm to the slope parameter $\gamma$. 
Likewise for $^{124}$Sn the values are 
0.14$\pm$0.03~fm from Ref.~\cite{STC03} and 0.16$\pm$0.02~fm from the 
present global analysis. 
Taking $^{208}$Pb as another example, in Ref.~\cite{KTJ07} the
rms difference is 0.16$\pm$0.04~fm whereas the present global  
expression leads to 0.18$\pm$0.02~fm. Similar agreements are found
in other cases.

\subsection{Radial sensitivity of X-ray data}
\label{sec:pbarsFD}

\begin{figure}
\includegraphics[scale=0.7,angle=0]{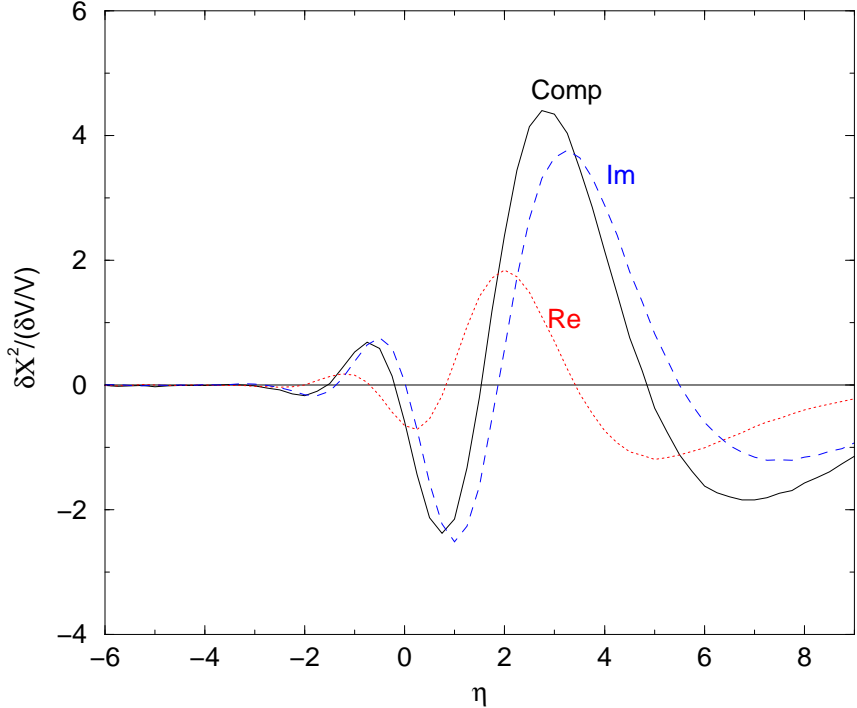}
\caption{Functional derivatives of the global best-fit $\chi ^2$ with respect
to relative changes in the full complex potential (solid curve), in 
the real part (dotted) and in 
the imaginary part (dashed) of the potential as function of the radial
position $\eta$, see Sec.~\ref{sec:piradsen}.}
\label{fig:pbarsFDX}
\end{figure}

Before proceeding to the radiochemical results which provide
information on the annihilation of atomic $\bar p$ at the extreme
periphery of the nucleus, it is instructive to examine the radial
sensitivity of X-ray data in order to 
get some idea on what are the nuclear regions that determine 
the potentials derived above. Following the 
preliminary results of the `notch test'
of Ref.~\cite{BFG97} where it was shown that 
$\bar p$ X-ray data are sensitive to the potential
at radii well outside of the nuclear surface,
we apply here the functional derivative method,
as discussed in Sec.~\ref{sec:piradsen}.
Figure~\ref{fig:pbarsFDX} shows the $\chi ^2$ FDs 
 for the best-fit potential with $\gamma$=1.0~fm,
$\delta =-$0.035~fm, a Gaussian range of 0.9~fm and with $b_0$=1.1+i1.8~fm.
The first conclusion from this figure is the dominance of the imaginary 
part of the potential as the FD with respect to it follows closely the 
FD with respect to the full complex potential. The other clear feature 
are the radial regions where the bulk of $|$FD$|$ is found,
indicating the regions to which the data are sensitive.
Strictly speaking, the FD refers to the optical potential and owing 
to the finite-range folding the relevant density regions are shifted 
to approximately 0.5~fm smaller radii, well outside of the half-density 
radius (at $\eta$=0), peaking between $\eta$=2 and $\eta$=6 where the 
densities are well below 10\% of the central nuclear density.

\subsection{Analysis of X-ray and radiochemical data}
\label{sec:xrc}

\begin{figure}
\includegraphics[scale=0.7,angle=0]{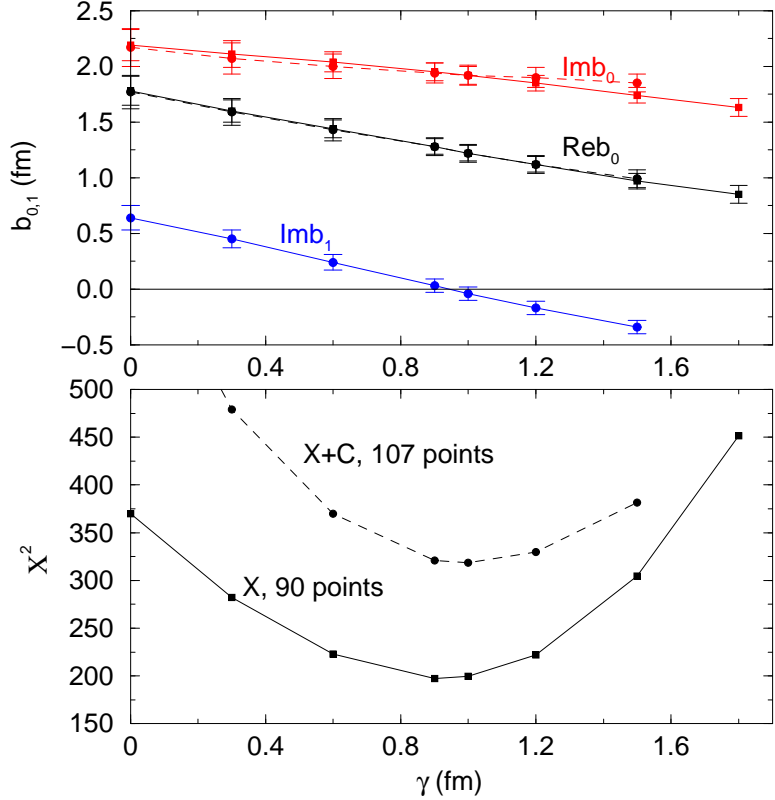}
\caption{Fits to the combined X-ray and
radiochemical data (X+C, dashed curves) compared to fits to the 
X-ray data (X, solid curves) only.}
\label{fig:pbarscomboXC}
\end{figure}

In the analysis of radiochemical data we adopt the approach of
Refs.~\cite{LJT98,SHK99,TJL01}, namely that the method is sensitive 
to the neutron to proton density ratio close to 2.5 fm outside of the
half-density radius of the charge density \cite{WSS96}.
In terms of the global parameter $\eta$ defined above, that corresponds
to $\eta \approx$ 5, which is within the region of sensitivity of 
the X-ray data but slightly
shifted towards larger radii, as seen from Fig.~\ref{fig:pbarsFDX}. 
It is therefore interesting to see
if analyses of the radiochemical data lead to results consistent
with what is obtained from the X-ray data. The experimental
ratios of absorption on neutrons to absorption on protons were therefore
compared to 
\begin{equation}
\label{eq:ratio}
\frac{{\rm Im}(b_0+b_1)I_n}{{\rm Im}(b_0-b_1)I_p}
\end{equation}
where $I_{n,p}$ are the volume integrals of the neutron and proton densities,
respectively, over an appropriate range. To check sensitivity to the 
chosen range of integration we have carried out the integration
either between 2.0  and 3.0 fm or between 2.5 and 3.5 fm
outside of the half-density radius of the charge density. 
For the finite-range potential used here the folded densities were used. 
Atomic wavefunctions were not included in the integrals because
their effect largely cancels out in the ratios. 
Moreover, we note that choosing the range
of integration
was guided by the conclusions of Ref.~\cite{WSS96} which were
based on properties of the atomic wavefunctions. With the 
potential parameter $b_1$ 
consistent with zero
the ratios Eq.~(\ref{eq:ratio}) become independent of the parameters
of the potential, but they are found to be 
sensitive to values of $r_n-r_p$
or to the parameter $\gamma$. Examining the $\chi ^2$ for the radiochemical
data as function of $\gamma$, it was found \cite{FGM05} that the minimum
occurred for $\gamma \approx$ 1.0 fm, as was the case for the X-ray data, 
if the integration range was 2.5 to 3.5 fm (but not 2 to 3~fm) 
outside of the half-density radius of the charge density. This result
confirms in a phenomenological way the theoretical conclusion of
Wycech et al.~\cite{WSS96} that most of the absorption takes place
close to 2.5 fm beyond the charge radius. Note that 
due to the exponential decrease of the densities
at such large radii the integrals are dominated by the densities close
to the lower limit of the range of integration.

Combining the results of the 
radiochemical technique with the X-ray data, fits were made
to the two kinds of data put together, a total of 107 points. 
From Fig.~\ref{fig:pbarscomboXC} showing results of fits to this
combined data set, in comparison with results from fits
to the X-ray only data, it is seen that
the overall picture is the same in both cases, 
with larger values of $\chi ^2$ per point for the combined data.  
In particular, with the minimum of $\chi ^2$ for the same value of $\gamma$, 
the same conclusions are reached regarding neutron densities.

Before closing this section we look into the broader perspective of neutron 
densities in nuclei obtained from antiprotonic atoms. The following 
conclusions may be made from the global analyses presented above: 

\begin{itemize} 

\item The potential parameters depend mostly on $r_n-r_p$ and not on the 
shape of the neutron densities, although the $\chi^2$ values do depend on 
the shape of $\rho _n$. 

\item The favored shape of $\rho _n$ is of the `halo' type,

\item The rms radii of $\rho _n$  are given on the average by 
Eq.~(\ref{eq:RMF}) with the parameter $\gamma \approx$~1.0~fm. 

\end{itemize} 

A possible difficulty regarding nuclear densities is that the sensitivity 
of $\bar p$ atom data is to extremely small densities, of the order 
of 5\% of the central nuclear density, where the proton densities too
are not determined well by the traditional methods of electron
scattering and muonic X-rays. In particular, the 2pF parameterization
need not be appropriate to describe the outer reaches of the 
proton densities $\rho _p$. On the other hand, the 
present analyses of $\bar p$ atomic data lead only to conclusions on 
{\it differences} between neutrons and protons in $N \neq Z$ nuclei, 
both on the differences
of rms radii and on differences in shapes. The preferred `halo' shape
in this context means that the diffuseness parameter $a_n$ is larger
than  the corresponding parameters for protons, 
which is quite reasonable considering the binding energies
of least bound nucleons and effects of the Coulomb potential.
The disagreement with pionic atoms regarding the shapes of $\rho _n$
is presumably due to the extreme simplification introduced in assuming
2pF parameterizations for the densities. Recall that pionic atom
data are sensitive to densities up to the full nuclear density. 
It is, therefore, 
no wonder that there are some differences in conclusions obtained from
experiments that are sensitive to
different density regions of the nucleus. The fact that the rms radii obtained
with the two methods are in full agreement with each other is not 
a coincidence. It was shown in Sec.~\ref{sec:neutdens} that  
potential parameters for pions depend mostly on $r_n-r_p$ and not
on the shape of the neutron densities. The same results are found for 
${\bar p}$ atoms, as emphasized above.

\subsection{Deeply bound antiprotonic atom states}
\label{sec:pbarsdeep}

\begin{figure}
\includegraphics[scale=0.7,angle=0]{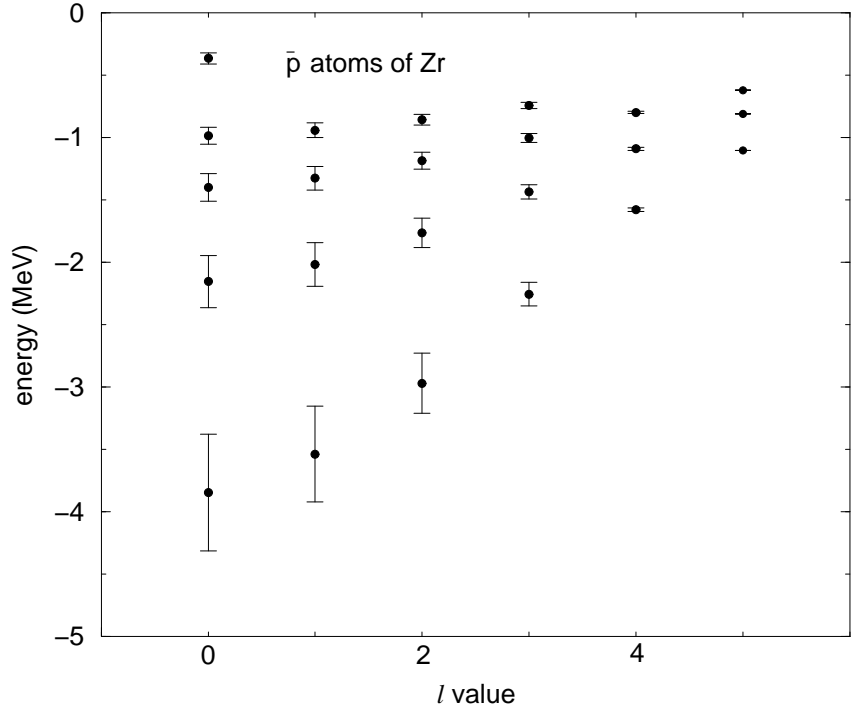}
\caption{Calculated energies of ${\bar p}$ atomic states in $^{90}$Zr.
The lowest energy for each $l$ value corresponds to $n=l+1$.
The bars represent the widths of the states.}
\label{fig:pbarsZrdeep}
\end{figure}

Much the same as with kaonic atoms, see Sec.~\ref{sec:Kdeep}, the optical
potential for antiprotonic atoms is dominated by its imaginary part
which is of the order of 100~MeV deep when extrapolated into the full
nuclear density. Such an absorption inevitably produces strong
suppression of the {\it atomic} wavefunction inside the nucleus, which
for a normalized atomic wavefunction located mostly outside of the nucleus
is equivalent to repulsion. The role of the phenomenological attractive real
part of the potential is more difficult to visualize. For sufficiently
attractive potentials there is the possibility of accommodating strongly
bound nuclear states, albeit very broad, as  their wavefunctions
are confined to the nuclear volume. Due to the orthogonality
requirement of nuclear and atomic wavefunctions having the same $l$-values,
the latter might be shifted considerably by the real potential, 
as demonstrated for kaonic
atoms in Fig.~\ref{fig:sat}, where large repulsion 
of the atomic wavefunction is observed as a result
of an attractive strong interaction. 
It is, therefore, not surprising, that
 the phenomenon of saturation of widths of atomic states
is observed \cite{FGa99a,FGa99b} 
in numerical calculations of antiprotonic atom spectra.

Figure~\ref{fig:pbarsZrdeep} shows a calculated energy spectrum for ${\bar p}$
atoms of Zr as a typical medium-weight nucleus \cite{FGa99b}.
The strong-interaction potential is taken from global fits to ${\bar p}$ 
atomic data. The saturation of the widths is easily seen, with the 
widths increasing very little when $l$ goes down towards $l$=0.

\subsection{Antiproton-nucleus interaction across threshold}
\label{sec:pbarsscatt}

With rather well-established phenomenology of the interaction of
antiprotons with nuclei in the subthreshold atomic regime, it is
of interest to see if the same picture prevails also above threshold. 
Indeed early analyses of elastic scattering of
47 MeV antiprotons on carbon showed \cite{BFL84} that very good fits to
the scattering data and to the then available ${\bar p}$ atom data
could be obtained with a common optical potential, dominated
by its imaginary part and  based on Gaussian folding with a range 
parameter of $\beta$=1.2~fm, quite similar to the present results.

Antiprotons offer a unique tool, compared to other exotic atoms,
for studying the interaction with nuclei very close to, 
but above threshold, in the form of
 ${\bar p}$ annihilation. At very low energies,
below the ${\bar p}p \rightarrow {\bar n}n$ charge-exchange threshold,
 the total ${\bar p}$ reaction cross section consists only
of ${\bar p}$ annihilation. Therefore measurements of annihilation
cross sections at such low energies may be compared with total reaction 
cross sections calculated with the optical potentials obtained from fits 
to antiprotonic atom data. A measurement of antiproton annihilation cross 
section at 57 MeV/c (1.7 MeV kinetic energy) on Ne was reported by 
Bianconi et al.~\cite{BBB00a} and was shown~\cite{GFB00} 
to agree with predictions made with potentials obtained
from global fits to ${\bar p}$ atom data. Comparisons for similar 
measurements of ${\bar p}$ annihilation on 
$^4$He showed that predictions of annihilation cross
sections made with ${\bar p}$ potentials obtained from fits to
atomic data for medium-weight and heavy nuclei, do not agree with
experiment. In contrast, when parameters of the potential were obtained 
from fits only to ${\bar p}$ atoms of
 $^{3,4}$He, then the predicted annihilation cross section
on $^4$He agreed with the measured one. 
From this example it may be concluded that
the ${\bar p}$-nucleus potentials cross smoothly the threshold from 
atomic states to the scattering regime. 
On the other hand the global
${\bar p}$ potentials which reproduce very well ${\bar p}$ atomic
data for targets heavier than $A \approx $10, fail to describe similar data
for the He isotopes.

A special case in this context is the annihilation of 
${\bar p}$ on the proton
very close to threshold. The ${\bar p}p$ total
annihilation cross section was measured at four momenta between
38 and 70~MeV/c \cite{ZBB99} 
and with the availability of strong interaction
shift, width and yield for the 1$s$ and 2$p$ levels in antiprotonic
hydrogen it is possible to study the 
${\bar p}p$ interaction across threshold \cite{BFG01}.
It is found that  a Gaussian potential with a range parameter
between 1 and 2~fm produces very good fits separately to the
annihilation cross sections and to the atomic ${\bar p}$H data,
but if both types of data are to be fitted 
together then the range parameter
turns out  to be $\beta$=1.5$\pm$0.15~fm with Re$b_0=-$0.15$\pm$0.15~fm,
Im$b_0$=1.80$\pm$0.06~fm. It is, therefore, possible to cross smoothly
the borderline of $E$=0 also for the ${\bar p}$H system. However,
the interaction parameters are different from those valid for $^{3,4}$He
and from those valid for target nuclei heavier than $A \approx $10. 
This demonstrates the limitations of using optical potentials down to the 
very light nuclear targets, where the energy dependence of those 
$\bar N N$ partial-wave amplitudes which may have quasibound 
states or resonances near threshold needs to be considered explicitly 
\cite{LWy05,WLo05}. 

Finally, it is interesting to note that the saturation of widths
predicted for  antiprotonic atom states is also predicted and 
{\it observed} above threshold in the form of saturation of
reaction cross sections \cite{BFG01}.
There is an interesting analogy between widths of bound states 
and total
reaction cross sections where for the Schr\"odinger equation the latter
is given by

\begin{equation} \label{eq:sigmaR}
\sigma_R = -\frac{2}{\hbar v} \int
{\rm Im} V_{{\rm opt}}(r) | \psi({\bf r}) |^2 d {\bf r}\quad,
\end{equation}
where $\psi({\bf r})$ is the $\bar p$-nucleus elastic scattering
wavefunction and $v$ is the c.m. velocity. Recall that the width of
a bound state, as discussed in Secs.~\ref{sec:deep} and \ref{sec:Kdeep} 
is given by
\begin{equation} \label{eq:gammapbars}
\Gamma= -2\frac{\int {\rm Im} V_{{\rm opt}}(r)
| \psi({\bf r}) | ^2  d {\bf r}}
{\int | \psi({\bf r}) | ^2  d {\bf r}}\quad,
\end{equation}
where $\psi({\bf r})$ is the $\bar p$ full atomic wavefunction.
The modification of this expression for the KG equation is mentioned in 
Sec.~\ref{sec:Kdeep}. It is therefore to be expected that large local 
variations of the wavefunction, in both cases, is a common mechanism 
behind departures from linear dependence on the imaginary potential. 

At very low antiproton energies 
where Coulomb focusing is effective, the annihilation
cross sections on nuclei are expected to scale as $ZA^{1/3}$ in
the perturbative regime \cite{BFG01}, but the experimental
annihilation cross sections on Ne and $^4$He \cite{BBB00a} 
differ strongly from this scaling law. 
This difference is a manifestation of saturation \cite{GFB00,BFG01}, 
confirming the general property of saturation of widths as discussed above.



\section{The repulsive ${\bf \Sigma}$ nuclear potential}
\label{sec:sigma}

\subsection{Preview} 
\label{sec:Sigprev} 
 
One Boson Exchange (OBE) models fitted to the scarce low-energy $YN$ 
scattering data produce within a $G$-matrix approach, with one exception 
(Nijmegen Model F), as much attraction 
for the $\Sigma$ nuclear potential as they 
do for the $\Lambda$ nuclear potential, see Ref.~\cite{DGa84} for a review 
of `old' models and Ref.~\cite{RYa06} for the latest state of the art for 
Nijmegen models. Indeed, the best-fit $t_{\rm eff}\rho$ potential for 
$\Sigma^-$ atoms was found by Batty et al.~\cite{Bat79,BGT83} to be attractive 
and absorptive, with central depths for the real and imaginary parts of 
25-30~MeV and 10-15~MeV, respectively. It took almost a full decade, 
searching for $\Sigma$ hypernuclear bound states at CERN, KEK and BNL, 
before it was realized that except for a special case for $^4_\Sigma$He, 
the observed continuum $\Sigma$ hypernuclear spectra indicate a very shallow, 
or even repulsive $\Sigma$ nuclear potential, 
as reviewed by Dover et al.~\cite{DMG89}. 
These indications have received firm support with the measurement of several 
$(K^-,\pi^{\pm})$ spectra at BNL \cite{BCF99} followed by calculations for 
$^9$Be~\cite{Dab99}. Recently, with measurements of the $\Sigma^-$ spectrum 
in the $(\pi^-,K^+)$ reaction taken at KEK across the periodic table 
\cite{NSA02,SNA04}, it has become established that the $\Sigma$ nuclear 
interaction is strongly repulsive. In parallel, analyses of $\Sigma^-$-atom 
in the early 1990s, allowing for density dependence or 
departure from the $t \rho$ prescription, motivated mostly by the precise 
data for W and Pb \cite{PEG93}, led to the conclusion 
that the {\it nuclear} interaction of $\Sigma$s is dominated by 
repulsion~\cite{BFG94a,BFG94b,MFG95}, as reviewed in Ref.~\cite{BFG97}. 
This might have interesting repercussions for the balance of 
strangeness in the inner crust of neutron stars~\cite{BGa97}, primarily by 
delaying the appearance of $\Sigma^-$ hyperons to higher densities, if at all. 
The inability of the Nijmegen OBE models, augmented by $G$-matrix 
calculations~\cite{RYa06}, to produce $\Sigma$ nuclear repulsion 
is a serious drawback for these models at present. This problem apparently 
persists also in the Juelich model approach~\cite{HMe05}. The only theoretical 
works that provide exception are SU(6) quark-model RGM calculations by the 
Kyoto-Nijata group~\cite{KFF00}, in which a strong Pauli repulsion appears in 
the $I=3/2,~{^3S_1}-{^3D_1}~\Sigma N$ channel, and Kaiser's SU(3) chiral 
perturbation calculation~\cite{Kai05} which yields repulsion of order 60 MeV. 

Below we briefly review and update the $\Sigma^-$ atom fits and the recent 
$(\pi^-,K^+)$ KEK results and their analysis. 

\subsection{Density dependent $\Sigma$ nuclear potentials from fits to 
$\Sigma^-$ atoms} 
\label{sec:SigDD} 

\begin{figure}
\includegraphics[scale=0.7]{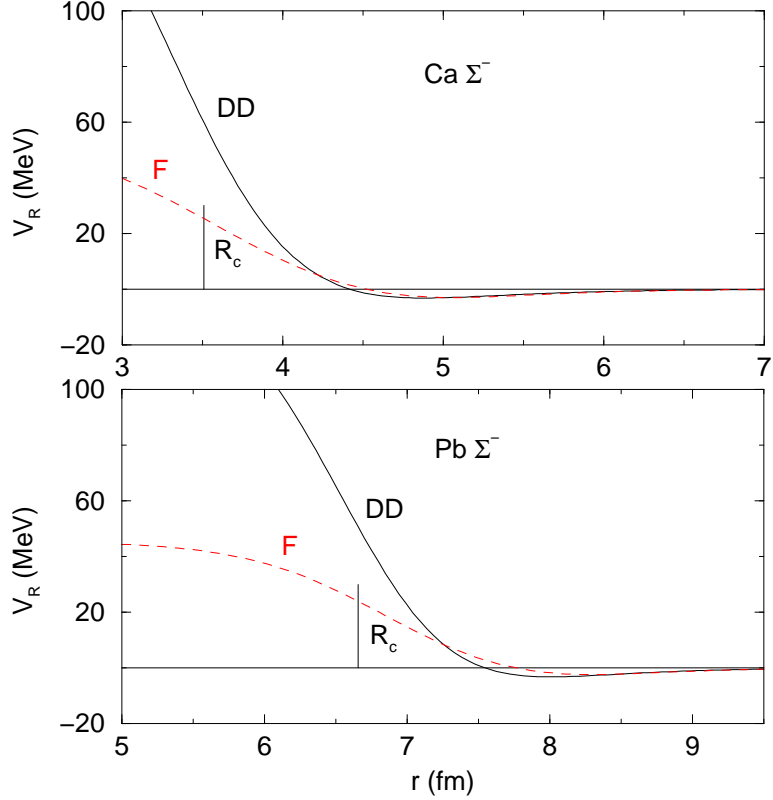} 
\caption{Re$V_{\rm opt}$  for DD (solid) and for the geometrical
model F (dashed)
$\Sigma^-$ nuclear potentials fitted to $\Sigma^-$ atomic 
data. 
Vertical bars indicate the half-density radius of 
the nuclear charge distribution.}
\label{fig:ddsig} 
\end{figure}

Batty et al.~\cite{BFG94a,BFG94b} analyzed the full data set
of $\Sigma^-$ atoms, consisting of strong-interaction level
shifts, widths and yields, in order to constrain the
density dependence of $V_\Sigma(r)$. By introducing a
phenomenological density dependent (DD) potential of the isoscalar 
form

\begin{equation}
\label{eq:DD}
V_\Sigma(r) \sim \left[ b_0 + B_0
\left({\rho(r)/\rho(0)}\right)^\alpha \right]
\rho(r) \quad , \qquad \alpha > 0 \quad ,
\end{equation}
and fitting the parameters $b_0, B_0$ and $\alpha$ to the
data, greatly improved fits to the data are obtained. 
Isovector components are readily included in Eq.~(\ref{eq:DD}) but
are found to have a marginal effect. Note, however, that the
absorption was assumed to take place only on protons. 
The complex parameter $b_0$ may be identified with the
spin-averaged $\Sigma^-N$ scattering length. For the
best-fit isoscalar potentials, Re$V_\Sigma$ is attractive
at low densities outside the nucleus, changing into
repulsion in the nuclear surface region.
The precise magnitude and shape of the repulsive
component within the nucleus is not determined by the
atomic data. The resulting potentials are shown in Fig.~\ref{fig:ddsig} 
(DD, solid lines),
where it is worth noting that the transition from attraction to 
repulsion occurs well outside of the nuclear radius, hence the 
occurrence of this transition should be largely model independent. 
To check this last point we have repeated the fits to the atomic data
with the `geometrical model' F of Sec.~\ref{sec:kbars}, using separate
$t \rho $ expressions in an internal and an external region, see
Eq.~(\ref{eq:DDF}). The neutron densities used in the fits were of the 
skin type, with the $r_n-r_p$ parameter Eq.~(\ref{eq:RMF}) $\gamma$=1.0~fm. 
The fits deteriorate significantly if the halo type is used for the neutron
density. The fit to the data is equally good  with this model as with 
the DD model, ($\chi ^2$ per degree of freedom of 0.9 here 
compared to 1.0 for the DD model)
and the potentials are shown as the dashed lines in Fig.~\ref{fig:ddsig}.
The half-density radius of the charge distribution is indicated in
the figure. It is clear that both models show weak attraction at
large radii, turning into repulsion approximately one fm outside of 
that radius.

\begin{figure}
\includegraphics[scale=0.7]{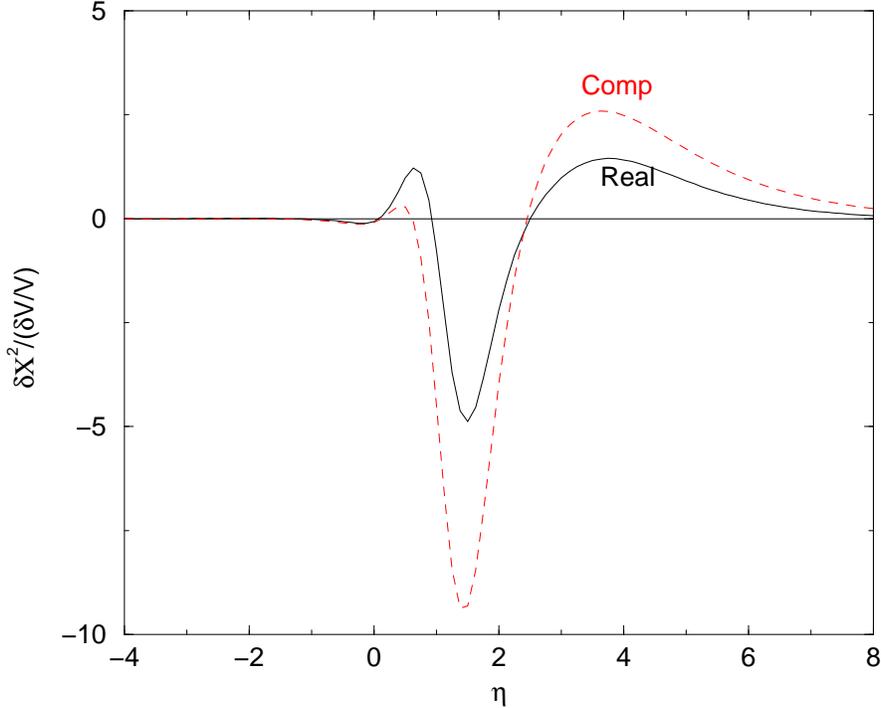}
\caption{Functional derivatives of $\chi ^2$ with respect to the real
(solid) and with respect to the full complex (dashed) optical potentials
for the best fit F potential.}
\label{fig:SigFD}
\end{figure}

Further insight into the geometry of the $\Sigma$-nucleus interaction
is gained by inspecting the functional derivatives (FD) of $\chi ^2$
with respect to the optical potentials, see Sec. \ref{sec:piradsen}. 
Figure~\ref{fig:SigFD} shows the FDs based on the best fit of 
the geometrical model F as
discussed above. From the differences between the FD with respect to the
full complex potential and the FD with respect to the real potential
it is concluded that both real and imaginary parts play similar roles
in the $\Sigma$-nucleus interaction.
The bulk of $|$FD$|$ is in the range of 0.5~$\leq~\eta~\leq$~6, 
covering the radial region where the weak attraction turns into repulsion. 
Obviously no information is obtained from $\Sigma^-$ atoms on the interaction
inside the nucleus.
It is also interesting to note quite generally that such 
 potentials do not produce bound states, and this
conclusion is in agreement with the experimental results from BNL 
\cite{BCF99} for the absence of $\Sigma$ hypernuclear peaks beyond He. 

Some semi-theoretical support for this finding of inner repulsion is given 
by RMF calculations by Mare{\v s} et al.~\cite{MFG95} who generated the 
$\Sigma$-nucleus interaction potential in terms of scalar ($\sigma$) and 
vector ($\omega,\rho$) meson mean field contributions, fitting its coupling 
constants to the relatively accurate $\Sigma^-$ atom shift and width data 
in Si and in Pb. The obtained potential fits very well the whole body 
of data on $\Sigma^-$ atoms. This potential, which is generally attractive 
far outside the nucleus, becomes repulsive at the nuclear surface and 
remains so inward in most of the acceptable fits, of order 10-20 MeV. 
The Pb data~\cite{PEG93} are particularly important in pinning 
down the isovector component of the potential which 
in this model is sizable and which, 
for $\Sigma^-$, acts against nuclear binding in core nuclei with $N-Z > 0$, 
countering the attractive Coulomb interaction. 
 On the other hand, 
for very light nuclear cores and perhaps only for $A = 4$ hypernuclei, 
this isovector component (Lane term) generates binding of 
$\Sigma^+$ configurations. In summary, the more modern fits to $\Sigma^-$ 
atom data~\cite{BFG94a,BFG94b,MFG95} and the present
fits with the geometrical model support the presence of a substantial 
repulsive component in the $\Sigma$-nucleus potential which 
excludes normal $\Sigma$-nuclear binding, except perhaps in very 
special cases such as $^4_\Sigma$He \cite{Hay89,HSA90,Nag98,Har98}. 

\subsection{Evidence from $\bf{(\pi^-,K^+)}$ spectra}
\label{sec:pi-K+spec} 

\begin{figure} 
\includegraphics[height=14cm,width=9cm]{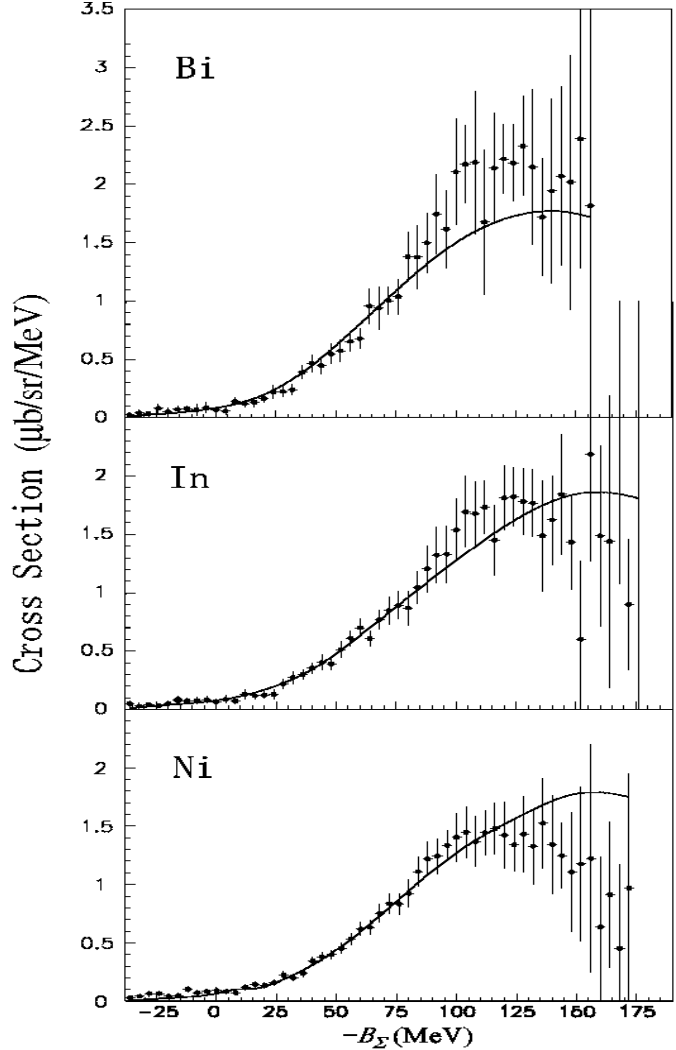} 
\caption{Inclusive $(\pi^-,K^+)$ spectra on Ni, In and Bi, fitted by 
a $\Sigma$-nucleus WS potential with depths $V_0 = 90$ MeV, $W_0=-40$ MeV 
{\protect\cite{SNA04}}.}
\label{fig:ninbispec} 
\end{figure}

\begin{figure}
\includegraphics[scale=0.8,angle=0]{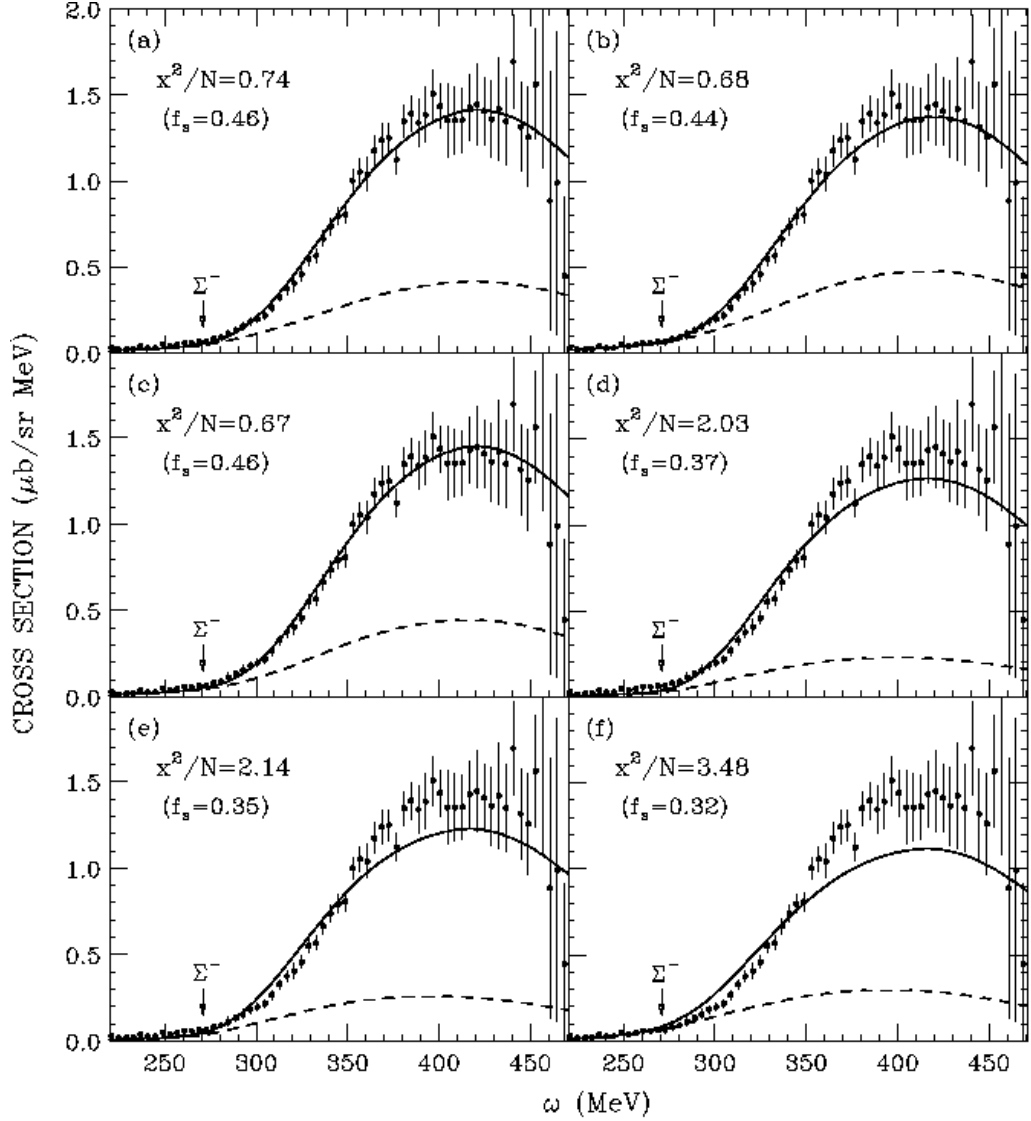}
\caption{Comparison between DWIA calculations~{\protect\cite{HHi05}}
and the measured $^{28}{\rm Si}(\pi^-,K^+)$ spectrum~{\protect\cite{SNA04}} 
using six $\Sigma$-nucleus potentials, (a)-(c) with inner repulsion,
(d)-(f) fully attractive. The solid and dashed curves denote the
inclusive and $\Lambda$ conversion cross sections, respectively.
Each calculated spectrum was normalized by a fraction $f_s$. The arrows
mark the ${\Sigma^-} - {^{27}{\rm Al}_{\rm g.s.}}$ threshold at
$\omega = 270.75$~MeV.}
\label{fig:haradaqf} 
\end{figure} 

A more straightforward information on the nature of the $\Sigma$-nuclear 
interaction has been provided by recent measurements of inclusive 
$(\pi^-,K^+)$ spectra on medium to heavy nuclear targets at KEK~\cite{NSA02,
SNA04}. The inclusive $(\pi^-,K^+)$ spectra on Ni, In and Bi are shown in 
Fig.~\ref{fig:ninbispec} together with a fit using Woods-Saxon potentials 
with depths $V_0 = 90$ MeV for the (repulsive) real part and $W_0=-40$ MeV 
for the imaginary part. These and other spectra measured on lighter targets 
suggest that a strongly {\it repulsive} $\Sigma$-nucleus potential is required 
to reproduce the shape of the inclusive spectrum, while the sensitivity to 
the imaginary (absorptive) component is secondary. The favored strength of 
the repulsive potential in this analysis is about 100 MeV, of the same order 
of magnitude reached by the DD $\Sigma^-$ atomic fit potential shown in 
Fig.~\ref{fig:ddsig} as it `enters' the nucleus inward. The general level of 
agreement in the fit shown in Fig.~\ref{fig:ninbispec} is satisfactory, but 
there seems to be a systematic effect calling for more repulsion, 
the heavier is the target. 
We conclude that a strong evidence has been finally established for the 
repulsive nature of the $\Sigma$-nucleus potential. 

More sophisticated theoretical analyses of these KEK $(\pi^-,K^+)$ spectra 
\cite{KFW04,KFW06,HHi05,HHi06} have also concluded that the $\Sigma$-nuclear 
potential is repulsive within the nuclear volume, although they yield 
a weaker repulsion in the range of 10-40 MeV. An example of a recent 
analysis of the Si spectrum is shown in Fig.~\ref{fig:haradaqf} from 
Ref.~\cite{HHi05} where six different $\Sigma$-nucleus potentials are tested 
for their ability within the Distorted Wave Impulse Approximation (DWIA) 
to reproduce the measured $^{28}{\rm Si}(\pi^-,K^+)$ 
spectrum~\cite{SNA04}. This particular DWIA version was tested 
on the well understood $^{28}{\rm Si}(\pi^+,K^+)$ quasi-free $\Lambda$ 
hypernuclear spectrum also taken at KEK with incoming pions of the same 
momentum $p_{\rm lab} = 1.2$~GeV/c. Potential (a) is the DD, type A' potential 
of Ref.~\cite{BFG94b}, (b) is one of the RMF potentials of Ref.~\cite{MFG95}, 
that with $\alpha_{\omega} = 1$, and (c) is a local-density approximation 
version of a $G$ matrix constructed from the Nijmegen model F. These three 
potentials are repulsive within the nucleus but differ considerably there 
from each other. Potentials (d)-(f) are all attractive within the nucleus, 
with (f) being of a $t_{\rm eff}\rho$ form. All of the six potentials are 
attractive outside the nucleus, as required by fits to the `attractive' 
$\Sigma^-$ atomic level shifts. The figure shows clearly, and judging by the 
associated $\chi^2/{\rm N}$ values, that fully attractive potentials are 
ruled out by the data and that only the {\it `repulsive'} $\Sigma$-nucleus 
potentials reproduce the spectrum very well, but without giving preference to 
any of these potentials (a)-(c) over the other ones in this group. 
It was shown by Harada and Hirabayashi~\cite{HHi06}, furthermore, 
that the $(\pi^-,K^+)$ data on targets with neutron excess, such as 
$^{209}$Bi, also lack the sensitivity to confirm the presence of a sizable 
(repulsive for $\Sigma^-$) isovector component of the $\Sigma$ nucleus 
interaction as found in the $\Sigma^-$-atom fits~\cite{BFG94a,BFG94b,MFG95}.   


\section{$\Xi$ hyperons}
\label{sec:xihyp} 

Experiments on exotic atoms of $\Xi^-$ hyperons have not been reported
so far but the possibilities of conducting such experiments have been
discussed by Batty et al.~\cite{BFG99} in some detail. Here we summarize 
only the main features of that study, within the broader context of the 
present Review.

\subsection{Preview} 
\label{sec:xiprev} 

Dedicated experiments with stopped $\Xi ^-$ hyperons had been proposed in 
Refs.~\cite{ZDG91,KKA95,YIk97} in order to produce some of the lightest 
$\Lambda \Lambda$ hypernuclei, $_{\Lambda \Lambda}^{~~6}$He and
$_{\Lambda \Lambda}^{~~4}$H (if the latter is particle stable), 
and $_{\Lambda \Lambda}^{~12}$B, respectively, by looking for a peak 
in the outgoing neutron spectrum in the two-body reaction 
\begin{equation} 
\label{eq:lamlam} 
\Xi ^- ~+~ ^AZ ~ \longrightarrow ~ _{\Lambda \Lambda}^A(Z-1) ~+~ n \,.
\end{equation} 
These proposals motivated the AGS experiment E885~\cite{Kha00a} on $^{12}$C,
using a diamond target to stop the $\Xi ^-$ hyperons resulting from the
quasi-free peak of the $p(K^-, K^+)\Xi ^-$ initial reaction. An upper bound 
of a few percent was established for the production of the 
$_{\Lambda \Lambda}^{~12}$B hypernucleus. The experimental evidence for 
$_{\Lambda \Lambda}^{~~6}$He and $_{\Lambda \Lambda}^{~~4}$H had to await 
different techniques~\cite{Tak01,Ahn01}, although the evidence for the latter 
species remains controversial. The stopped $\Xi ^-$ reaction in deuterium, 
$(\Xi ^- d)_{\rm atom} \rightarrow Hn$, was used in the AGS experiment E813 
to search for the doubly strange $H$ dibaryon, yielding a negative result 
\cite{Mer01}. A similar search by the KEK E224 collaboration, stopping 
$\Xi ^-$ on a scintillating fiber active carbon target, also yielded 
a negative result \cite{Ahn96}. On the positive side, following the 
discovery of a double-$\Lambda$ hypernucleus \cite{Aok91} in light emulsion 
nuclei by the KEK stopped $\Xi^-$ experiment E176, and its interpretation 
due to $_{\Lambda \Lambda}^{~13}$B~\cite{DMG91}, this experiment gave 
evidence for several events, each showing a decay into a pair of known 
single $\Lambda$ hypernuclei~\cite{ABC93,ABC95}. 
One could then attempt to use these events in order to deduce properties 
of the initial $\Xi^-$ atomic states. However, the typical error of 100 keV 
incurred in emulsion work is three orders of magnitude larger than the 
anticipated sensitivity of strong-interaction shifts and 
widths of $\Xi^-$ atomic levels to the $\Xi$-nucleus strong interaction. 
This simple argument provides a major justification for pursuing a program 
of measuring $\Xi^-$ X rays, in parallel to more conventional 
strong-interaction reactions involving $\Xi$ hyperons, as discussed in 
Sec.~\ref{sec:xiat}. 

Very little is established 
experimentally or phenomenologically on the interaction of $\Xi$ hyperons 
with nuclei. Dover and Gal \cite{DGa83}, analyzing old emulsion data which 
had been interpreted as due to $\Xi^-$ hypernuclei, obtained an attractive
$\Xi$-nucleus interaction with a nuclear potential well depth of
$V_{0}^{(\Xi)}$= 21-24~MeV. This range of values agreed well with their 
theoretical prediction~\cite{DGa84} for $\Xi$ in nuclear 
matter, using model D of the Nijmegen group~\cite{NRD77} to describe 
baryon-baryon interactions in an SU(3) picture, in contrast with the 
$\Xi$-nucleus repulsion obtained using model F~\cite{NRD79}. 
Similar predictions were subsequently made with more detailed $G$ matrix 
evaluations by Yamamoto et al.~\cite{YMH94,Yam95a,Yam95b} who argued for 
a considerable $A$ dependence of $V_{0}^{(\Xi)}$, such that the well depth 
for light and medium weight nuclei is significantly lower than for heavy 
nuclei where it approaches the value calculated for nuclear matter. 
It should be noted, however, that the predictions of the Nijmegen hard-core 
models D and F are extremely sensitive to the values assumed for the 
hard-core radius. The confidence in the predictive power of model D in 
strangeness $-$2 hypernuclear physics was to a large extent due to its 
success to yield the substantial attractive $\Lambda \Lambda$ interaction 
which was deemed necessary to reproduce the three known $\Lambda \Lambda$ 
binding energies in the 1990s. This picture has changed during the last 
decade for several reasons, as follows. 

\begin{itemize} 

\item 
Inclusive $(K^-,K^+)$ spectra taken at the KEK-PS and at the BNL-AGS 
accelerators on $^{12}$C, Refs.~\cite{Fuk98,Kha00b} respectively, 
when fitted near the $\Xi^-$-hypernuclear threshold yield more moderate 
values for the attractive $\Xi$ well depth, $V_{0}^{(\Xi)} \sim 15$~MeV. 

\item 
The uniquely identified $_{\Lambda \Lambda}^{~~6}$He hypernucleus 
\cite{Tak01} implies a considerably weaker $\Lambda \Lambda$ interaction than 
produced by reasonable versions of Model D. In particular, the Nijmegen 
soft-core potentials NSC97~\cite{SRi99} provide a more realistic framework 
for reproducing the weaker strength of the $\Lambda \Lambda$ interaction, 
as discussed in Refs.~\cite{FGa02a,FGa02b}. 

\item 
New versions of Nijmegen extended soft-core potentials 
ESC04~\cite{RYa06,RYa07} predict a weak $\Xi$-nucleus interaction with 
a delicate pattern of spin and isospin dependence. Similar conclusions 
are also reached in spin-flavor SU(6) quark models by Fujiwara et 
al.~\cite{FSN07,FKS07}.  

\end{itemize} 

Looking ahead at the prospects of further research in this strangeness 
$-2$ sector, it is safe to argue that if the interaction of $\Xi$ hyperons 
with nuclei is sufficiently attractive to cause binding, as has been 
repeatedly argued since the original work of Dover and Gal~\cite{DGa83}, 
then a rich source of spectroscopic information would become available 
and the properties of the in-medium $\Xi N$ interaction could be extracted. 
Bound states of $\Xi$ hypernuclei would also be useful as a gateway to form 
double $\Lambda$ hypernuclei \cite{MDG94,DGM94,IFM94,YMF94}. Finally, 
a minimum strength for $V_{0}^{(\Xi)}$ of about 15 MeV is required to 
realize the exciting possibility of `strange hadronic matter'~\cite{SBG00}, 
where protons, neutrons, $\Lambda$s and $\Xi$s are held together to form 
a system which is stable against strong-interaction decay. 
The study of $\Xi$-nuclear interactions, as part of studying strangeness 
$-2$ hadronic and nuclear physics, is high on the agenda of two forthcoming 
major high-intensity hadron facilities. 

\begin{itemize} 

\item 
At J-PARC, Japan, the main accelerator ring is a 50-GeV proton synchronton 
and the proton beam, with 30 GeV energy and $9\mu{\rm A}$ current initially, 
will produce various high-intensity beams of secondary particles. Strangeness 
$-2$ physics will be explored with a $K^-$ beam at $p_{\rm lab}=1.8$~GeV/c. 
An approved day-1 experiment is {\it Spectroscopic study of the 
$\Xi$-hypernucleus $_{\Xi}^{12}$Be via the $^{12}{\rm C}(K^-,K^+)$ reaction} 
(T.~Nagae, Spokesperson~\cite{Nag07}). The overall energy resolution 
in the $\Xi^-$ bound-state region is expected to be better than 3~MeV at FWHM, 
using an improved version of the existing SKS spectrometer at KEK. 
Another J-PARC approved experiment, although not prioritized as `day-1', 
is {\it Measurement of X rays from $\Xi^-$ atoms} 
(K.~Tanida, Spokesperson~\cite{Nag07}), the physics considerations and 
the experimental concerns of which are discussed in Sec.~\ref{sec:xiat} 
below, following the work of Batty et al.~\cite{BFG99}. 

\item 
A major component of the upgraded GSI facility in Darmstadt, Germany, 
will be the High Energy Storage Ring (HESR) for high-intensity, phase-space 
cooled antiprotons between 1.5 and 15 GeV/c. A general purpose detector 
PANDA (Proton ANtiproton at DArmstadt) will be set up at the HESR. 
PANDA is scheduled to provide access to high-resolution spectroscopy of 
$S=-2$ hypernuclei and hyper-atoms by producing abundantly $\Xi^-$ hyperons 
via the reactions~\cite{Poc05} 
\begin{equation} 
{\bar p}~+~p \rightarrow \Xi^- ~+~{\bar \Xi}^+ \,, ~~~~~~~~
{\bar p}~+~n \rightarrow \Xi^- ~+~{\bar \Xi}^0 \,, 
\label{eq:xiprod} 
\end{equation} 
occurring on a nuclear target at $p_{\rm lab} \sim 3$~GeV/c. The trigger 
for these reactions will be based on the detection of  high-momentum 
$\bar \Xi$ antihyperons at small angles or of $K^+$ mesons  produced by 
the absorption of antihyperons in the primary target nuclei. The $\Xi^-$ 
hyperons will be slowed down and captured in a secondary nuclear target. 
One expects in this way to reconstruct approximately 3000 stopped 
$\Xi^-$ hyperons  per day in PANDA. A recent simulation is found in 
Ref.~\cite{FAI07}. 
  
\end{itemize}

\subsection{$\Xi^-$ atoms} 
\label{sec:xiat}

Conventional measurements of particle energies to investigate $\Xi^-$ 
hypernuclei suffer from insufficient accuracy for providing detailed 
quantitative information on the interaction of $\Xi^-$ hyperons with nuclei.
Complementarily, the usual precision for measuring the energies of X-rays 
from transitions between levels of exotic atoms offers the possibility 
of obtaining further information. The successful observation and reasonably 
precise measurement of strong interaction effects in $\Sigma^-$ atoms, 
which had provided significant clues to the interaction of $\Sigma^-$ 
hyperons with nuclei, may serve as a guide in assessing the feasibility of 
experiments on exotic atoms of $\Xi^-$. Recall that the $\Xi^-$ and the 
$\Sigma^-$ hyperons have very similar masses and lifetimes, namely 1321.32 
vs. 1197.34 MeV, and 0.1642 vs. 0.1482 nsec, for $\Xi^-$ and $\Sigma^-$ 
respectively. Full atomic cascade calculations were performed for $\Sigma^-$ 
and $\Xi^-$ atoms \cite{BFG99} and confirmed that the processes within 
these two hadronic atoms are very similar. The remaining major differences 
are in the production reactions. Whereas relatively slow $\Sigma^-$ hyperons 
are produced by the $p(K^-,\pi^+)\Sigma^-$ stopped $K^-$ reaction, 
the $p(K^-,K^+)\Xi^-$ in-flight reaction produces relatively fast $\Xi^-$ 
hyperons, thus causing non-negligible decay losses during 
the slowing down time of the $\Xi^-$ hyperon. 
Prior to such an experiment it is necessary to optimize 
the experimental setup, which includes a hydrogen production target, 
a heavy moderator such as Pb or W, the target to be studied and the 
detectors, both for X-rays and for the detection of the outgoing 
$K^+$, which is essential in order to reduce background. 

When selecting targets for  possible experiments on $\Xi^-$ atoms, 
it must be assumed that such experiments will probably not be feasible 
on more than very few targets, and one must therefore ask whether it is 
at all likely that  useful information on the interaction of $\Xi^-$ with 
nuclei will be obtained from the resulting rather limited range of data. 
It was shown \cite{Fri98} that the main features of the interaction 
of $K^-$ and $\Sigma^-$ with nuclei, as found from analyses of all the 
available data, may in fact, be obtained by analyzing a small fraction 
of the available hadronic atom data, if the target nuclei are carefully 
selected. A key point here is to have target nuclei over as wide a range 
of the periodic table as possible. This observation suggests that 
experiments on $\Xi^-$ atoms may provide useful information.

\begin{figure}
\includegraphics[scale=0.7,angle=0]{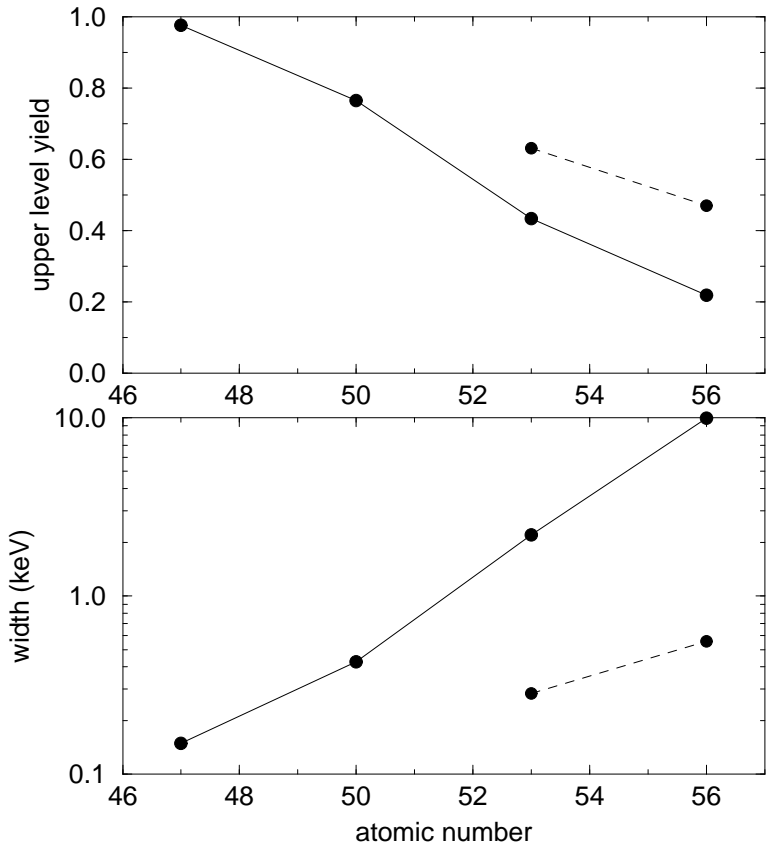}
\caption{Solid curves: calculated strong interaction widths and upper level
relative yields for the $7i$ level in medium-heavy $\Xi^-$ atoms as function
of the atomic number $Z$. The dashed curves are for
$b_0=-0.25+{\rm i}0.04$~fm, i.e. a repulsive real potential.}
\label{fig:XiSn}
\end{figure}

For estimating strong interaction effects in $\Xi^-$ atoms the $t \rho$ 
potential Eq.~(\ref{eq:Vopt}) was adopted with Re$b_0=0.25$~fm which yields 
a potential depth of about 20 MeV inside nuclei, and Im$b_0=0.04$~fm, 
yielding for the imaginary potential a depth of about 3 MeV. 
Whereas the real potential may be regarded as `typical', according to the 
above discussion, the imaginary potential is about twice as large as 
estimated \cite{YMH94} in model D. Reducing the imaginary potential will 
only cause the calculated widths of the states to decrease by roughly the 
same proportion, and the relative yields (see below) of transitions to 
become larger. This will not, however, change the last observed atomic level. 
In choosing criteria for the suitability of a transition as a source of 
information on the $\Xi$ nucleus interaction, one is guided by experience 
with other hadronic atoms \cite{BFG97} and select X-ray transitions 
$(n+1,l+1) \rightarrow (n,l)$ between circular atomic states $(n=l+1)$ 
with energies greater than 100 keV, where the strong interaction shift for 
the `last' $(n,l)$ level is at least 0.5 keV and the width less than about 
10 keV. The `upper' level relative yield, defined as the ratio of the 
intensity of the $(n+1,l+1) \rightarrow (n,l)$ X-ray transition to the 
summed intensity of all X-ray transitions feeding the $(n+1,l+1)$ state, 
is also required to be at least 10\%.

Strong interaction shifts and widths of $\Xi^-$ atomic levels have been 
calculated using the above optical potential for a large number of nuclei. 
As the overlap of atomic wavefunctions with nuclei vary smoothly with charge 
number, it is to be expected that generally shifts, widths and yields will 
vary smoothly along the periodic table. Figure~\ref{fig:XiSn} shows 
calculated widths and `upper' level relative yields for the $7i$ state in 
medium-heavy $\Xi^-$ atoms and it is seen that a suitable target may be found 
near Sn or I. The dashed lines in this figure are obtained by reversing the 
sign of the real potential used for calculating the solid curves. 
It is seen that in such a case the range of suitable targets will move to 
between I and  Ba, where the strong interaction width and relative yield 
are more acceptable. The sign of the strong interaction shift will be 
reversed in this case, but it has no experimental consequences. 
This exemplifies a general property of hadronic atoms, which are dominated 
by the Coulomb interaction, namely, that large variations in the strong 
interaction potential will move the proposed targets only a few units of 
charge along the periodic table. 

\begin{figure}
\includegraphics[scale=0.7,angle=0]{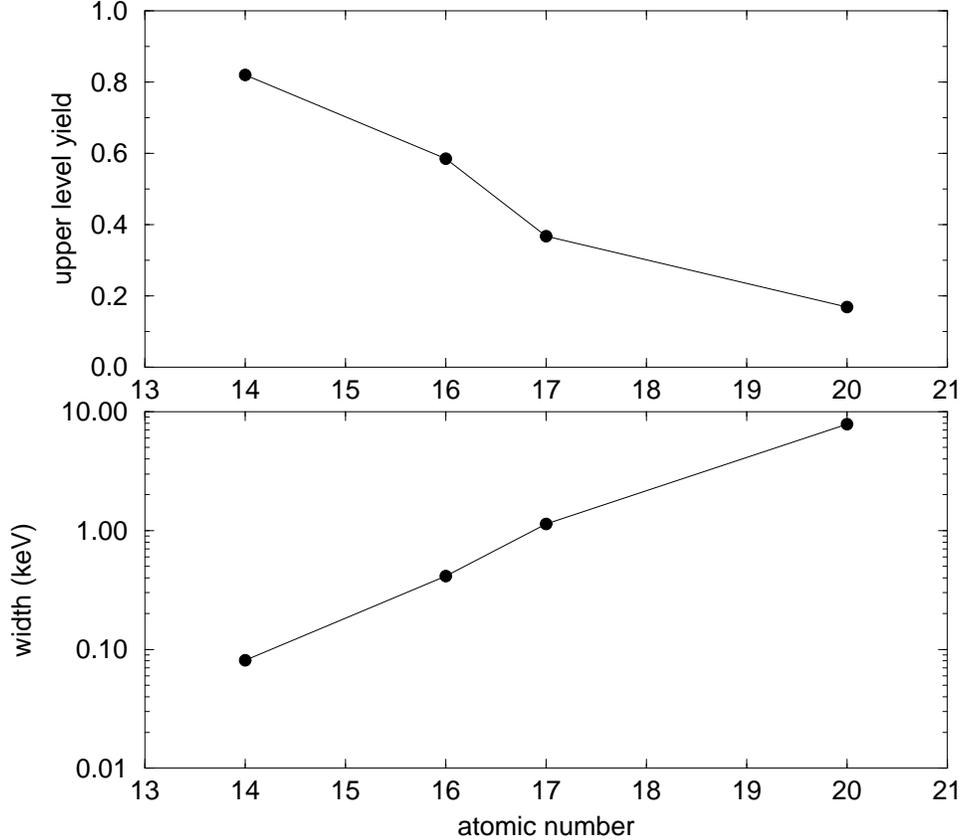}
\caption{Calculated strong interaction widths and upper level relative
yields for the $4f$ level in $\Xi^-$ atoms.}
\label{fig:XiCl}
\end{figure}

\begin{figure}
\includegraphics[scale=0.7,angle=0]{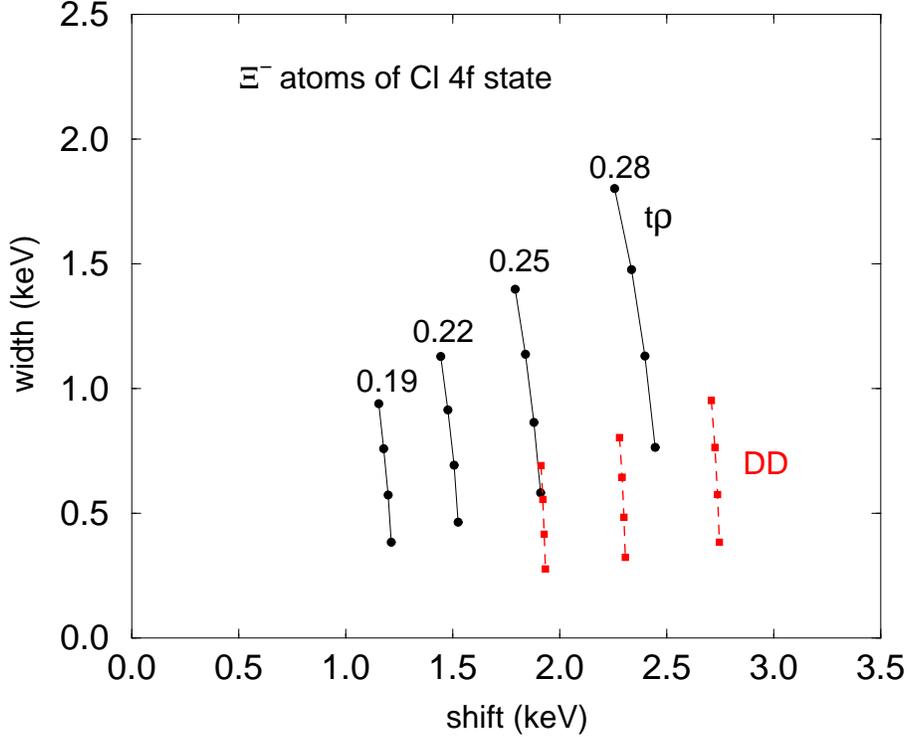}
\caption{Calculated strong interaction shifts and widths for the $4f$
level in $\Xi^-$ atoms of Cl for different optical potentials, see text
for detail.}
\label{fig:XiClDD}
\end{figure}

Figure \ref{fig:XiCl} shows results for the $4f$ state of $\Xi^-$ atoms, 
where it is seen that for a Si target the effects could be too small to 
measure, whereas for Ca the width could be too large and the relative yield 
too small. In this region a Cl target may be appropriate, perhaps in the 
form of the liquid CCl$_4$. More detailed results for Cl are shown in 
Fig.~\ref{fig:XiClDD} where the sensitivities to assumptions regarding the 
optical potential are also typical of results for other targets. 
The solid curves connect points obtained within the $t\rho$ potential 
for fixed values of Re$b_0$, listed above the lines. The four points 
along each line correspond to values of Im$b_0$ from 0.05 fm down to 
0.02 fm in steps of 0.01 fm. Departures from this $t\rho$ potential are 
represented by the dashed lines, calculated from phenomenological density 
dependent (DD) real potentials similar to those found from analyses of 
experimental results for $K^-$ and $\Sigma^-$ atoms, as discussed in 
Secs.~\ref{sec:kbars} and \ref{sec:sigma}, respectively. The imaginary part 
of the potential is of the $t\rho$ type and the points along the dotted 
lines correspond to the same values of Im$b_0$ as above. The real potentials 
in these calculations are similar to the real potential for $\Sigma^-$ atoms, 
having an attractive pocket about 5-10 MeV deep outside the nuclear surface, 
with a repulsive potential of about 20-30 MeV in the nuclear interior. 
The results in the figure serve only to illustrate the expected range 
of strong interaction effects. If the actual values of shift and width 
turn out to be within the area covered by the lines, these effects will 
most likely be measurable. 

\begin{table}
\caption{Predictions for likely targets for a $\Xi^-$ atoms experiment.
Calculations are based on a $t\rho$ potential with $b_0=0.25+{\rm i}0.04$ fm.
E$_x$ is the transition energy, $Y$ is the upper level relative yield.}
\label{tab:preds}
\begin{tabular}{lccccc}
\hline
target & F & Cl & Sn & I & Pb \\  \hline
transition &~~ $4f\rightarrow3d$~~ &~~ $5g\rightarrow4f$~~&~~
$8j\rightarrow7i$~~&
         ~~  $8j\rightarrow7i$~~ &~~ $10l\rightarrow9k$~~ \\
E$_x$ (keV) & 131.29 & 223.55 & 420.25 & 474.71 & 558.47 \\
 $Y$ & 0.31 & 0.37 & 0.76 & 0.43 & 0.58 \\
shift (keV) & 1.56 & 1.84 & 0.67 & 2.79 & 1.73 \\
width (keV) & 0.99 & 1.14 & 0.43 & 2.21 & 1.26 \\ \hline
\end{tabular}
\end{table}

Table \ref{tab:preds} summarizes results for possible targets for $\Xi^-$ 
atom experiments~\cite{BFG99}. It should be kept in mind that due to the 
discrete nature of quantum numbers it is not always possible to `fine-tune' 
one's choice of a target, considering widths and yields, in spite of their 
smooth variation with $Z$. 
Furthermore, as mentioned above, the main difficulty is likely to be 
associated with the efficient slowing down of $\Xi^-$ hyperons.

\begin{acknowledgments} 
We would like to acknowledge and thank our long-time collaborators 
Chris Batty and Ji\v{r}\'{\i} Mare\v{s} for making significant contributions 
to the present Review, and Wolfram Weise for stimulating discussions 
and valuable criticism in recent years. We are pleased to acknowledge 
the helpful advice and useful communications obtained on various related 
topics from Tullio Bressani, Bob Chrien, Toru Harada and Paul Kienle. 
Special thanks go to Elisabeth Friedman for her dedicated proofreading 
of the manuscript. This work was supported in part by the Israel Science 
Foundation, Jerusalem, grant 757/05. 
\end{acknowledgments} 


\end{document}